\documentclass[a4paper,10pt]{article}
\pdfoutput=1

\usepackage{graphics,epsfig}
\usepackage{xspace,colortbl}
\usepackage{cite} 
\usepackage{hyperref} 

\newcommand{\bea}{\begin{eqnarray}}
\newcommand{\eea}{\end{eqnarray}}
\newcommand{\beq}{\begin{equation}}
\newcommand{\eeq}{\end{equation}}

\newcommand{\mydelta}{\mathbf{\Delta}}
\def\slash#1{\mbox{$\not\!\! #1$}}

\def\simge{\mathrel{\rlap{\raise 0.511ex \hbox{$>$}}{\lower 0.511ex
 \hbox{$\sim$}}}}
\def\simle{\mathrel{\rlap{\raise 0.511ex \hbox{$<$}}{\lower 0.511ex
 \hbox{$\sim$}}}}
\def\slash#1{\setbox0=\hbox{$#1$}\dimen0=\wd0 \setbox1=\hbox{/} \dimen1=\wd1
 \ifdim\dimen0>\dimen1 \rlap{\hbox to \dimen0{\hfil/\hfil}} #1
 \else \rlap{\hbox to \dimen1{\hfil$#1$\hfil}} / \fi}

\def\rmii{a}
\def\infntv{b}
\def\rmiii{c}
\def\infntre{d}
\def\rmi{e}
\def\rmiff{f}
\def\lpt{g}

\newcommand{\sla}[1]%
        {\kern .25em\raise.18ex\hbox{$/$}\kern-.6em #1}

\newcommand{\gol}{\raisebox{-0.0\totalheight}{\includegraphics[scale=.4]{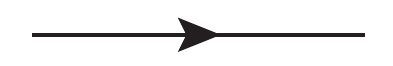}}}
\newcommand{\goi}{\raisebox{-0.3\totalheight}{\includegraphics[scale=.4]{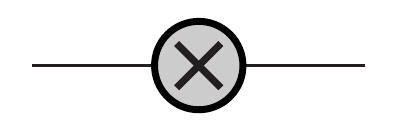}}}
\newcommand{\goip}{\raisebox{-0.3\totalheight}{\includegraphics[scale=.4]{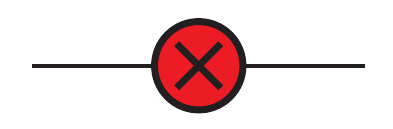}}}

\newcommand{\golvert}{\raisebox{-0.0\totalheight}{\includegraphics[scale=.4]{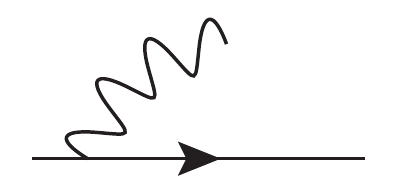}}}
\newcommand{\golself}{\raisebox{-0.0\totalheight}{\includegraphics[scale=.4]{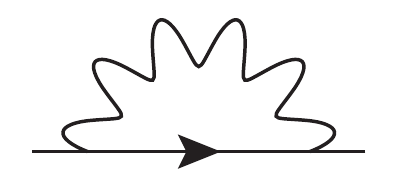}}}
\newcommand{\goidisc}{\raisebox{-0.0\totalheight}{\includegraphics[scale=.4]{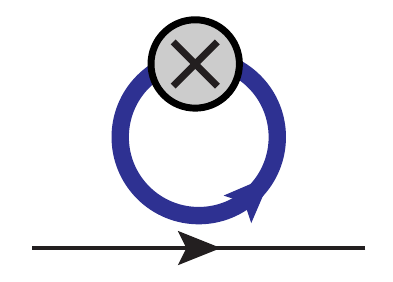}}}
\newcommand{\goipdisc}{\raisebox{-0.0\totalheight}{\includegraphics[scale=.4]{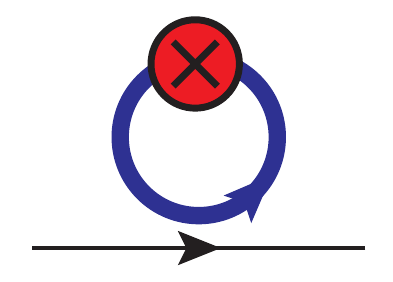}}}
\newcommand{\goiplaq}{\raisebox{-0.0\totalheight}{\includegraphics[scale=.4]{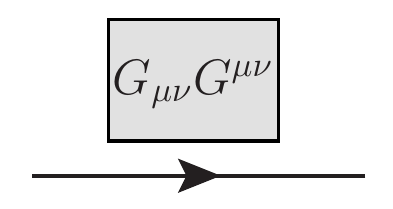}}}

\newcommand{\goltad}{\raisebox{-0.0\totalheight}{\includegraphics[scale=.4]{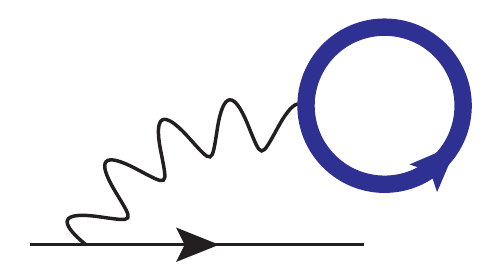}}}
\newcommand{\golltad}{\raisebox{-0.0\totalheight}{\includegraphics[scale=.4]{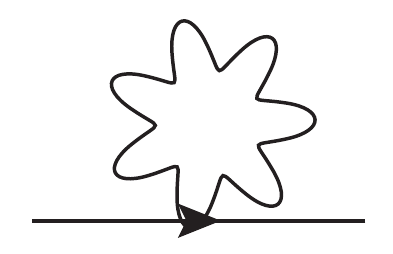}}}
\newcommand{\golvp}{\raisebox{-0.0\totalheight}{\includegraphics[scale=.4]{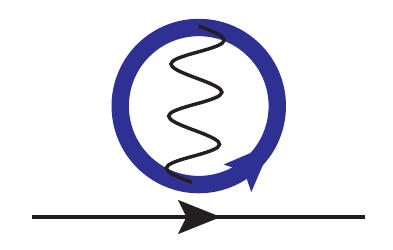}}}
\newcommand{\golvpp}{\raisebox{-0.0\totalheight}{\includegraphics[scale=.4]{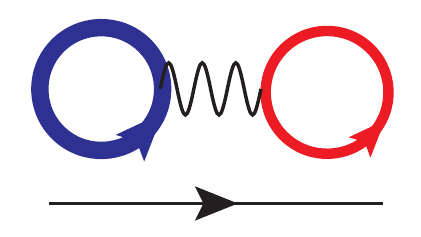}}}
\newcommand{\golvppp}{\raisebox{-0.0\totalheight}{\includegraphics[scale=.4]{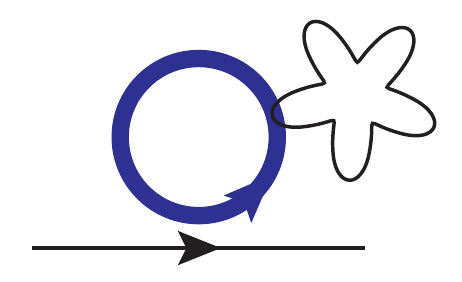}}}

\newcommand{\plaq}{\raisebox{-0.3\totalheight}{\includegraphics[scale=.45]{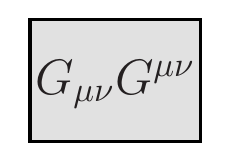}}}
\newcommand{\vpz}{\raisebox{-0.3\totalheight}{\includegraphics[scale=.4]{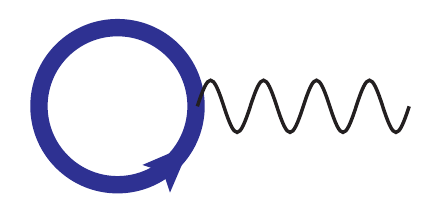}}}
\newcommand{\vp}{\raisebox{-0.3\totalheight}{\includegraphics[scale=.4]{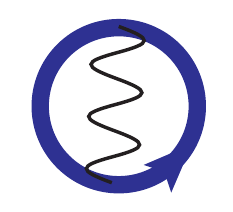}}}
\newcommand{\vpp}{\raisebox{-0.3\totalheight}{\includegraphics[scale=.4]{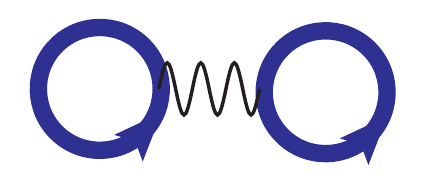}}}
\newcommand{\vppp}{\raisebox{-0.3\totalheight}{\includegraphics[scale=.4]{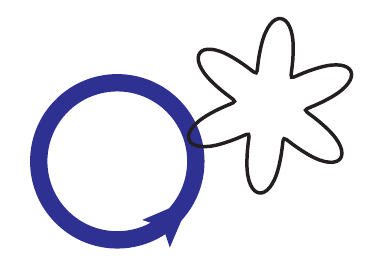}}}

\newcommand{\gdll}{\raisebox{-0.4\totalheight}{\includegraphics[scale=.3]{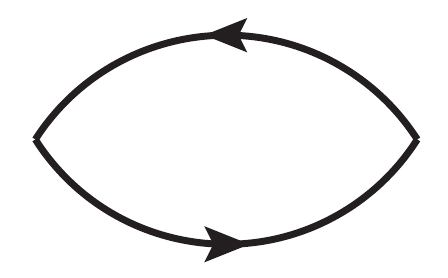}}}
\newcommand{\gdsi}{\raisebox{-0.4\totalheight}{\includegraphics[scale=.3]{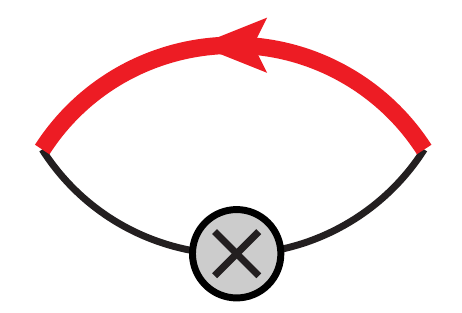}}}
\newcommand{\gdsip}{\raisebox{-0.4\totalheight}{\includegraphics[scale=.3]{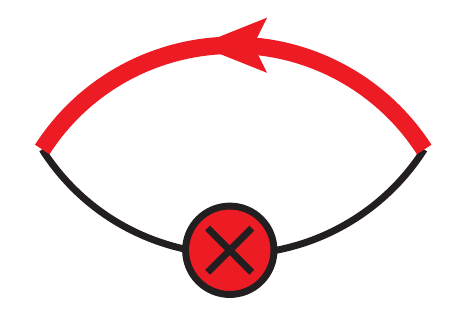}}}
\newcommand{\gdsil}{\raisebox{-0.4\totalheight}{\includegraphics[scale=.3]{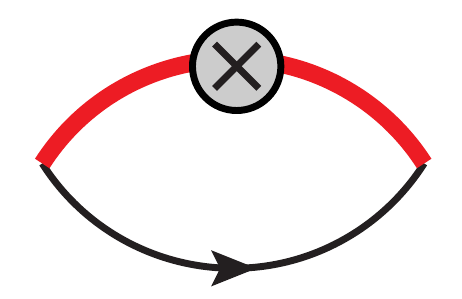}}}
\newcommand{\gdsipl}{\raisebox{-0.4\totalheight}{\includegraphics[scale=.3]{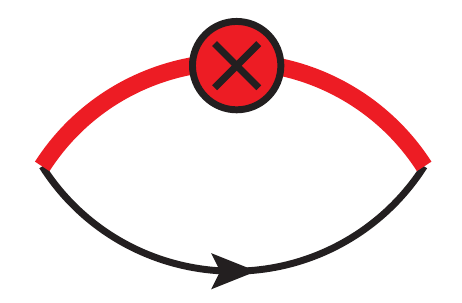}}}

\newcommand{\gdli}{\raisebox{-0.4\totalheight}{\includegraphics[scale=0.3]{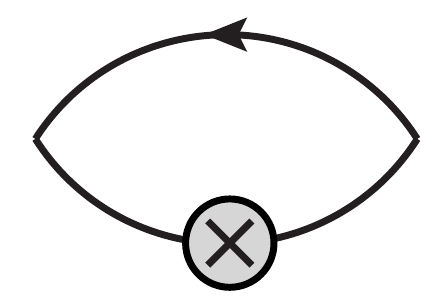}}}
\newcommand{\gdlip}{\raisebox{-0.4\totalheight}{\includegraphics[scale=0.3]{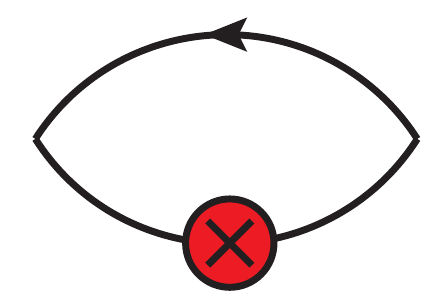}}}

\newcommand{\gdsl}{\raisebox{-0.4\totalheight}{\includegraphics[scale=.3]{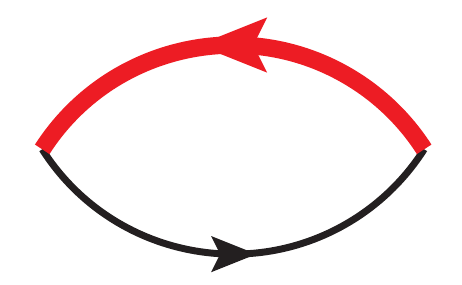}}}

\newcommand{\gdslselfs}{\raisebox{-0.22\totalheight}{\includegraphics[scale=.3]{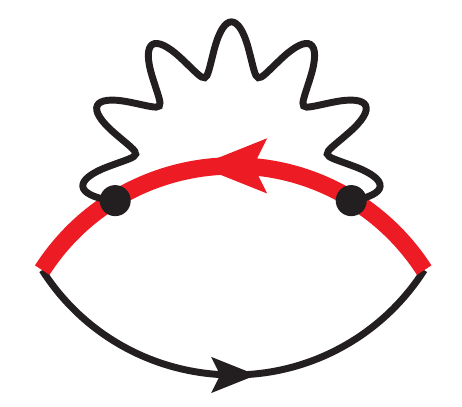}}}
\newcommand{\gdslselfl}{\raisebox{-0.6\totalheight}{\includegraphics[scale=.3]{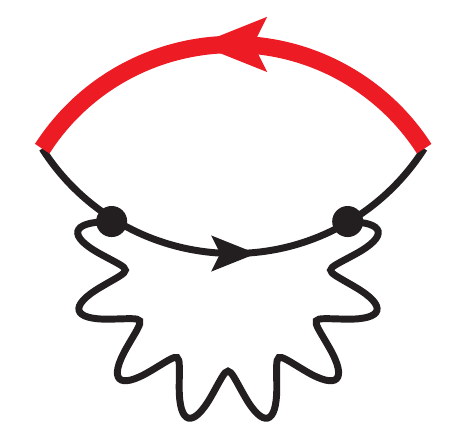}}}
\newcommand{\gdslexch}{\raisebox{-0.4\totalheight}{\includegraphics[scale=.3]{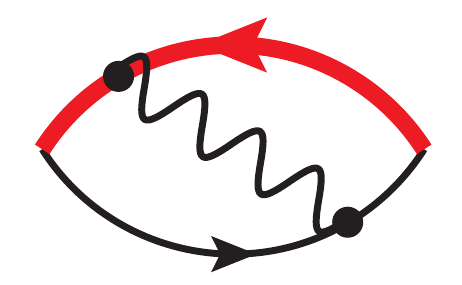}}}

\newcommand{\gdllself}{\raisebox{-0.4\totalheight}{\includegraphics[scale=.3]{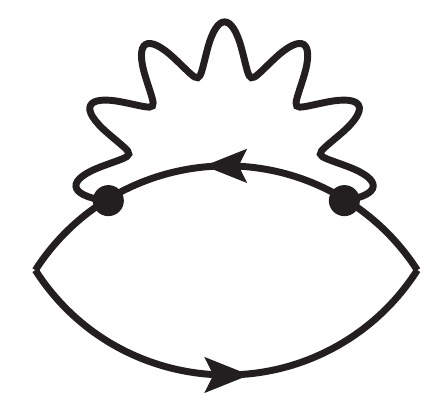}}}
\newcommand{\gdllexch}{\raisebox{-0.4\totalheight}{\includegraphics[scale=.3]{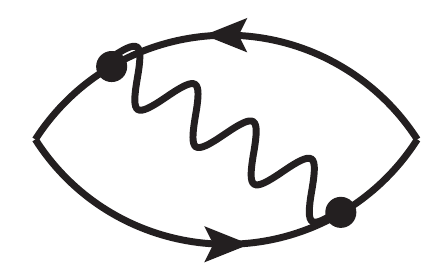}}}
\newcommand{\discgdllexch}{\raisebox{-0.2\totalheight}{\includegraphics[scale=0.4]{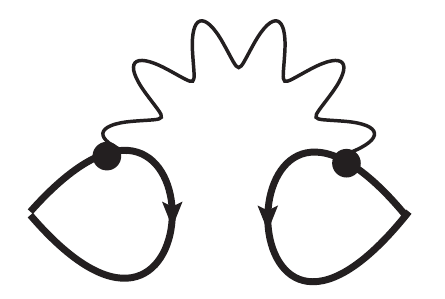}}}

\newcommand{\gdlltadf}{\raisebox{-0.4\totalheight}{\includegraphics[scale=.3]{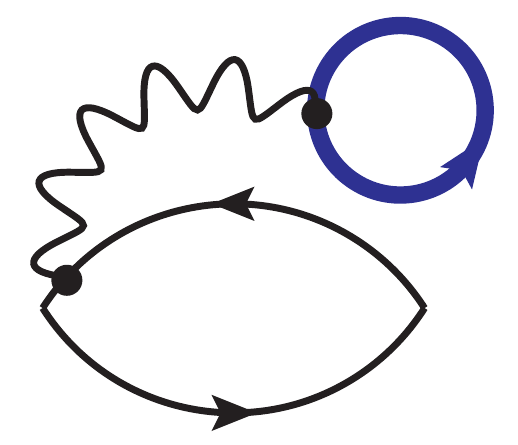}}}
\newcommand{\gdllphtad}{\raisebox{-0.3\totalheight}{\includegraphics[scale=.3]{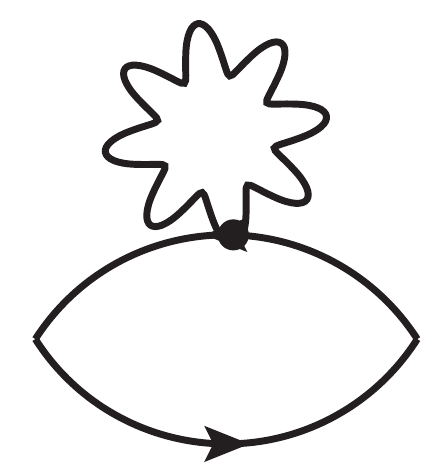}}}

\newcommand{\gvdll}{\raisebox{-0.4\totalheight}{\includegraphics[scale=.3]{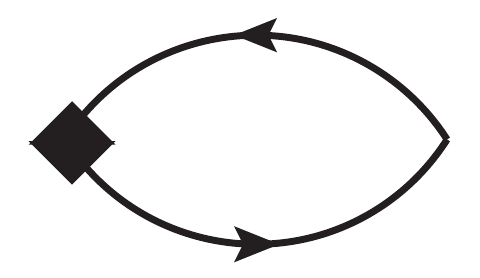}}}
\newcommand{\gvdlip}{\raisebox{-0.4\totalheight}{\includegraphics[scale=0.3]{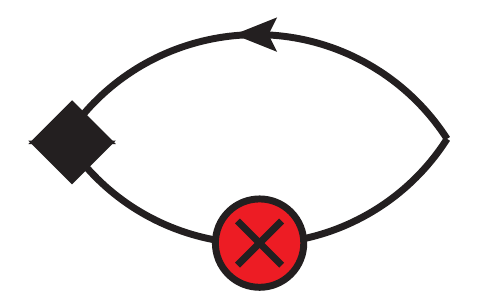}}}
\newcommand{\gvdllself}{\raisebox{-0.3\totalheight}{\includegraphics[scale=.3]{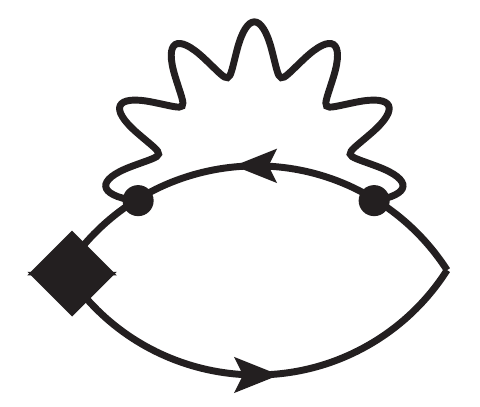}}}
\newcommand{\gvdllexch}{\raisebox{-0.4\totalheight}{\includegraphics[scale=.3]{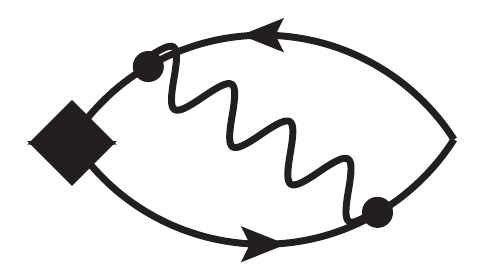}}}
\newcommand{\gvdllphtad}{\raisebox{-0.25\totalheight}{\includegraphics[scale=.3]{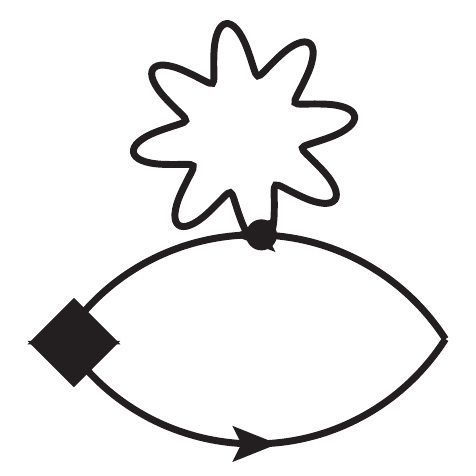}}}

\newcommand{\gdsltadf}{\raisebox{-0.4\totalheight}{\includegraphics[scale=.3]{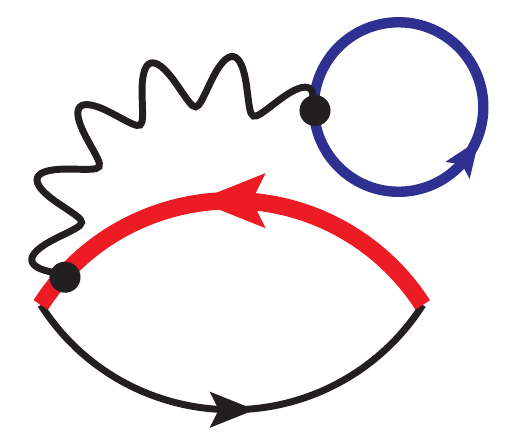}}}

\newcommand{\gdslltadf}{\raisebox{-0.4\totalheight}{\includegraphics[scale=.3]{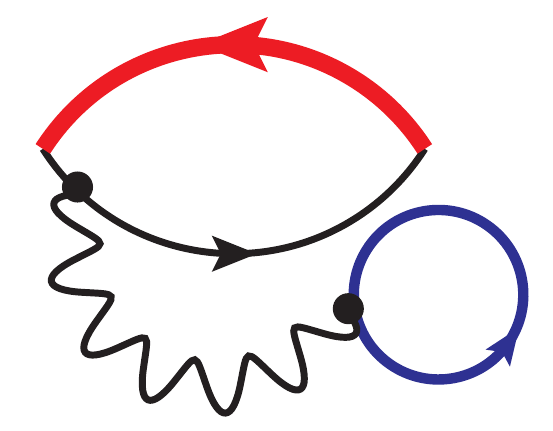}}}
\newcommand{\gdslphtads}{\raisebox{-0.2\totalheight}{\includegraphics[scale=.3]{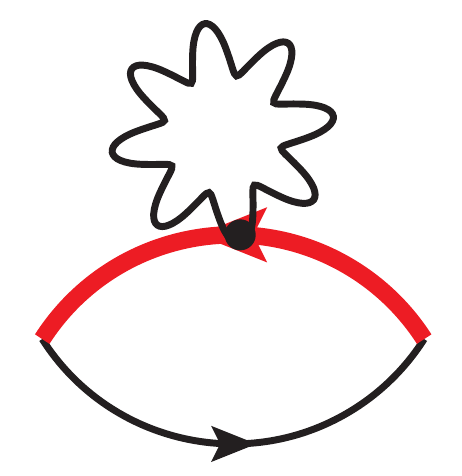}}}
\newcommand{\gdslphtadl}{\raisebox{-0.6\totalheight}{\includegraphics[scale=.3]{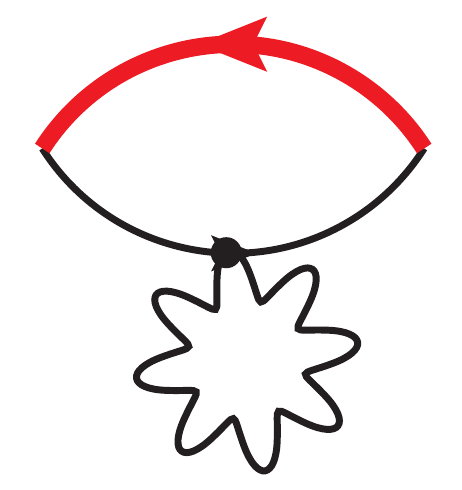}}}

\newcommand{\bear}[1]{\begin{equation}\begin{array}{#1}}
\newcommand{\eear}{ \end{array}\end{equation}}
\newcommand{\barr}[1]{\begin{array}{#1}}
\newcommand{\earr}{\end{array}}

\newcommand{\ket}[1]{\ensuremath{| {#1} \rangle }}
\newcommand{\bra}[1]{\ensuremath{\langle {#1} |}}

\usepackage[title,titletoc]{appendix}

\newcommand{\deltampi}{1.44(13)(16) \times 10^{3} \ \mbox{MeV}^2}
\newcommand{\epsilongamma}{0.79(18)(18)}
\newcommand{\deltamkQED}{2.26(23)(23) \times 10^{3} \ \mbox{MeV}^2}
\newcommand{\deltamkQCD}{-6.16(23)(23) \times 10^{3} \ \mbox{MeV}^2}

\newcommand{\deltamq}{2.39(8)(17)~\mbox{MeV}}
\newcommand{\rmq}{0.50(2)(3)}
\newcommand{\Rsu}{38(2)(3)}
\newcommand{\Qsu}{23(1)(1)}
\newcommand{\mqu}{2.40(15)(17)~\mbox{MeV}}
\newcommand{\mqd}{4.80(15)(17)~\mbox{MeV}}

\newcommand{\deltaf}{-0.0040(3)(2)}
\newcommand{\deltamp}{2.9(6)(2)~MeV}

\begin{document}

\begin{titlepage}

\vspace{-1cm} 
\begin{flushright}\footnotesize
RM3-TH/13-2\\
ROM2F/2013/01\\
\end{flushright}
\vspace{1cm}

\begin{center}
{\LARGE\sc leading isospin breaking effects on the lattice} \\
\end{center}

\vspace{2cm}
\baselineskip 20pt plus 2pt minus 2pt

\begin{center}
{\sc \Large RM123 Collaboration}\\
\vspace{0.5cm}
{\it
G.M.~de Divitiis$^{(\rmii,\infntv)}$, 
R.~Frezzotti$^{(\rmii,\infntv)}$,
V.~Lubicz$^{(\rmiii,\infntre)}$,
G.~Martinelli$^{(\rmi,\rmiff)}$,
R.~Petronzio$^{(\rmii,\infntv)}$,\\   
G.C.~Rossi$^{(\rmii,\infntv)}$, 
F.~Sanfilippo$^{(\lpt,\rmiff)}$,
S.~Simula$^{(\infntre)}$, 
N.~Tantalo$^{(\rmii,\infntv)}$
}
\end{center}

\begin{center}
\begin{footnotesize}
\noindent \vspace{0.2cm}

$^{(\rmii)}$ Dip. di Fisica, Universit{\`a} di Roma ``Tor Vergata", 
Via della Ricerca Scientifica 1, I-00133 Rome, Italy

\vspace{-6pt}
$^{(\infntv)}$ INFN, Sez. di Roma ``Tor Vergata",
Via della Ricerca Scientifica 1, I-00133 Rome, Italy

\vspace{-6pt}
$^{(\rmiii)}$ Dip. di Matematica e Fisica, Universit{\`a} Roma Tre, Via della Vasca Navale
84, I-00146 Rome, Italy

\vspace{-6pt}
$^{(\infntre)}$ INFN, Sez. di Roma Tre, Via della Vasca Navale 84, I-00146 Rome,
Italy

\vspace{-6pt}
$^{(\rmi)}$ SISSA - Via Bonomea 265 - 34136, Trieste - Italy 

\vspace{-6pt}
$^{(\rmiff)}$  INFN, Sezione di Roma, P.le A.~Moro 5, I-00185 Rome, Italy

\vspace{-6pt}
$^{(\lpt)}$ Laboratoire de Physique Th\'eorique (B\^at. 210),
   Universit\'e Paris Sud, F-91405 Orasay-Cedex, France

\end{footnotesize}
\end{center}

\vspace{1.0cm}

\begin{abstract}
We present a method to evaluate on the lattice the leading isospin breaking effects due to both the small mass difference between the up and down quarks and the QED interaction. Our proposal is applicable in principle to any QCD+QED gauge invariant hadronic observable which can be computed on the lattice. It is based on the expansion of the path--integral in powers of the small parameters $(\hat m_d - \hat m_u)/\Lambda_{QCD}$ and $\hat \alpha_{em}$, where $\hat m_f$ is the renormalized quark mass and $\hat \alpha_{em}$ the renormalized fine structure constant. In this paper we discuss in detail the general strategy of the method and the conventional, although arbitrary, separation of QCD from QED isospin breaking corrections. We obtain results for the pion mass splitting,  $M_{\pi^+}^2-M_{\pi^0}^2=\deltampi$, for the Dashen's theorem breaking parameter $\varepsilon_{\gamma}=\epsilongamma$, for the light quark masses, $[\hat m_d - \hat m_u](\overline{MS},2\mbox{ GeV})= \deltamq$, $[\hat m_u/\hat m_d](\overline{MS},2\mbox{ GeV})=\rmq$ and for the flavour symmetry breaking parameters $R$ and $Q$. We also update our previous results for the QCD isospin breaking corrections to the $K_{\ell 2}$ decay rate and for the QCD contribution to the neutron--proton mass splitting.
\end{abstract}

\end{titlepage}

\setlength\abovedisplayskip{22pt plus 3pt minus 7pt}
\setlength\belowdisplayskip{22pt plus 3pt minus 7pt}

\section{Introduction}
\label{sec:intro}
One of the primary goals of lattice QCD is to calculate non--perturbatively hadronic observables at the level of accuracy required for phenomenological applications. In the flavour physics sector, for instance, the combined efforts of the lattice QCD community resulted in calculations of quantities such as the $K_{\ell 2}$ and $K_{\ell 3}$ decay rates with relative overall uncertainties of the order of half a percent (see ref.~\cite{Colangelo:2010et} for a recent review). These results have been obtained, in most of the cases, within the isosymmetric theory, i.e. by neglecting the difference of the up and down quark masses together with the QED interaction and by taking into account the corresponding effects by relying on chiral perturbation theory or on model--dependent approximations. At the level of precision presently achieved for some flavour physics observables isospin breaking effects cannot be neglected any longer. For example, by neglecting the pion mass difference ($3\%$) and the kaon mass difference ($1\%$) a systematic error is unavoidably introduced on the corresponding determination of the $K_{\ell 3}$ decay rate or on any dimensional quantity if these masses are used to calibrate the lattice.

In ref.~\cite{deDivitiis:2011eh} we provided a method to calculate the leading QCD isospin breaking effects, i.e. the ones associated with the difference of the up and down quark masses, and we checked the validity of the proposed procedure by computing the kaon and nucleon mass difference, the difference of the decay constants ratio $F_{K^+}/F_{\pi^+}$ with respect to the value of the isosymmetric theory and estimated QCD isospin breaking effects on the $K_{\ell 3}$ decay rate. The results just mentioned were obtained by relying on the estimates of QED isospin breaking effects, often based on model dependent approximations, provided by other groups. With the purpose of removing this approximation we briefly discussed in ref.~\cite{deDivitiis:2011eh} how order $\hat \alpha_{em}$\footnote{Through all the paper we indicate the renormalized couplings with an ``hat", for example $\hat \alpha_{em}$, to distinguish them from the corresponding bare quantities, for example $\alpha_{em}$.} QED corrections can be calculated on the lattice.

In this paper we develop a method to calculate leading isospin breaking effects on the lattice by including those associated with QED interactions. These are tiny because very small factors, $(\hat m_d - \hat m_u)/\Lambda_{QCD}$ and $\hat \alpha_{em}$, multiply sizable matrix elements of hadronic operators. Our approach consists in a combined expansion of the lattice path--integral in powers of $\hat m_d -  \hat m_u$ and $\hat \alpha_{em}$. We consider the two expansion parameters of the same order of magnitude, $(\hat m_d - \hat m_u)/\Lambda_{QCD} \sim \hat \alpha_{em} \sim \varepsilon$, and neglect in this work terms of $O(\varepsilon^2)$. In this sense we talk of ``leading isospin breaking'' (LIB) effects. A great advantage of our method with respect to other approaches (see for example refs.~\cite{Duncan:1996xy,Basak:2013iw,Blum:2010ym,Portelli:2012pn,Ishikawa:2012ix}) is that, by working at fixed order in a perturbative expansion, we are able to factorize the small coefficients and to get relatively large numerical signals. For the same reason, we do not need to perform simulations at unphysical values of the electric charge, thus avoiding extrapolations of the lattice data with respect to $\hat \alpha_{em}$.

The expansion of the lattice path--integral in powers of $\hat \alpha_{em}$ leads to correlators containing the integral over the whole space--time lattice volume of two insertions of the quark electromagnetic currents multiplied by the lattice photon propagator. These quantities have both infrared and ultraviolet divergences that must be removed by providing an infrared safe finite volume definition of the lattice photon propagator and by imposing suitable renormalization conditions. In this paper we discuss in detail these issues.

The main results of the paper are
\begin{eqnarray}
&&M_{\pi^+}^2-M_{\pi^0}^2=\deltampi \;,
\nonumber \\
\nonumber \\
&&\left[M_{K^+}^2-M_{K^0}^2\right]^{QED}=\deltamkQED \;,
\nonumber \\
\nonumber \\
&&\left[M_{K^+}^2-M_{K^0}^2\right]^{QCD}=\deltamkQCD \;,
\nonumber \\
\nonumber \\
\varepsilon_\gamma=
&&\frac{\left[M_{K^+}^2-M_{K^0}^2\right]^{QED}-\left[M_{\pi^+}^2-M_{\pi^0}^2\right]^{QED}}{M_{\pi^+}^2-M_{\pi^0}^2}
=\epsilongamma\;,
\nonumber \\
\nonumber \\
&&\left[\hat m_d - \hat m_u\right](\overline{MS},2\mbox{ GeV})= \deltamq\; ,
\nonumber \\
\nonumber \\
&&\frac{\hat m_u}{\hat m_d}(\overline{MS},2\mbox{ GeV}) = \rmq \;,
\nonumber \\
\nonumber \\
&&\hat m_u(\overline{MS},2\mbox{ GeV}) = \mqu \;,
\nonumber \\
\nonumber \\
&&\hat m_d(\overline{MS},2\mbox{ GeV}) = \mqd \;,
\nonumber \\
\nonumber \\
&&R(\overline{MS},2\mbox{ GeV})=\left[\frac{\hat m_s -\hat m_{ud}}{\hat m_d - \hat m_u}\right](\overline{MS},2\mbox{ GeV})
 = \Rsu \;,
\nonumber \\
\nonumber \\
&&Q(\overline{MS},2\mbox{ GeV})=\left[\sqrt{\frac{\hat m_s^2 -\hat m_{ud}^2}{\hat m_d^2 - \hat m_u^2}}\right](\overline{MS},2\mbox{ GeV})
 = \Qsu \;,
\nonumber \\
\nonumber \\
&&\left[ \frac{F_{K^+}/F_{\pi^+}}{F_K/F_\pi}-1 \right]^{QCD}=
\deltaf \; ,
\nonumber \\
\nonumber \\
&&\left[ M_n - M_p \right]^{QCD} = \deltamp \;.
\nonumber
\end{eqnarray}
and have been obtained in the $n_f=2$ theory. The numbers in the first parentheses correspond to the statistical errors while those in the second parentheses are the systematic errors, mainly due to chiral, continuum and infinite volume extrapolations. The results for the quark masses, $F_{K^+}/F_{\pi^+}$ and the neutron--proton mass splitting are an update of our previous results obtained for these quantities in ref.~\cite{deDivitiis:2011eh}. Note that, because of the QED interactions, ratios of quark masses of different electric charges are renormalization scheme and scale dependent.

At first order in $\hat m_d- \hat m_u$ and $\hat \alpha_{em}$ the pion mass difference is neither affected by QCD isospin breaking corrections nor by electromagnetic isospin breaking effects coming from the dynamical sea quarks. For this reason the result for $M_{\pi^+}^2-M_{\pi^0}^2$ is a particularly clean theoretical prediction, though our results were obtained by neglecting a quark disconnected contribution to $M_{\pi^0}$ of $O(\hat \alpha_{em} \hat m_{ud})$, see eq.~(\ref{eq:pionmasses}). From the phenomenological point of view this contribution is expected to be very small, i.e. of the same order of magnitude of the other $O(\hat \alpha_{em} [\hat m_{d}-\hat m_{u}])$ contributions neglected in this paper.

The kaon mass splitting includes both strong and electromagnetic isospin breaking effects. With our method these can be conveniently separated by implementing the renormalization prescription discussed in detail in section~\ref{sec:separating} (see also ref.~\cite{Gasser:2003hk}) and the notation $\mathcal{O}^{QED,QCD}$ also means that the corresponding numerical result depends upon this prescription. The results for $M_{K^+}^2-M_{K^0}^2$, together with those for the Dashen's theorem breaking parameter $\varepsilon_\gamma$ and for the light quark masses have been obtained within the electro--quenched approximation, i.e. by considering dynamical sea quarks as neutral with respect to electromagnetism. 

The paper is organized as follows: in section~\ref{sec:dependence} we discuss the general aspects of our method by assuming that the regulated theory retains all the symmetries of the continuum theory. In section~\ref{sec:noncompact} we provide an infrared safe definition of the lattice photon propagator and discuss a convenient stochastic method to calculate electromagnetic corrections to lattice correlators. In section~\ref{sec:wilsonfermions} we enter into the details associated with the regularization of the fermion action used in this work. In particular we discuss the issue of the determination of the electromagnetic contributions to the critical masses of Wilson quarks. In section~\ref{sec:pathintegral} we discuss all the details needed in order to derive the isospin breaking corrections for a given correlator and obtain explicit results for the pion and kaon two--point functions. In section~\ref{sec:pionmasses} we show our results for the pion mass difference. In section~\ref{sec:tuningk} we discuss the numerical determination of the electromagnetic critical masses of the quarks. In section~\ref{sec:separating} we discuss the separation of QED from QCD isospin breaking corrections to the kaon mass difference and in section~\ref{sec:extrapolations} we discuss chiral, continuum and infinite volume extrapolations of our lattice data. We draw our conclusions in section~\ref{sec:theend}.

\section{Electromagnetic corrections to hadronic observables}
\label{sec:dependence}
In this section we illustrate the general strategy underlying the method that we have devised in order to calculate LIB corrections to hadronic observables.
The method presented here is a generalization of the one presented in ref.~\cite{deDivitiis:2011eh} and is based on a combined perturbative expansion of the full theory\footnote{We call ``full'' the theory with both QCD and QED interactions switched on and (consequently) with $\hat m_d \neq \hat m_u$ while we call ``isosymmetric QCD'' or simply ``isosymmetric'' the theory without electromagnetic interactions and with $\hat m_d=\hat m_u$.} lattice path--integral in the small parameters $\hat \alpha_{em}$ and $(\hat m_d-\hat m_u)/\Lambda_{QCD}$. The two parameters are considered of the same order of magnitude,
\begin{eqnarray}
\frac{\hat m_d-\hat m_u}{\Lambda_{QCD}} \quad \sim\quad \hat \alpha_{em} 
\quad \sim \quad O(\varepsilon) \;,
\end{eqnarray}
and contributions of $O(\varepsilon^2)$ are neglected in the present paper. By using this method it is possible to calculate LIB corrections by starting from gauge configurations generated with the isosymmetric QCD action. 
All the details associated with the lattice regularization used in this work will be given in the coming sections together with the formulae necessary to compute LIB corrections to specific observables.

In order to calculate $O(\hat \alpha_{em})$ corrections to a given physical quantity we have to cope with correlators containing two insertions of the electromagnetic current multiplied by the photon propagator and integrated over the space--time volume. More precisely, the correction to a given correlator is proportional to 
\begin{eqnarray}
T \langle \mathcal{O}(x_i) \rangle
&\longrightarrow&
T\int d^4y d^4z\ D_{\mu\nu}(y-z)\; \langle \mathcal{O}(x_i)
J^\mu(y) J^\nu(z) 
\rangle  \; ,
\label{eq:schematiccorrection}
\end{eqnarray}
where $T\langle \mathcal{O}(x_i) \rangle$ is the $T$--product of a certain number of local operators, $D_{\mu\nu}(y-z)$ is the photon propagator in a fixed QED gauge and $J^\mu(x)$ is the sum of the electromagnetic currents of all the flavours. There are two important issues that have to be addressed in order to give a physical meaning to the previous expression. The first, the ``infrared" problem, concerns a proper definition of the finite volume lattice photon propagator. In section~\ref{sec:noncompact} we provide a solution to this problem by discussing in detail how the convolution integrals appearing into eq.~(\ref{eq:schematiccorrection}) can be calculated numerically.
In the remaining part of this section we illustrate the ``ultraviolet" problem associated with eq.~(\ref{eq:schematiccorrection}), i.e. the appearance of divergent contributions generated by the contact interactions of the electromagnetic currents. 
The problem is illustrated here in continuum--like notation, i.e. by assuming the existence of a non-perturbative regularization that retains all the symmetries of the continuum action (think for example of Overlap lattice Dirac operators), whereas we shall enter into the details specific to the lattice regularization used in this paper in the next sections. In particular, in section~\ref{sec:wilsonfermions} we shall discuss the delicate issue of the cancellation of the linear divergences associated with the shift of the quark critical masses induced by electromagnetism. 

We are interested in the calculation of the electromagnetic corrections to hadron masses and we do not discuss the renormalization of the operators $\mathcal{O}(x_i)$. These are needed in order to interpolate the external states and, in general, are not QED gauge invariant (see sec.~\ref{sec:hadronmasses} for an extended discussion of this point). The appearance in eq.~(\ref{eq:schematiccorrection}) of ultraviolet divergent contributions associated with the contact interactions of the quark electromagnetic currents is understood by considering the short distance expansion of their product, which reads
\begin{eqnarray}
J^\mu(x)J_\mu(0) &\sim&
c_1(x) \mathtt{1} + \sum_f c_m^f(x)  m_f \bar \psi_f \psi_f+c_{g_s}(x) G_{\mu\nu}G^{\mu\nu}
+\cdots \; .
\label{eq:opevv}
\end{eqnarray}
The ``counter--term" coefficients $c_1$, $c_m^f$ and $c_{g_s}$ are divergent quantities that must be fixed by specifying appropriate renormalization prescriptions. In particular, the terms proportional to $c_m^f$ can be reabsorbed by a redefinition of each quark mass $m_f$ in the full theory with respect to isosymmetric QCD, the term proportional to $c_{g_s}$ can be reabsorbed by a redefinition of the strong coupling constant (i.e. of the lattice spacing) while the term proportional to $c_1$ corresponds to the vacuum polarization and the associated divergence cancels by taking the fully connected part of the right hand side of  eq.~(\ref{eq:schematiccorrection}).

In order to take into account the dependence of the parameters of the theory, for example $m_f$, with respect to $\alpha_{em}$ and to absorb the divergences originating from electromagnetic interactions, one can include in the correlator $T\langle \mathcal{O}(x_i) \rangle$ explicit insertions of the corresponding operators, for example of $\bar \psi_f \psi_f$. To put the discussion on a concrete basis, let us consider a generic ``physical" observable $\mathcal{O}$ in the full theory,
\begin{eqnarray}
\mathcal{O}(\vec g) \ =\ \mathcal{O}(e^2,g_s^2,m_u,m_d,m_s)
\ =\ \langle \mathcal{O} \rangle^{\vec g}\; ,
\label{eq:genericobs}
\end{eqnarray}
where we have used the following compact vector notation for the bare parameters of the theory
\begin{eqnarray}
\vec g=\Big(e^2,g_s^2,m_{u},m_{d},m_s\Big)
\end{eqnarray}
and where the notation $\langle \cdot \rangle^{\vec g}$ means that the path--integral average is performed in the full theory (see section~\ref{sec:pathintegral}).
In the previous expressions we listed the bare mass parameters of the three lightest quarks, but the discussion can be easily generalized to include heavier quarks. We have called $g_s$ the bare strong coupling constant and $e$ the bare electric charge. Note that physical observables are QED and QCD gauge invariant and depend on $e^2$ and $g_s^2$. Our method consists in expanding any observable $\mathcal{O}(\vec g)$ with respect to the isosymmetric QCD result $\mathcal{O}(\vec g^0)$ according to
\begin{eqnarray}
\mathcal{O}(\vec g)
&=& \mathcal{O}(\vec g^0) +
\left. \left\{
e^2 \frac{\partial}{\partial e^2} +
\left[g_s^2-(g_s^0)^2\right] \frac{\partial}{\partial g_s^2} +
\left[m_f-m_f^0\right] \frac{\partial}{\partial m_f}
\right\}\mathcal{O}(\vec g)\right\vert_{\vec g=\vec g^0}
\nonumber \\\
\nonumber \\
&=& 
\langle \mathcal{O}\rangle^{\vec g^0} + \mydelta \mathcal{O}\; ,
\label{eq:strategy}
\end{eqnarray}
where 
\begin{eqnarray}
\vec{g}^0=\Big(0,(g_s^0)^2,m_{ud}^0,m_{ud}^0,m_s^0\Big) \; .
\end{eqnarray}
The notation $\langle \cdot \rangle^{\vec g^0}$ means that the path--integral average is performed in the isosymmetric theory and the expression of $\mydelta \mathcal{O}$ in terms of $\langle \cdot \rangle^{\vec g^0}$ and the appropriate reweighting factor is given in eq.~(\ref{eq:deltarew}). 

The bare parameters $\vec g^0$ of the isosymmetric theory can be fixed independently from the parameters $\vec g$ by using an hadronic scheme in order to renormalize isosymmetric QCD, i.e. by performing a ``standard" QCD simulation, by using a suitable number of hadronic inputs to calibrate the isosymmetric lattice and by assuming that isospin breaking effects are negligible. The corrections $\mydelta \mathcal{O}$ to the physical observables that have been used to calibrate the isosymmetric lattice vanish by construction with this prescription while, obviously, $\mydelta \mathcal{O}$ is  different from zero for any other predictable quantity.
On the other hand, by performing simulations of the full theory, the parameters $\vec g^0$ can also be fixed by matching the renormalized couplings of the two theories at a given scale $\mu^\star$~\cite{Gasser:2003hk}. More precisely, once the renormalized parameters $\hat g_i(\mu)=Z_i(\mu) g_i$ have been fixed by using an hadronic prescription, the renormalized couplings of the isosymmetric theory $\hat g_i^0(\mu)=Z_i^0(\mu) g_i^0$ at the scale $\mu^\star$ are fixed  by imposing the following matching conditions
\begin{eqnarray}
&&\hat g_s^0(\mu^\star)=\hat g_s(\mu^\star) \; ,
\nonumber \\
\nonumber \\
&&\hat m_{ud}^0(\mu^\star)=\hat m_{ud}(\mu^\star)=\frac{\hat m_d(\mu^\star)+\hat m_u(\mu^\star)}{2} \; ,
\nonumber \\
\nonumber \\
&&\hat m_{s}^0(\mu^\star)=\hat m_{s}(\mu^\star) \; .
\label{eq:matching}
\end{eqnarray}
In this work we rely on this prescription by matching the couplings renormalized in the $\overline{MS}$ scheme at $\mu^\star=2\mbox{ GeV}$.

It is important to realize that a physical observable is a Renormalization Group Invariant (RGI) quantity,
\begin{eqnarray}
\mathcal{O}(g_i)=\mathcal{O}(\hat g_i) \; ,
\qquad
\qquad
\mathcal{O}(g_i^0)=\mathcal{O}(\hat g_i^0) \; .
\end{eqnarray}
By using these properties, the perturbative expansion of eq.~(\ref{eq:strategy}) can be expressed in terms of the renormalized couplings according to
\begin{eqnarray}
\mathcal{O}(\hat g_i)
= \mathcal{O}\left(\hat g^0_i\right)
+
\left. \left\{
\hat e^2 \frac{\partial}{\partial \hat e^2} +
\left[\hat g_s^2- \left(\frac{Z_{g_s}}{Z^0_{g_s}}\hat g_s^0\right)^2\right]
\frac{\partial}{\partial \hat g_s^2}+
\left[\hat m_f-\frac{Z_{m_f}}{Z^0_{m_f}}\hat m_f^0\right]
\frac{\partial}{\partial \hat m_f}
\right\}\mathcal{O}(\hat g_i)\right\vert_{\hat g_i= \frac{Z_i}{Z_i^0}\hat g^0_i}  \; .
\label{eq:strategyren}
\end{eqnarray}
From the comparison of the previous equation with eq.~(\ref{eq:opevv}) we find in the differential operator language the divergent terms proportional to $Z_{m_f}/Z^0_{m_f}$ and $Z_{g_s}/Z^0_{g_s}$ that correspond to the short distance expansion counter--terms $c_m^f$ and $c_{g_s}$ respectively. In practice, these counter--terms do appear because the renormalization constants (the bare parameters) of the full theory are different from the corresponding quantities of isosymmetric QCD, the theory in which we perform the numerical simulations. Once the counter--terms have been properly tuned, our procedure can be interpreted as the expansion of the full theory in the renormalized parameters $\hat \alpha_{em}$ and $\hat m_d-\hat m_u$.

\section{Non-compact QED on the lattice at $\mathcal{O}(\alpha_{em})$}
\label{sec:noncompact}
In this section we discuss the non--compact formulation of lattice QED, the issues associated with the expansion of the quark action with respect to the electric charge and address the ``infrared" problem mentioned in section~\ref{sec:dependence}, i.e. we provide an infrared safe definition of the finite volume lattice photon propagator that can be conveniently used in numerical calculations by working directly in coordinate space. 

Non--compact lattice QED has been used also in ref.~\cite{Duncan:1996xy}, where the effects of electromagnetism have been computed non--perturbatively on the lattice for the first time, and in all the other computations subsequently performed (see refs.~\cite{Basak:2013iw,Blum:2010ym,Portelli:2012pn,Ishikawa:2012ix} for recent works on the subject). In practice, the non--compact formulation consists in treating the gauge potential $A_\mu(x)$ in a fixed QED gauge as a dynamical variable. The quarks covariant derivatives are then defined by introducing the QED links through exponentiation,
\begin{eqnarray}
A_\mu(x) &\longrightarrow& E_\mu(x)=e^{-ieA_\mu(x)}\; ,
\end{eqnarray}
and by multiplying the QCD links for the appropriate $U(1)_{em}$ factors,
\begin{eqnarray}
\mathcal{D}_\mu^+[U,A]\; \psi_f(x) &=&
[E_\mu(x)]^{e_f}\ U_\mu(x) \psi_f(x+\mu)-\psi_f(x) \; .
\end{eqnarray} 
In previous expressions $e_f$ is the fractional electric charge of the quark of flavour $f$, i.e. $e_f$ is $2/3$ for up--type quarks and $-1/3$ for down--type quarks. Given our conventions, exact gauge invariance is obtained if the fields are transformed as follows
\begin{eqnarray}
\psi_f(x) \longrightarrow e^{i e_fe\lambda(x)} \psi_f(x) \; ,
\qquad
\bar \psi_f(x) \longrightarrow \bar \psi_f(x) e^{-i e_f e\lambda(x)}  \; ,
\qquad
A_\mu(x) \longrightarrow A_\mu(x)+\nabla^+_\mu \lambda(x)  \; ,
\end{eqnarray}
and we define
\begin{eqnarray}
\nabla_\mu^+ f(x)= f(x+\hat \mu) - f(x) \;,
\qquad
\nabla_\mu^- f(x)= f(x) - f(x-\hat \mu) \;,
\qquad
\nabla_\mu = \frac{\nabla_\mu^+ + \nabla_\mu^-}{2} \; .
\end{eqnarray}
We want to treat electromagnetism at fixed order with respect to $\hat \alpha_{em}$ and, to this end, we need to expand the quarks action in powers of $e$. 
This procedure is performed by starting from the explicit expression of the lattice Dirac operator $D_f[U,A;\vec g]$ to be used in numerical simulations and by calculating 
\begin{eqnarray}
&&\sum_x \bar \psi_f(x) \Big\{ D_f[U,A;\vec g]-D_f[U,0;\vec g]\Big\} \psi_f(x)
\nonumber \\
\nonumber \\
&&\qquad \qquad \qquad \qquad \qquad =
\sum_{x,\mu}\left\{ (e_fe) A_\mu(x) V^\mu_f(x) + \frac{(e_fe)^2}{2}A_\mu(x)A_\mu(x) T_f^\mu(x)+\dots \right\} \;,
\label{eq:qaexp}
\end{eqnarray}
where $V^\mu_f(x)$ is the conserved vector current corresponding to the quark $f$ while $T_f^\mu(x)$ is the ``tadpole" vertex. Both the conserved vector current and the tadpole vertex depend upon the particular choice made for the discretization of the fermion action and we shall provide the explicit expressions for $V^\mu_f$ and $T^\mu_f$ corresponding to the regularization used in this paper in the following sections, see eqs.~(\ref{eq:explicitvt}). Note that tadpole insertions, a feature of lattice discretization, cannot be neglected because these play a crucial role in order to preserve gauge invariance at order $ e^2$. The electromagnetic current and the tadpole vertex to be inserted in correlators are the sums over all the quarks of $V^\mu_f$ and $T^\mu_f$ with the corresponding charge factors,  
\begin{eqnarray}
J^\mu(x)&=&\sum_f e_f e\ V^\mu_f(x)
\ =\ \sum_f e_f e\ \bar \psi_f\, \Gamma_V^\mu[U]\, \psi_f(x) \; ,
\nonumber \\
\nonumber \\
T^\mu(x)&=&\sum_{f} (e_f e)^2\ T^\mu_f(x)
\ =\ \sum_{f} (e_f e)^2\  \bar \psi_f\, \Gamma_T^\mu[U]\, \psi_f(x)\; .
\label{eq:qaexp2}
\end{eqnarray}

From the validity on the lattice of exact gauge Ward--Takahashi identities (WTI) and from the fact that the fermion action is by construction renormalization group invariant, it follows that the expansion of eq.~($\ref{eq:qaexp}$) is perfectly well defined and that it can be re--expressed in terms of the renormalized electric charge and gauge potential fields by replacing $e\rightarrow \hat e/ Z_e$ and $A^\mu\rightarrow  \hat A^\mu  Z_e$. Furthermore, at the $O(\hat \alpha_{em})$ at which we are working, there is no need to renormalize the electric charge, a problem that has to be faced instead at higher orders.

Once the fermion action has been expanded, the leading QED corrections to a given lattice correlator are obtained by considering the time product of the original operators with two integrated insertions of the combination $A_\mu(x) J^\mu(x)$ or with a single integrated insertion of $\sum_\mu A_\mu(x) A_\mu(x) T^\mu(x)$. As anticipated in section~\ref{sec:dependence} the corrected correlator is expressed in terms of the photon propagator. To give an example, let us consider the electromagnetic corrections to the kaon two--point correlator. Among other contributions discussed in detail in the following sections, in this case one has to calculate
\begin{eqnarray}
-e_s e_u  e^2\gdslexch 
&=& 
e_s e_u  e^2\left\langle\; \sum_{x,y} A_\mu(x)A_\nu(y)\
T \bra{0}\, [\bar u\gamma_5 s](t)\, 
V^\mu_{s}(x)\, V^\nu_{u}(y)\, 
[\bar s \gamma_5 u](0) \, \ket{0} \right\rangle^{A}
\nonumber \\
\nonumber \\
&=& 
e_s e_u  e^2\sum_{x,y} D_{\mu\nu}(x-y)\
T \bra{0}\, [\bar u\gamma_5 s](t)\, 
V^\mu_{s}(x)\, V^\nu_{u}(y)\, 
[\bar s \gamma_5 u](0) \, \ket{0} \; ,
\label{eq:kexample}
\end{eqnarray}
where the notation $\langle \cdot \rangle^A$ represents the path integral average over the gauge potential $A_\mu$ (see eq.~(\ref{eq:funca})), $D_{\mu\nu}(x-y)$ is the lattice photon propagator and we have ignored the quark disconnected contributions coming from the contractions of the vector currents $V^\mu_{s}(x)$ and $V^\nu_{u}(y)$ among themselves. 

In order to define the lattice photon propagator we start by considering the lattice action of the QED gauge field in Feynman gauge
\begin{eqnarray}
S_{gauge}[A]
&=&
\frac{1}{2}\sum_{x,\mu,\nu}
A_\mu(x)\left[-\nabla^-_\nu\nabla^+_\nu\right] A_\mu(x)
\ =\
\frac{1}{2}\sum_{k,\mu,\nu}
\tilde A_\mu^\star(k)\left[2\sin(k_\nu/2)\right]^2 \tilde A_\mu(k) \; ,
\label{eq:sqed}
\end{eqnarray}
where $A_\mu(x)$ is a real field while $\tilde A_\mu(k)$ denotes its Fourier transform that is a complex field satisfying the condition $\tilde A_\mu^\star(k)=\tilde A_\mu(-k)$. In the previous expression we have explicitly shown the QED action in momentum space to highlight a well known problem with the definition of the lattice photon propagator, i.e. the infrared divergence associated with the zero momentum mode. The $A_\mu$ propagator is defined as the inverse of the kinetic term and, in order to define the inverse of the lattice Laplace operator $-\nabla^-_\nu\nabla^+_\nu$, one has to provide a prescription to cope with its kernel.
Any ``derivative" gauge fixing condition does not constrain the zero momentum mode of the electromagnetic gauge potential,
\begin{eqnarray}
\nabla^-_\mu \left[ A_\mu(x) +c \right] = \nabla^-_\mu A_\mu(x)\; ,
\end{eqnarray}
and, as a consequence, the gauge fixing has to be ``completed" by giving a prescription to regularize the associated infrared divergence. 

One possibility, widely used in the literature after the original proposal made in ref.~\cite{Duncan:1996xy}, is to make the zero momentum mode to vanish identically by sampling the gauge potential in momentum space. It can be shown that this prescription results into finite volume effect on physical observables, see section~\ref{sec:extrapolations}. 
We also follow this strategy and set $\tilde A(k=0)=0$ with the difference that we work directly in coordinate space, thus avoiding Fourier transforms, and calculate the infrared regularized photon propagator stochastically. More precisely, by introducing the operator $\mathtt{P^\perp}$ projecting a given field on the subspace orthogonal to the zero momentum mode,
\begin{eqnarray}
\mathtt{P^\perp} \phi(x)=\phi(x) - \frac{1}{V}\sum_y \phi(y) \; ,
\end{eqnarray}
we calculate the regularized photon propagator
\begin{eqnarray}
D_{\mu\nu}^\perp(x-y) = \left[
\frac{\delta_{\mu\nu}}{-\nabla^-_\rho\nabla^+_\rho}\; \mathtt{P^\perp}
\right](x-y) \; ,
\end{eqnarray}
by following the procedure outlined here below:
\begin{itemize}
\item we extract four independent real fields $B_\mu(x)$ distributed according to a real $Z_2$ noise,
\begin{eqnarray}
\left\langle B_\mu(x) B_\nu(y) \right\rangle^B =\delta_{\mu\nu}\ \delta(x-y)\; ;
\end{eqnarray}

\item for each field $B_\mu(x)$ we solve numerically the equation of motion in Feynman gauge,
\begin{eqnarray}
[-\nabla^-_\rho\nabla^+_\rho] C_\mu[B;x] = \mathtt{P^\perp}\; B_\mu(x) \; ;
\label{eq:keyequation}
\end{eqnarray}
the solution is
\begin{eqnarray}
C_\mu[B;x] = 
\left[ \frac{\delta_{\mu\nu}}{-\nabla^-_\rho\nabla^+_\rho}\; \mathtt{P^\perp}\right] B_\nu(x) 
= \sum_z D_{\mu\nu}^\perp(x-z) B_\nu(z) \; ,
\end{eqnarray}
and the field $C_\mu[B;x]$ is a functional of $B_\mu$;

\item the photon propagator is thus obtained by using the properties of the $Z_2$ noise according to
\begin{eqnarray}
\left\langle B_\mu(y) C_\nu[B;x] \right\rangle^{B} = \sum_z {D_{\nu\rho}^\perp(x-z) \left\langle B_\mu(y) B_\rho(z) \right\rangle^{B}}  = D_{\mu\nu}^\perp(x-y)\; .
\end{eqnarray}

\end{itemize}
This procedure relies on the actual possibility of obtaining a numerical solution of eq.~(\ref{eq:keyequation}). By working in double precision we have been able to obtain a stable and efficient numerical bi--conjugate gradient stabilized ({\tt bicgstab}) inverter. The solution is obtained with about hundred iterations on lattice volumes as large as $V=96\times 48^3$.  
Coming back to the example discussed previously, we have that the infrared regularized version of eq.~(\ref{eq:kexample}) can be calculated on the lattice according to
\begin{eqnarray}
-\gdslexch 
&=& 
\sum_{x,y} D_{\mu\nu}^\perp(x-y)\
T \bra{0}\, [\bar u\gamma_5 s](t)\, 
V^\mu_{s}(x)\, V^\nu_{u}(y)\, 
[\bar s \gamma_5 u](0) \, \ket{0} 
\nonumber \\
\nonumber \\
&=& 
\left\langle \sum_{x,y} B_\mu(x)C_\nu[B;y]\
T \bra{0}\, [\bar u\gamma_5 s](t)\, 
V^\mu_{s}(x)\, V^\nu_{u}(y)\, 
[\bar s \gamma_5 u](0) \, \ket{0} \right\rangle^B 
\nonumber \\
\nonumber \\
&=& 
-\left\langle \sum_{x,y} B_\mu(x)C_\nu[B;y]\
\mbox{Tr}\Big\{
\gamma_5 S_s[U;t-x] \Gamma_V^\mu S_s[U;x] 
\gamma_5 S_{ud}[U;-y] \Gamma_V^\nu S_{ud}[U;y-t]
\Big\} \right\rangle^{B}, 
\nonumber \\
\end{eqnarray}
where we used the compact notation $S_f[U]$ to indicate the isosymmetric QCD lattice quark propagator $S_f[U;\vec g^0]$ obtained by inverting the Dirac operator $D_f[U]=D_f[U,\vec g^0]$ (see eq.~(\ref{eq:diracoperator1iso}) below).
The problem of the numerical calculation of the diagram appearing on the left hand side of the previous expression is thus reduced to the calculation of two sequential propagators. More precisely, one can solve the following two systems
\begin{eqnarray}
\left\{ D_f[U]\; \Psi_B^f \right\}(x) &=& \sum_\mu B_\mu(x) \Gamma_V^\mu S_f[U;x]  \; ,
\nonumber \\
\nonumber \\
\left\{ D_f[U]\; \Psi_C^f\right\} (x) &=& \sum_\mu C_\mu[B;x] \Gamma_V^\mu S_f[U;x]  \; ,
\end{eqnarray}
for different values of the $B_\mu(x)$ and $C_\mu[B;x]$ fields (we have used $3$ electromagnetic stochastic sources per QCD gauge configuration) and then calculate the corrected correlator according to
\begin{eqnarray}
-e_s e_u  e^2\gdslexch &=& 
-e_s e_u  e^2\left\langle\ \mbox{Tr}\left\{\ [\Psi_C^{ud}]^\dagger(t)\ 
\Psi_B^s(t)\ \right\} \ \right\rangle^{B} \; .
\end{eqnarray}
In the previous expressions we have been assuming that the lattice Dirac operator, and consequently the conserved vector current, satisfies the property $S_f[U;x]^\dagger = \gamma_5 S_f[U;-x] \gamma_5$. Generalizations of the previous procedure to calculate all the other correlators (fermionic Wick contractions) appearing in this paper can be readily obtained.

\section{Fermionic lattice action}
\label{sec:wilsonfermions}
In this section we enter into the details of the fermionic lattice action used in this work, namely the maximally twisted Wilson action~\cite{Frezzotti:2000nk,Frezzotti:2003ni}. In order to minimize cutoff effects and statistical errors we have been working within a mixed--action approach~\cite{Frezzotti:2004wz,Constantinou:2010qv}. In particular, the results described in the following sections have been obtained with the action $S=S_{sea}+S_{val}$. The sea quark action is given by
\begin{eqnarray}
S_{sea} &=& \sum_x\left\{
\bar q_u D^+_u[U,A] q_u
+
\bar q_d D^-_d[U,A] q_d
\right\} \; ,
\end{eqnarray}
where $q_f$ are fermionic variables and the $D_f^\pm[U,A]$ lattice Dirac operators are
\begin{eqnarray}
D_f^\pm[U,A]\; \psi(x)=m_f\psi(x) \; \pm\; i\gamma_5(m^{cr}_f+4)\psi(x)
&-&\sum_\mu{\frac{\pm i\gamma_5-\gamma_\mu}{2}U_\mu(x)[E_\mu(x)]^{e_f}\psi(x+\mu)}
\nonumber \\
\nonumber \\
&-&
\sum_\mu{\frac{\pm i\gamma_5+\gamma_\mu}{2}U_\mu^\dagger(x-\mu)[E_\mu^\dagger(x-\mu)]^{e_f}\psi(x-\mu)} \; .
\nonumber \\
\label{eq:diracoperator1}
\end{eqnarray}
Note that the operators $D_f^\pm[U,A]$ depend upon the bare parameters of the full theory and we used the compact notation $D_f^\pm[U,A]=D_f^\pm[U,A;\vec g]$. The corresponding operators $D_f^\pm[U]=D_f^\pm[U;\vec g^0]$ of the isosymmetric theory are
\begin{eqnarray}
D_f^\pm[U]\; \psi(x)=m_f^0\psi(x) \; \pm\; i\gamma_5(m^{cr}_0+4)\psi(x)
&-&\sum_\mu{\frac{\pm i\gamma_5-\gamma_\mu}{2}U_\mu(x)\psi(x+\mu)}
\nonumber \\
\nonumber \\
&-&
\sum_\mu{\frac{\pm i\gamma_5+\gamma_\mu}{2}U_\mu^\dagger(x-\mu)\psi(x-\mu)} \; .
\nonumber \\
\label{eq:diracoperator1iso}
\end{eqnarray}
Concerning the content of the valence sector, we have considered a doublet of fermionic fields for each flavour, $\psi_f^T=(\psi_f^+,\psi_f^-)$, and a corresponding doublet of bosonic fields (pseudo--quarks), $\phi_f^T=(\phi_f^+,\phi_f^-)$. The fields within the same doublet have the same mass $m_f$, the same electric charge $e_f$ but opposite chirally rotated Wilson terms. Calling $\tau^i$ the Pauli matrices acting on a flavour doublet and defining
\begin{eqnarray}
D_f[U,A]=\frac{1+\tau^3}{2} D_f^+[U,A] +\frac{1-\tau^3}{2} D_f^-[U,A] \; ,
\label{eq:diracoperator2}
\end{eqnarray}
we can write the action for the matter fields in the valence sector in the following compact notation
\begin{eqnarray}
S_{val} &=& \sum_{f,x}\left\{
\bar \psi_f D_f[U,A] \psi_f
+
\bar \phi_f D_f[U,A] \phi_f
\right\} \; .
\end{eqnarray}
As far as the mass splitting of the pions (or of the nucleons) is concerned, the mixed action setup used in this paper allows to compute observables with $O(a^2)$ cutoff effects at the price of introducing unitarity violations that disappear when the continuum limit is performed (at matched sea and valence renormalized quark masses the resulting continuum theory is unitary). For each correlator, by possibly replicating some of the valence matter fields, the choice made for the action allows to consider only the fermionic Wick contractions that would arise in the continuum theory, thus avoiding the introduction of (finite) isospin breaking lattice artifacts. The resulting diagrams are then discretized by using for each quark propagator a convenient choice of the sign of the twisted Wilson term. In practice we consider for any meson interpolating operators of the form
\begin{eqnarray}
O_H=\bar \psi_{f_1}^+ \Gamma \psi_{f_2}^- \; .
\label{eq:plusminusoh}
\end{eqnarray}
The resulting correlators have reduced cutoff effects and smaller statistical errors with respect to the other possible choices of $O_H$, as for example $\bar \psi_{f_1}^+ \Gamma \psi_{f_2}^+$, see refs.~\cite{Frezzotti:2004wz,Constantinou:2010qv}. In the case of the connected fermionic Wick contraction arising in the neutral pion two--point functions we use $O_{\pi^0}^{conn}=(\bar u^+ \gamma^5 u^- - \bar d^+ \gamma^5 d^-)/\sqrt{2}$.

In the following we have also computed isospin breaking corrections to the kaon masses. In this case the strange quark is quenched and the theory violates unitarity also in the continuum limit. In the calculation of the kaon mass splitting additional violations of unitarity will be introduced when we shall consider what we can call the``electro--quenched" approximation. This approximation consists in forcing the sea quarks to be neutral with respect to electromagnetic interactions and is implemented by replacing $S_{sea}$ with
\begin{eqnarray}
S_{sea}^{e=0} &=& \sum_x\left\{
\bar q_u D^+_u[U] q_u
+
\bar q_d D^-_d[U] q_d
\right\} \; .
\label{eq:electroquenched}
\end{eqnarray}

\subsection{Critical mass counter--terms}
\label{sec:kmasses}
We now discuss the problem of the (re)tuning of the critical masses necessary in the presence of electromagnetic interactions when the lattice fermionic action includes a Wilson term. In this case eq.~(\ref{eq:opevv}) is modified both on the left--hand side, to take into account the presence of the tadpole vertices of the different quarks, and on the right--hand side, because of the appearance  of additional divergent contributions that have to be reabsorbed by a redefinition of the critical masses. The lattice version of eq.~(\ref{eq:opevv}) corresponding to the regularization used in this paper is thus given by\footnote{
The twisted lattice action, even in the presence of electromagnetic interactions, enjoys the symmetry $P\times \mathcal{D}_d\times (m_f\mapsto -m_f)$ where $P$ is the ordinary lattice parity and $\mathcal{D}_d$ is the lattice version of the transformation that replaces a generic operator $\mathcal{O}(x)$ of dimension $d$ with $e^{id\pi}\mathcal{O}(-x)$. It follows (see ref.~\cite{Frezzotti:2003ni,Frezzotti:2005gi}) that parity--even physical observables are automatically $O(a)$ improved and that the operator $G_{\mu\nu}\tilde{G}^{\mu\nu}$ does not appear in eq.~(\ref{eq:opevvlatt}), though cutoff effects proportional to the insertions of (certain combinations of) the parity--odd operators $\hat m_f G_{\mu\nu}\tilde{G}^{\mu\nu}$ are not forbidden by the symmetries of the theory.}
\begin{eqnarray}
J^\mu(x)J_\mu(0) &+&\sum_\mu T^\mu(x)  
\nonumber \\
\nonumber \\
&\sim&
c_1(x) \mathtt{1} \ +\ 
\sum_f c_k^f(x) \bar \psi_f i\gamma_5 \tau^3 \psi_f \ +\
\sum_f c_m^f(x)  m_f \bar \psi_f \psi_f \ +\
c_{g_s}(x) G_{\mu\nu}G^{\mu\nu}
\ +\ \cdots ,
\nonumber \\
\label{eq:opevvlatt}
\end{eqnarray}
where (see eqs.~(\ref{eq:qaexp2}) above) the exactly conserved vector current and the tadpole vertex corresponding to the Dirac operators of eqs.~(\ref{eq:diracoperator1}) and~(\ref{eq:diracoperator2}) can be obtained by expanding the lattice quark action according to eq.~(\ref{eq:qaexp}) and are given by
\begin{eqnarray}
V^\mu_f(x) 
&=&
i\left[\bar \psi_f(x)\frac{i\tau^3\gamma_5-\gamma_\mu}{2}U_\mu(x)\psi_f(x+\mu)
-
\bar \psi_f(x+\mu)\frac{i\tau^3\gamma_5+\gamma_\mu}{2}U_\mu^\dagger(x)\psi_f(x)
\right]\; ,
\nonumber \\
\nonumber \\
\nonumber \\
T^\mu_f(x) 
&=&
\quad \bar \psi_f(x)\frac{i\tau^3\gamma_5-\gamma_\mu}{2}U_\mu(x)\psi_f(x+\mu)
+
\bar \psi_f(x+\mu)\frac{i\tau^3\gamma_5+\gamma_\mu}{2}U_\mu^\dagger(x)\psi_f(x)
\; .
\label{eq:explicitvt}
\end{eqnarray}
Note the presence in eq.~(\ref{eq:opevvlatt}) of the critical mass counter--term coefficients $c_k^f(x)$. In case of standard (untwisted) Wilson fermions one would get a similar expression with the scalar operators $\bar \psi_f \psi_f$  instead of the pseudoscalar parity--odd operators $\bar \psi_f i \gamma_5 \tau^3\psi_f$. 

We have considered two different strategies to determine the counter--terms associated with the electromagnetic shift of the critical masses, both based on the use of the WTI of the continuum theory. The first strategy can be used with both standard and twisted Wilson fermions and is based on the Dashen's theorem, a consequence of chiral WTI valid in the massless theory, i.e. the theory with
\begin{eqnarray}
\hat m_f=\{ \hat m_u,\hat m_d,\hat m_s\} = 0\; .
\end{eqnarray}
According to the theorem, even in the presence of electromagnetic interactions, the neutral pion and the neutral kaons are non--singlet Goldstone's bosons, so that
\begin{eqnarray}
\hat m_f = 0
\qquad \rightarrow \qquad
M_{\pi^0}=M_{K^0}=0 \; .
\label{eq:kcfirststrategy}
\end{eqnarray}
Furthermore, from the vector flavor symmetries which remain valid in the massless theory even in the presence of electromagnetic interactions it follows that 
\begin{eqnarray}
 \hat m_f = 0
\qquad \rightarrow \qquad
M_{\pi^+}=M_{K^+}
\label{eq:dashencharged}
\end{eqnarray}
and that the critical mass parameters of the down and of the strange quarks are equal.
By using this observation, after a detailed discussion of the corrections to kaon and pion masses, we shall give explicit formulae to determine the critical mass counter--terms by imposing the validity of eqs.~(\ref{eq:kcfirststrategy}), see section~\ref{sec:tuningk}.

The second approach to tune the critical mass parameters is peculiar to chirally twisted Wilson fermions and is commonly used to implement the maximal twist condition in simulations of isosymmetric QCD, see ref.~\cite{Baron:2009wt}. By starting from the explicit expression of the lattice Dirac operator for a given valence flavour doublet, eqs.~(\ref{eq:diracoperator1}) and~(\ref{eq:diracoperator2}), one realizes that the critical mass of each valence quark can be separately tuned by working in the theory with electromagnetic interactions and with massive non--degenerate quarks ($\hat m_s> \hat m_d > \hat m_u> 0$) and by imposing the validity of the following vector WTI\footnote{A similar procedure to tune the critical masses can be used with standard Wilson fermions by starting from the PCAC relation and by using the knowledge of the PCAC quark mass in the isosymmetric theory.}
\begin{eqnarray}
W_f(\vec g)=\nabla_\mu \langle \;
 \left[\bar \psi_f \gamma^\mu \tau^1 \psi_f \right](x) \;
\left[\bar \psi_f \gamma^5 \tau^2 \psi_f \right](0) 
\; \rangle^{\vec g}
 =0  \; , 
 \qquad \qquad
 f=\{u,d,s\}\; .
\label{eq:tmvwi}
\end{eqnarray}
Also in the case of this WTI we shall derive explicit formulae corresponding to its expansion in powers of $e$ in section~\ref{sec:tuningk}.

Before closing this section we discuss the generalization of eq.~(\ref{eq:strategy}) required to take into account the dependence of a generic lattice observable on the critical masses $m^{cr}_f$. More precisely, by enlarging the parameter space of the full theory,
\begin{eqnarray}
\mathcal{O}(\vec g) \ =\ \mathcal{O}(e^2,g_s^2,m_{u},m_{d},m_s,m^{cr}_{u},m^{cr}_{d},m^{cr}_s)\; ,
\qquad
\qquad
\vec g=\Big(e^2,g_s^2,m_{u},m_{d},m_s,m^{cr}_{u},m^{cr}_{d},m^{cr}_s\Big) \;,
\end{eqnarray}
and by calling $m^{cr}_0$ the single critical mass parameter of the symmetric theory, we see that isosymmetric QCD simulations correspond to
\begin{eqnarray}
\vec g^0=\Big(0,(g_s^0)^2,m_{ud}^0,m_{ud}^0,m_s^0,m^{cr}_0,m^{cr}_0,m^{cr}_0\Big) \; .
\end{eqnarray}
The value of $m^{cr}_0$ has been precisely determined in ref.~\cite{Baron:2009wt} in the isosymmetric theory by requiring the validity of the vector Ward--Takahashi identity of eq.~(\ref{eq:tmvwi}) with $m_f^0=m_{ud}^0$,
\begin{eqnarray}
W_{ud}(\vec g^0)=0 
\qquad
\longrightarrow
\qquad
m^{cr}_0
\; .
\end{eqnarray}
Our gauge ensembles have been generated at this well defined value of critical mass for each $\beta^0=6/(g_s^0)^2$ (see Appendix~\ref{sec:tm}) and, for this reason, we cannot tune the value of the different $m^{cr}_f$ by looking at the dependence of $W_f$, or of $M_{\pi^0}$ and $M_{K^0}$, on the quark critical masses used in the simulations. On the other hand, the LIB corrections to any observable can be obtained by making an expansion, at fixed lattice spacing, with respect to the differences $m^{cr}_f-m^{cr}_0$ which represents a regularization specific isospin breaking effect induced by the electromagnetic interactions. The generalization of eq.~(\ref{eq:strategy})  to be used on the lattice with Wilson fermions is
\begin{eqnarray}
\mydelta \mathcal{O}
&=&
\left. \left\{
e^2 \frac{\partial}{\partial e^2} +
\left[g_s^2- (g_s^0)^2\right] 
\frac{\partial}{\partial g_s^2}+
[m_f-m_f^0] \frac{\partial}{\partial m_f}+
[m^{cr}_f-m^{cr}_0] \frac{\partial}{\partial m^{cr}_f}
\right\}\mathcal{O}(\vec g)  \right\vert_{\vec g=\vec g^0} \; .
\label{eq:latstrategy1}
\end{eqnarray}
In the next section we discuss in detail how eq.~$(\ref{eq:latstrategy1})$ can be used to expand the lattice path--integral and to derive explicit formulae for the calculation of the LIB corrections.

\section{Expansion of the lattice path--integral at $O(\hat \alpha_{em})$}
\label{sec:pathintegral}
In this section, by following the strategy outlined in the previous sections, we discuss the details concerning the derivation of the formulae necessary to calculate the LIB corrections to specific observables. The starting point is the path--integral representation of the observable in the full theory,
\begin{eqnarray}
\mathcal{O}(\vec g)\ =\ \langle \mathcal{O} \rangle^{\vec g}&=&
\frac{
\int{
dAe^{-S_{gauge}[A]}\; dU\; e^{-\beta S_{gauge}[U]}\;
\prod_{f=1}^{n_f}\det\left(D_f^\pm[U,A;\vec g]\right)\; \mathcal{O}[U,A;\vec g]
}}
{
\int{
dAe^{-S_{gauge}[A]}\; dU\; e^{-\beta S_{gauge}[U]}\;
\prod_{f=1}^{n_f}\det\left(D_f^\pm[U,A;\vec g]\right)
}} \; ,
\label{eq:pathstrategy1}
\end{eqnarray}
where $S_{gauge}[A]$ has been given in eq.~(\ref{eq:sqed}) and is a functional of the gauge potential $A_\mu$, $S_{gauge}[U]$ is the QCD gauge action ($\beta=6/g_s^2$) and is a functional of the link variables $U_\mu(x)$, $D_f^\pm[U,A;\vec g]$ are the Dirac operators defined in eq.~(\ref{eq:diracoperator1}). Note that when the masses and the charges of the light quarks are different the product of the determinants of the up and of the down is not positive definite unless one resorts to lattice regularizations in which the determinant of each quark is separately positive.
We want to express the observable $\mathcal{O}(\vec g)$ in terms of the path--integral average in the isosymmetric theory, i.e.
\begin{eqnarray}
\mathcal{O}(\vec g^0)\ =\ \langle \mathcal{O} \rangle^{\vec g^0}&=&
\frac{
\int{
dU\; e^{-\beta^0 S_{gauge}[U]}\;
\prod_{f=1}^{n_f}\det\left(D_f^\pm[U;\vec g^0]\right)\;  \mathcal{O}[U]
}}
{
\int{
dU\; e^{-\beta^0 S_{gauge}[U]}\;
\prod_{f=1}^{n_f}\det\left(D_f^\pm[U;\vec g^0]\right)
}} \; ,
\label{eq:pathstrategy2}
\end{eqnarray}
where the Dirac operators $D_f^\pm[U;\vec g^0]$ have been defined in eq.~(\ref{eq:diracoperator1iso}). This can be done by introducing the appropriate reweighting factor and the functional average $\langle \cdot \rangle^{A}$ with respect to the free photon field, 
\begin{eqnarray}
&&R[U,A;\vec g]=e^{-(\beta-\beta^0) S_{gauge}[U]}\ r[U,A;\vec g]\; ,
\nonumber \\
\nonumber \\
&&r[U,A;\vec g]
=
\prod_{f=1}^{n_f}r_f[U,A;\vec g] 
\ =\
\prod_{f=1}^{n_f}\frac{\det\left(D_f^\pm[U,A;\vec g]\right)}{\det\left(D_f^\pm[U;\vec g^0]\right)} \; ,
\nonumber \\
\nonumber \\
\nonumber \\
&&\langle \mathcal{O} \rangle^A = \frac{\int{dA\; e^{-S_{gauge}[A]}\; \mathcal{O}[A]}}
{\int{dA\; e^{-S_{gauge}[A]}}} \; .
\label{eq:funca}
\end{eqnarray}
Eq.~(\ref{eq:pathstrategy1}) can be conveniently rewritten as follows
\begin{eqnarray}
\langle \mathcal{O} \rangle^{\vec g}
&=&
\frac{\big\langle R \mathcal{O}  \big\rangle^{A,\vec g^0}}
{\big\langle R \big\rangle^{A,\vec g^0}}
\ =\
\frac{\Big\langle\; \big\langle\; R[U,A;\vec g]\; O[U,A;\vec g] \; \big\rangle^A \; \Big\rangle^{\vec g^0}}
{\Big\langle\; \big\langle\; R[U,A;\vec g]\; \big\rangle^A \; \Big\rangle^{\vec g^0}}\; ,
\label{eq:pathstrategy}
\end{eqnarray}
and leading order isospin breaking corrections can now be obtained by applying the differential operator $\mydelta $ defined in eq.~(\ref{eq:latstrategy1}) to the observable $\mathcal{O}$ defined in eq.~(\ref{eq:pathstrategy}). More precisely,
\begin{eqnarray}
&&\mydelta  \mathcal{O}=
\big\langle \mydelta (R\mathcal{O}) \big\rangle^{A,\vec g^0}-
\big\langle \mydelta R \big\rangle^{A,\vec g^0}
\big\langle \mathcal{O} \big\rangle^{\vec g^0} =
\nonumber \\
\nonumber \\
&&
\big\langle \mydelta  \mathcal{O}[U,A;\vec g]\big\vert_{\vec g=\vec g^0}\big\rangle^{A,\vec g^0}
+
\left\{
\big\langle\mydelta \left( R \mathcal{O}-\mathcal{O}\right)[U,A;\vec g]\big\vert_{\vec g=\vec g^0}\big\rangle^{A,\vec g^0}-
\big\langle\mydelta  R[U,A;\vec g]\big\vert_{\vec g=\vec g^0}\big\rangle^{A,\vec g^0}
\big\langle \mathcal{O}[U;\vec g^0]\big\rangle^{\vec g^0}
\right\}  .
\nonumber \\
\nonumber \\
\label{eq:deltarew}
\end{eqnarray}
In the previous expression we have separated the term $\langle \mydelta  \mathcal{O} \rangle^{A,\vec g^0}$, representing the correction to the given observable, from the contributions in curly brackets coming from the corrections to the reweighting factor and, consequently, to the sea quark determinants. In the following we shall call these contributions ``vacuum polarization terms'' or ``disconnected terms''.

Once the quark fields have been integrated out from the path--integral, as implicitly done in the expressions above, in order to calculate the LIB corrections we must be able to apply the differential operator $\mydelta $ to the Dirac operator and to the quark propagator. To this end, it is useful to observe that
\begin{eqnarray}
\left. \frac{\partial \big\langle \mathcal{O} \big\rangle^A\left(e^2\right)}{\partial (e^2)} \right\vert_{e^2=0} &=&
\left\langle \left. \frac{1}{2}\frac{\partial^2\mathcal{O}[A;e]}{\partial e^2}\right\vert_{e=0} \right\rangle^A \;,
\label{eq:frezzotti}
\end{eqnarray}
and to consider the following expressions and related graphical representations
\begin{eqnarray}
\frac{1}{2}\frac{\partial^2S_f}{\partial  e^2} =
S_f \frac{\partial D_f}{\partial  e} S_f \frac{\partial D_f}{\partial  e} S_f
-\frac{1}{2}S_f \frac{\partial^2D_f}{\partial  e^2} S_f 
&=& 
e_f^2 \golself + e_f^2 \golltad
\; ,
\nonumber \\
\nonumber \\
\frac{\partial S_f}{\partial m_f} = -S_f \frac{\partial D_f}{\partial m_f} S_f &=&
-\goi \; ,
\nonumber \\
\nonumber \\
\frac{\partial S_f^\pm}{\partial m^{cr}_f} = -S_f^\pm \frac{\partial D_f^\pm}{\partial m^{cr}_f} S_f^\pm 
&=& \mp \goip
\; .
\label{eq:graphicalnotation2}
\end{eqnarray}
The graphical representation given in the last of the previous formulae, corresponding to the derivative of the quark propagator with respect to the critical mass, is specific to the lattice Dirac operators used in this work and the $\mp$ signs correspond respectively to $D_f^\pm$ defined into eq.~(\ref{eq:diracoperator1}). In the case of standard Wilson fermions red and grey ``blobs" would coincide. All the disconnected contributions coming from the reweighting factor can be readily obtained by using eqs.~(\ref{eq:graphicalnotation2}). For example,
\begin{eqnarray}
\frac{\partial R}{\partial g_s^2}
&=& \frac{6}{(g_s^0)^4}S_{gauge}[U] = \plaq \; ,
\nonumber \\
\nonumber \\
\nonumber \\
\frac{1}{2}\frac{\partial^2 r_f}{\partial e^2}
&=&
\frac{1}{2}\mbox{Tr}\left(S_f \frac{\partial^2 D_f}{\partial e^2}\right)
-
\frac{1}{2}\mbox{Tr}\left(S_f \frac{\partial D_f}{\partial e}S_f \frac{\partial D_f}{\partial e}\right)
+
\frac{1}{2}\mbox{Tr}\left(S_f \frac{\partial D_f}{\partial e}\right)
\mbox{Tr}\left(S_f \frac{\partial D_f}{\partial e}\right)
\nonumber \\
\nonumber \\
&=&
-{\color{blue}e_f^2}\vp-{\color{blue}e_f^2}\vppp+{\color{blue}e_f^2}\vpp
\; .
\label{eq:graphicalnotation1}
\end{eqnarray}
In writing eqs.~(\ref{eq:graphicalnotation2}) and~(\ref{eq:graphicalnotation1}) we assumed that the derivatives have been evaluated at $\vec g=\vec g^0$ and that the functional integral $\langle \cdot \rangle^A$ with respect to the photon field has already been performed.  
Note however that, in order to apply the operator $\mydelta $ to the product $(R[U,A;\vec g]\; \mathcal{O}[U,A;\vec g])$ (see eqs.~(\ref{eq:deltarew}) and~(\ref{eq:frezzotti})  above), at fixed QED gauge background one also needs the following expressions for the first order derivatives of the quark propagators and of the quark determinants with respect to $e$ 
\begin{eqnarray} 
\frac{\partial S_f}{\partial  e} &=& -S_f \frac{\partial D_f}{\partial  e} S_f \ =\ e_f \golvert \; ,
\nonumber \\
\nonumber \\
\frac{\partial r_f}{\partial e}
&=&
\mbox{Tr}\left(S_f \frac{\partial D_f}{\partial e}\right)
\ =\ -{\color{blue}e_f}\vpz \; .
\end{eqnarray}

A concrete example of application of the formulae given in eqs.~(\ref{eq:graphicalnotation2}) and~(\ref{eq:graphicalnotation1}) is represented by the correction to the $S_f^\pm$ quark propagators worked out below
\begin{eqnarray}
&&\mydelta  \gol^\pm =
\nonumber \\
\nonumber \\
&&
(e_f  e)^2 \golself + (e_f  e)^2 \golltad 
-[m_f-m_f^0] \goi
\mp [m^{cr}_f-m^{cr}_0] \goip
\nonumber \\
\nonumber \\
&&
-e^2 e_f\sum_{f_1}{\color{blue}e_{f_1}} \goltad
-e^2\sum_{f_1}{\color{blue}e_{f_1}^2} \golvp 
-e^2 \sum_{f_1}{\color{blue} e_{f_1}^2} \golvppp
+e^2\sum_{f_1 f_2}{\color{blue}e_{f_1}}
{\color{red}e_{f_2}} \golvpp
\nonumber \\
\nonumber \\
&&
+\sum_{f_1}{\color{blue} \pm [m^{cr}_{f_1}-m^{cr}_0]} \goipdisc
+\sum_{f_1}{\color{blue} [m_{f_1}-m_{f_1}^0]} \goidisc
+\left[g_s^2-(g_s^0)^2\right]\goiplaq   \; .
\label{eq:qprop2}
\end{eqnarray}
Here quarks propagators of different flavours have been drawn with different colors and different lines.

The formulae above have been explicitly displayed not only because they represent the building blocks of the derivation of the LIB corrections to the hadron masses discussed in the following, but also for illustrating the implications of the electro--quenched approximation (see eq.~(\ref{eq:electroquenched}) above).  
This approximation is not required in the calculation of the pion mass splitting because
the quark disconnected diagrams containing sea quark loops are exactly canceled in the difference of $\mydelta  M_{\pi^+}$ and $\mydelta  M_{\pi^0}$ (see eq.~(\ref{eq:pionmasses}) below). This does not happen in the case of the kaon mass difference, see eq.~(\ref{eq:kaonmasses}). Quark disconnected diagrams are noisy and difficult to calculate and, for this reason, we have derived the numerical results for $M_{K^+}-M_{K^0}$ within the electro--quenched approximation. The perturbative expansion of the electro--quenched theory, i.e. the theory corresponding to the action $S_{sea}^{e=0}$ for the sea quarks, is obtained in practice by setting $g_s=g_s^0$ and
\begin{eqnarray}
r_f[U,A,\vec g_0]=1 \; .
\end{eqnarray}
In the electro-quenched approximation all quark disconnected contributions are absent. It follows that in this theory eq.~(\ref{eq:qprop2}) simply becomes
\begin{eqnarray}
\mydelta  \gol^\pm &=&
(e_f  e)^2 \left[ \golself + \golltad \right]
-[m_f-m_f^0] \goi
\mp[m^{cr}_f-m^{cr}_0] \goip 
\; .
\nonumber \\
\label{eq:qprop2eq}
\end{eqnarray}

\subsection{LIB corrections to hadron correlators}
\label{sec:hadronmasses}
In order to extract the mass of a given hadron $H$, by including electromagnetic interactions and QCD isospin breaking corrections, we start by considering in the full theory the two-point correlator of an interpolating operator $\mathcal{O}_H(t,\vec p=0)$ having the appropriate quantum numbers,
\begin{eqnarray}
C_{HH}(t;\vec g)&=&\langle\ \mathcal{O}_H(t)\ \mathcal{O}_H^\dagger(0)\ \rangle^{\vec g}
\ =\ Z_H e^{-tM_H}\ + \ \cdots \; ,
\nonumber \\
\nonumber \\
e^{M_H} &=& \frac{C_{HH}(t-1;\vec g)}{C_{HH}(t;\vec g)}\ + \ \cdots \; ,
\label{eq:corrchh}
\end{eqnarray}
where the dots represent non leading exponential contributions to the correlator.
It is important to stress that, if $H$ is an electrically charged particle, the correlator $C_{HH}(t;\vec g)$ is $not$ invariant under $U(1)_{em}$ gauge transformations. 
For this reason it is not possible, in general, to extract physical informations directly from $Z_H$, the residue of the pole corresponding to the hadron $H$ (see also ref.~\cite{Gasser:2010wz} concerning this point). On the other hand, the mass of the hadron $M_H$ is gauge invariant and \emph{finite} in the continuum limit, provided the parameters $\vec g$ of the action have been properly tuned. It follows that, at large times and at any given order in a perturbative expansion in any of the parameters of the action, the ratio $C_{HH}(t-1;\vec g)/C_{HH}(t;\vec g)$ is both gauge and renormalization group invariant (up to discretization effects and exponentially suppressed contributions). From eqs.~(\ref{eq:corrchh}) it follows
\begin{eqnarray}
&&C_{HH}(t;\vec g) = C_{HH}(t;\vec g^0)\left[ 1 + \frac{\mydelta  C_{HH}(t)}{C_{HH}(t;\vec g^0)}
+\dots \right] \; ,
\nonumber \\
\nonumber \\
\nonumber \\
&&\mydelta  M_H=M_H -M_H^0= -\partial_t \frac{\mydelta  C_{HH}(t)}{C_{HH}(t;\vec g^0)}+\dots  \; ,
\label{eq:gencorr}
\end{eqnarray}
where we have defined
\begin{eqnarray}
\partial_t f(t) = f(t)-f(t-1) \;
\label{eq:partialt}
\end{eqnarray}
and $\mydelta $ is defined in eq.~(\ref{eq:latstrategy1}).

In our lattice simulations we have enforced periodic (anti--periodic) boundary conditions for the gauge (matter) fields along the time direction. For this reason, we have extracted the correction to pseudoscalar meson masses by fitting the ratio $\mydelta  C_{PP}(t)/C_{PP}(t;\vec g^0)$ of corrected over uncorrected pseudoscalar--pseudoscalar two--point functions according to the following functional form 
\begin{eqnarray}
\frac{\mydelta  C_{PP}(t)}{C_{PP}(t;\vec g^0)}
= const. + \mydelta  M_{P}(T/2-t)\tanh\left[ M_{P}^0(T/2-t)\right] +\cdots \; ,
\label{eq:latticepartialt}
\end{eqnarray}
where the constant term contains the correction to the residue of the pole corresponding to the lightest state of mass $M_{P}$ and $T$ is the extension of the time direction of the lattice. The formula above is obtained by noting that a pseudoscalar--pseudoscalar correlator is even under the symmetry $t\mapsto T-t$ and by using ordinary perturbation theory in order to predict the time dependence of corrected correlators, see ref.~\cite{deDivitiis:2011eh} for further details concerning this point. In the following we continue to use the symbol $\partial_t$ but, when referred to lattice correlators, we actually mean the operation that allows extracting the coefficient $\mydelta  M_{P}$ by a fit of the numerical correlators according to eq.~(\ref{eq:latticepartialt}).

As explained in section~\ref{sec:wilsonfermions},
in order to minimize cutoff effects and optimize the numerical signal, we work in a mixed action setup and extract both charged and neutral meson masses from two--point correlators of twisted Wilson quarks having opposite chirally rotated Wilson terms. In practice, the use of interpolating pseudoscalar operators of the form $\bar \psi_{f_1}^+ \gamma^5 \psi_{f_2}^-$ (see eq.~(\ref{eq:plusminusoh}) above) corresponds to   
\begin{eqnarray}
&&\gdsl \longrightarrow 
\begin{array}{c}
+\\
\gdsl\\
-
\end{array}
=\mbox{Tr}\left\{\; 
\gamma^5\; {\color{red}S_{f_1}^+[U;\vec g^0]} \;
\gamma^5\; S_{f_2}^-[U;\vec g^0] \;
\right\} \; .
\end{eqnarray}
The same assignment of Wilson parameter signs has been used also when the two quark propagators correspond to the same physical flavour (for example $f_1=f_2=u$), see appendix~\ref{sec:tm} for further details.

\subsection{Pion two--point functions}
\label{sec:pionstwopt}
Given the observations made in the previous subsection, we now derive the leading isospin breaking corrections to pion masses by using the same technique employed to obtain the corrections to the quark propagator. In the case of the charged pions we can start from the full theory correlator
\begin{eqnarray}
C_{\pi^+\pi^-}(t;\vec g) =
\langle\ [\bar u^+ \gamma^5 d^-](t,\vec p= 0)\ [\bar d^- \gamma^5 u^+](0)\ \rangle^{\vec g} \;,   
\end{eqnarray}
and apply the differential operator $\mydelta $ defined in eqs.~(\ref{eq:latstrategy1}). We get
\begin{eqnarray}
\mydelta  M_{\pi^+}=
&-&e_ue_d  e^2 \partial_t\frac{\gdllexch}{\gdll}
-(e_u^2+e_d^2) e^2\partial_t\frac{\gdllself+\gdllphtad}{\gdll}
+2[m_{ud}-m_{ud}^0]\partial_t\frac{\gdli}{\gdll}
\nonumber \\
\nonumber \\
&+&(e_u+e_d) e^2 \sum_{f=sea}{{\color{blue}e_f}\partial_t\frac{\gdlltadf}{\gdll}}
-(m^{cr}_u+m^{cr}_d-2m^{cr}_0)\partial_t\frac{\gdlip}{\gdll}
+ \mbox{[isosym. vac. pol.]} \; ,
\nonumber \\
\label{eq:mpip}
\end{eqnarray}
where $m_{ud}=(m_d+m_u)/2$ is the bare isosymmetric light quark mass.
In the case of the neutral pion we obtain
\begin{eqnarray}
\mydelta  M_{\pi^0}=
&-&\frac{e_u^2+e_d^2}{2}  e^2\partial_t\frac{\gdllexch}{\gdll}
-(e_u^2+e_d^2) e^2\partial_t\frac{\gdllself+\gdllphtad}{\gdll}
+2[m_{ud}-m_{ud}^0]\partial_t\frac{\gdli}{\gdll}
\nonumber \\
\nonumber \\
&+&(e_u+e_d) e^2 \sum_{f=sea}{{\color{blue}e_f}\partial_t\frac{\gdlltadf}{\gdll}}
-(m^{cr}_u+m^{cr}_d-2m^{cr}_0)\partial_t\frac{\gdlip}{\gdll}
\nonumber \\
\nonumber \\
&+&\frac{(e_u-e_d)^2}{2}e^2\partial_t\frac{\discgdllexch}{\gdll}
+ \mbox{[isosym. vac. pol.]} \; .
\label{eq:mpi0}
\end{eqnarray}
The sea quark propagators have been drawn in blue (and with a different line) and the isosymmetric vacuum polarization diagrams have not been displayed explicitly. By combining the previous expressions we find the elegant formula
\begin{eqnarray}
M_{\pi^+}-M_{\pi^0}
&=&
\frac{(e_u-e_d)^2}{2} e^2\partial_t\frac{\gdllexch-\discgdllexch}{\gdll}\; .
\label{eq:pionmasses}
\end{eqnarray}
All the isosymmetric vacuum polarization diagrams cancel by taking the difference of $\mydelta  M_{\pi^+}$ and $\mydelta  M_{\pi^0}$ together with the disconnected sea quark loop contributions explicitly shown in eqs.~(\ref{eq:mpip}) and~(\ref{eq:mpi0}).
Note, in particular, the cancellation of the corrections/counter--terms corresponding to the variation of the  symmetric up--down quark mass $m_{ud}-m_{ud}^0$ and to the variation of the strong coupling constant $g_s^2-(g_s^0)^2$. This is a general feature: at first order of the perturbative expansion in $\hat \alpha_{em}$ and $\hat m_d-\hat m_u$, the isosymmetric corrections coming from the variation of the stong gauge coupling (the lattice spacing), of $m_{ud}$ and of the heavier quark masses do not contribute to observables that vanish in the isosymmetric theory, like the mass splitting $M_{\pi^+}-M_{\pi^0}$. Furthermore, as already stressed, the electric charge does not need to be renormalized at this order and, for all these reasons, the expression for the pion mass splitting can be considered a ``clean" theoretical prediction. 

On the other hand, the lattice calculation of the disconnected diagram present in eq.~(\ref{eq:pionmasses}) is a highly non trivial numerical problem and we shall neglect this contribution in this paper. Relying on the same arguments that lead to the derivation of the flavor $SU(3)$ version of the Dashen's theorem, see eq.~(\ref{eq:kcfirststrategy}), it can be shown that the neutral pion mass has to vanish in the limit $\hat m_u=\hat m_d=0$ for arbitrary values of $e_u$, $e_d$ as well as the masses $\hat m_f$ and the electric charges $e_f$ of the heavier quarks. This happens because the electric charge operator is diagonal in the up--down space and commutes with the isospin generator $\tau^3$. Once the critical mass counter--terms $m^{cr}_{u,d}-m^{cr}_0$ have been properly tuned, the contributions to eq.~(\ref{eq:mpi0}) can be separated by the dependence with respect to $e_u$, $e_d$ and $e_f$ of the different coefficients. It follows that the disconnected diagram of eq.~($\ref{eq:pionmasses}$) vanishes in the $SU(2)$ chiral limit and, consequently, it is of $O(\hat \alpha_{em} \hat m_{ud})$ . 
Neglecting this $O(\hat \alpha_{em} \hat m_{ud})$ diagram we are thus introducing a small systematic error that, from the phenomenological point of view, can be considered of the same order of magnitude of the other $O(\hat \alpha_{em} [\hat m_{d}-\hat m_{u}])$ contributions neglected in this paper.

\subsection{Kaon two--point functions}
\label{sec:kaonstwopt}
By repeating the analysis performed for the pions in the case of kaon two--point functions, we obtain the following result for the LIB corrections to $M_{K^+}$
\begin{eqnarray}
\mydelta  M_{K^+}=&+&
[m_u-m_{ud}^0] \partial_t \frac{\gdsi}{\gdsl}-e_ue_s  e^2 \partial_t \frac{\gdslexch}{\gdsl}
-e_u^2  e^2 \partial_t \frac{\gdslselfl+\gdslphtadl}{\gdsl}
\nonumber \\
\nonumber \\
&+&e_u e^2 \sum_f{ {\color{blue} e_f} \partial_t \frac{\gdslltadf}{\gdsl}}
-[m^{cr}_u-m^{cr}_0]\partial_t \frac{\gdsip}{\gdsl}
+[m^{cr}_s-m^{cr}_0]\partial_t \frac{\gdsipl}{\gdsl}
\nonumber \\
\nonumber \\
&+&
e_s e^2 \sum_f{ {\color{blue} e_f} \partial_t \frac{\gdsltadf}{\gdsl}}
-e_s^2  e^2 \partial_t \frac{\gdslselfs+\gdslphtads}{\gdsl}
+[m_s-m_{s}^0] \partial_t \frac{\gdsil}{\gdsl}
\nonumber \\
\nonumber \\
&+&
 \mbox{[isosymmetric vac. pol.]} \; ,
\end{eqnarray}
where the strange quark propagator is drawn in red and the sea quark propagators in blue (with different lines). The leading corrections to $M_{K^0}$ are given by
\begin{eqnarray}
\mydelta  M_{K^0}=&+&
[m_d-m_{ud}^0]  \partial_t \frac{\gdsi}{\gdsl}-e_de_s  e^2 \partial_t \frac{\gdslexch}{\gdsl}
-e_d^2  e^2 \partial_t \frac{\gdslselfl+\gdslphtadl}{\gdsl}
\nonumber \\
\nonumber \\
&+&e_d e^2 \sum_f{ {\color{blue} e_f} \partial_t \frac{\gdslltadf}{\gdsl}}
-[m^{cr}_d-m^{cr}_0]\partial_t \frac{\gdsip}{\gdsl}
+[m^{cr}_s-m^{cr}_0]\partial_t \frac{\gdsipl}{\gdsl}
\nonumber \\
\nonumber \\
&+&
e_s e^2 \sum_f{ {\color{blue} e_f} \partial_t \frac{\gdsltadf}{\gdsl}}
-e_s^2  e^2 \partial_t \frac{\gdslselfs+\gdslphtads}{\gdsl}
+[m_s-m_{s}^0] \partial_t \frac{\gdsil}{\gdsl}
\nonumber \\
\nonumber \\
&+&
 \mbox{[isosymmetric vac. pol.]} \; .
\label{eq:mk0}
\end{eqnarray}
By taking the difference of the last two expressions we get
\begin{eqnarray}
M_{K^+}-M_{K^0}&=&
(e_u^2-e_d^2)  e^2\partial_t \frac{\gdslexch}{\gdsl}
-(e_u^2-e_d^2)  e^2 \partial_t \frac{\gdslselfl+\gdslphtadl}{\gdsl}
\nonumber \\
\nonumber \\
&-&2\Delta  m_{ud} \partial_t \frac{\gdsi}{\gdsl}
-(\Delta m^{cr}_u-\Delta m^{cr}_d)\partial_t \frac{\gdsip}{\gdsl}
+(e_u-e_d)  e^2 \sum_f{ {\color{blue} e_f} \partial_t \frac{\gdslltadf}{\gdsl}}
\; ,
\nonumber \\
\label{eq:kaonmasses}
\end{eqnarray}
where we defined
\begin{eqnarray}
\Delta m_{ud}=\frac{m_d-m_u}{2} \; , \qquad\qquad \Delta m^{cr}_f= m^{cr}_f-m^{cr}_0 \;,
\end{eqnarray}
and used the relation $e_s=-(e_u+e_d)$.
Also in the kaon sector, by taking the difference $(\mydelta  M_{K^+}-\mydelta  M_{K^0})$, all the isosymmetric vacuum polarization diagrams cancel as well as the corrections/counter--terms corresponding to the variation of the  symmetric up--down quark mass $m_{ud}-m_{ud}^0$ and of the strange quark mass $m_{s}-m_{s}^0$.
The ``sea--tadpole" diagrams don't cancel and contribute to $M_{K^+}-M_{K^0}$. These terms, absent in the case of the pion mass difference, vanish however in the $SU(3)$ chiral limit and/or within the electro--quenched approximation that we shall employ in sections~\ref{sec:tuningk} and~\ref{sec:separating} to obtain our numerical results for the kaon mass difference.

\section{Pion masses}
\label{sec:pionmasses}
\begin{figure}[!t]
\begin{center}
\includegraphics[width=0.49\textwidth]{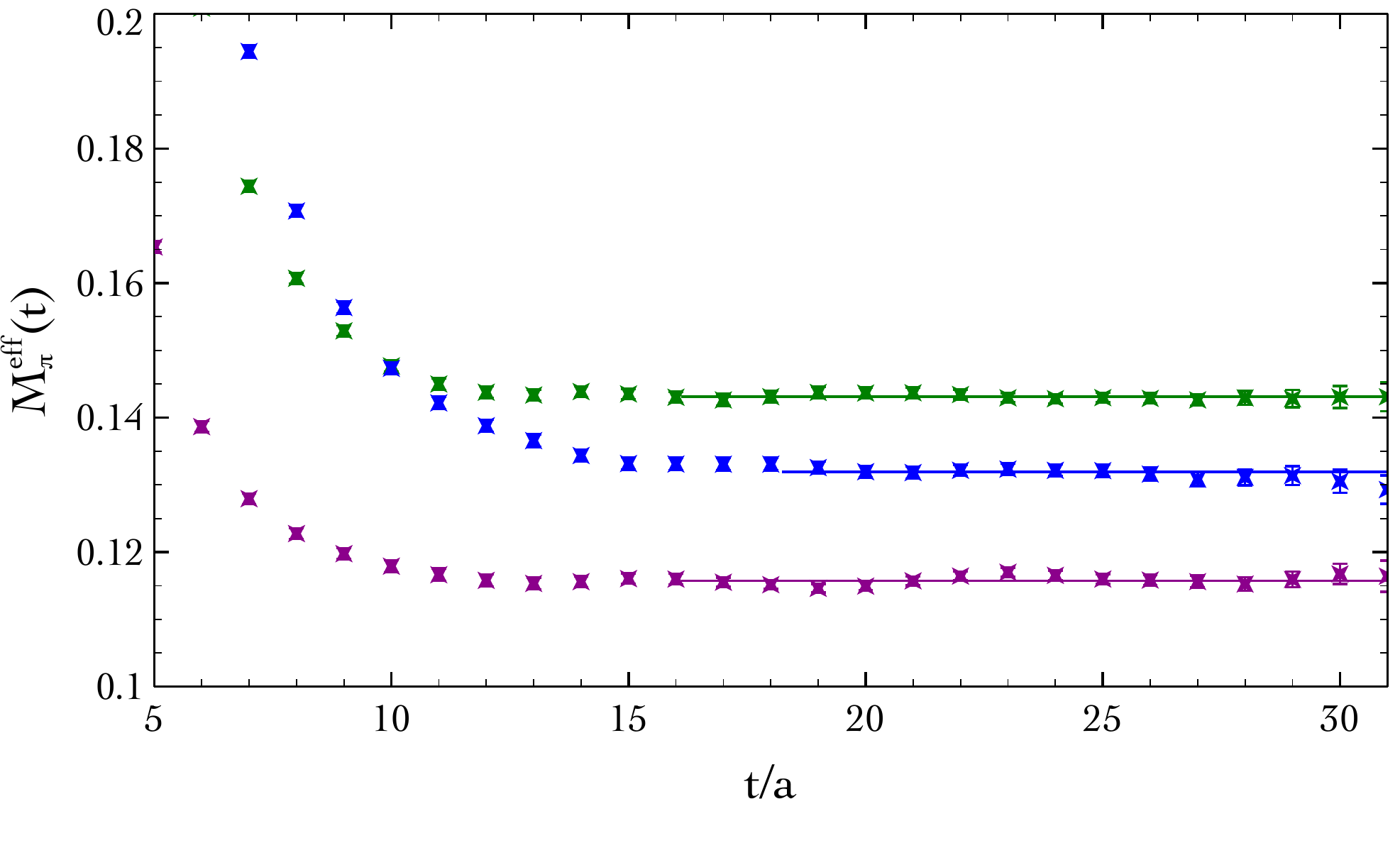}\hfill
\includegraphics[width=0.49\textwidth]{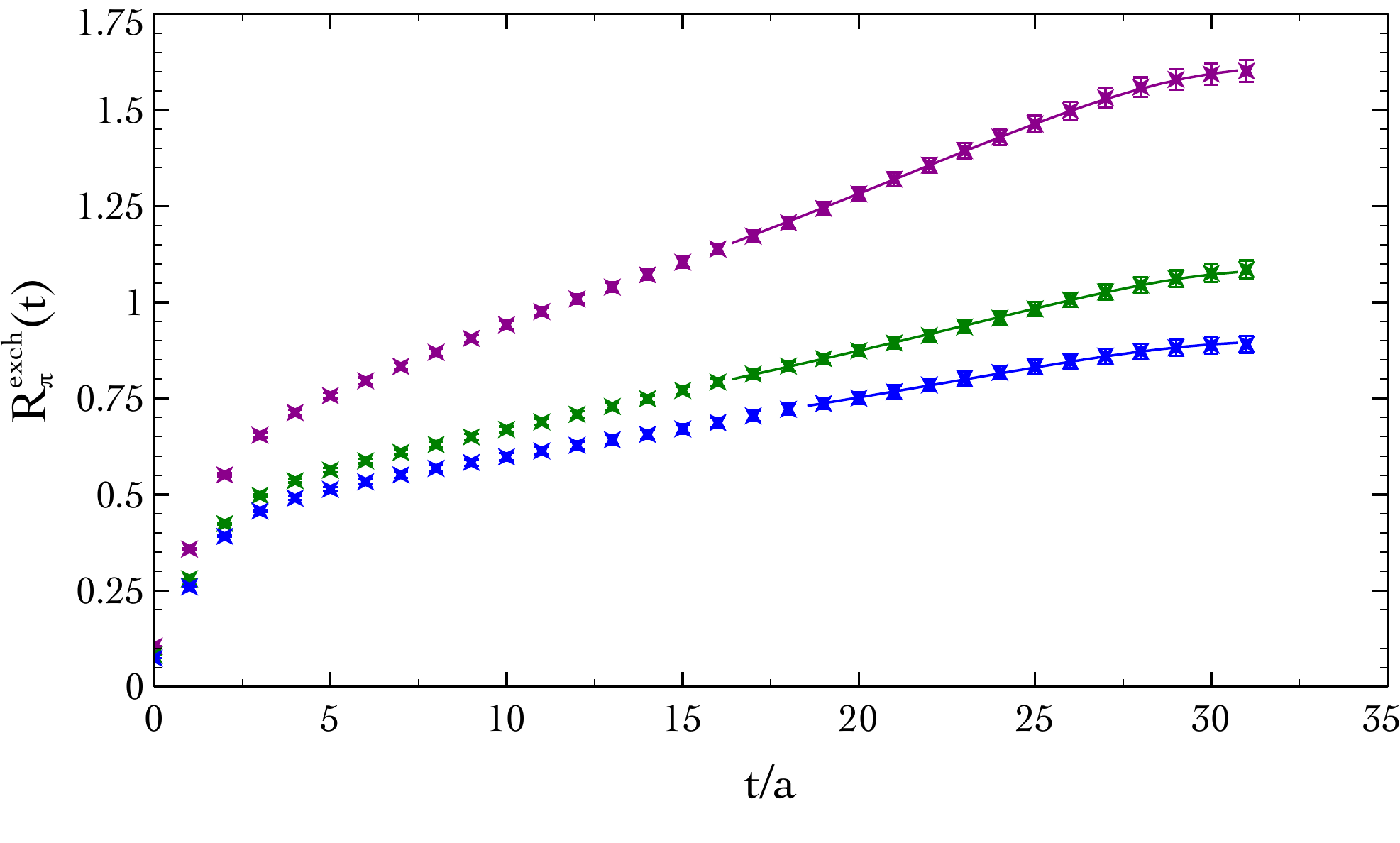}
\caption{\label{fig:PIslopes} \footnotesize
{\it Left panel}: extraction of the pion mass in the isosymmetric theory from $C_{\pi\pi}(t;\vec g^0)$.
{\it Right panel}: fits of $R^{exch}_{\pi}(t)$ according to eq.~(\ref{eq:latticepartialt}). The dark magenta points correspond to $\beta=3.90$ and $(am_{ud})^0=0.0030$, the green points to $\beta=4.05$ and $(am_{ud})^0=0.0060$ while the blue points correspond to $\beta=4.20$ and $(am_{ud})^0=0.0065$ (see Appendix~\ref{sec:tm}). Data are in lattice units.
}
\end{center}
\end{figure}

In this section we discuss our results for the physical pion mass splitting. The starting point is eq.~(\ref{eq:pionmasses}) that, by neglecting the disconnected diagram coming from the neutral pion, can be conveniently rewritten as
\begin{eqnarray}
R^{exch}_{\pi}(t)=\frac{\gdllexch}{\gdll}
\qquad
\longrightarrow
\qquad
M_{\pi^+}^2-M_{\pi^0}^2 =
(e_u-e_d)^2  e^2 M_\pi\  
\partial_t R^{exch}_{\pi}(t) \; .
\label{eq:pionmasseseq}
\end{eqnarray}
In the left panel of Figure~\ref{fig:PIslopes} we show the mass of the pions in the isosymmetric theory, $M_\pi$, as extracted from the unperturbed correlators $C_{\pi\pi}(t;\vec g^0)$.  The data are shown for different values of the lattice spacing and for different values of the symmetric light quark mass $m_{ud}^0$. In the right panel of Figure~\ref{fig:PIslopes} we show the fits of the ratio of correlators $R^{exch}_{\pi}(t)$ according to eq.~(\ref{eq:latticepartialt}). As can be seen, we are able to obtain a good numerical signal by using three photon stochastic sources per QCD gauge configuration. The data follow the expected behavior as a function of the time variable.  

\begin{figure}[!t]
\begin{center}
\includegraphics[width=0.6\textwidth]{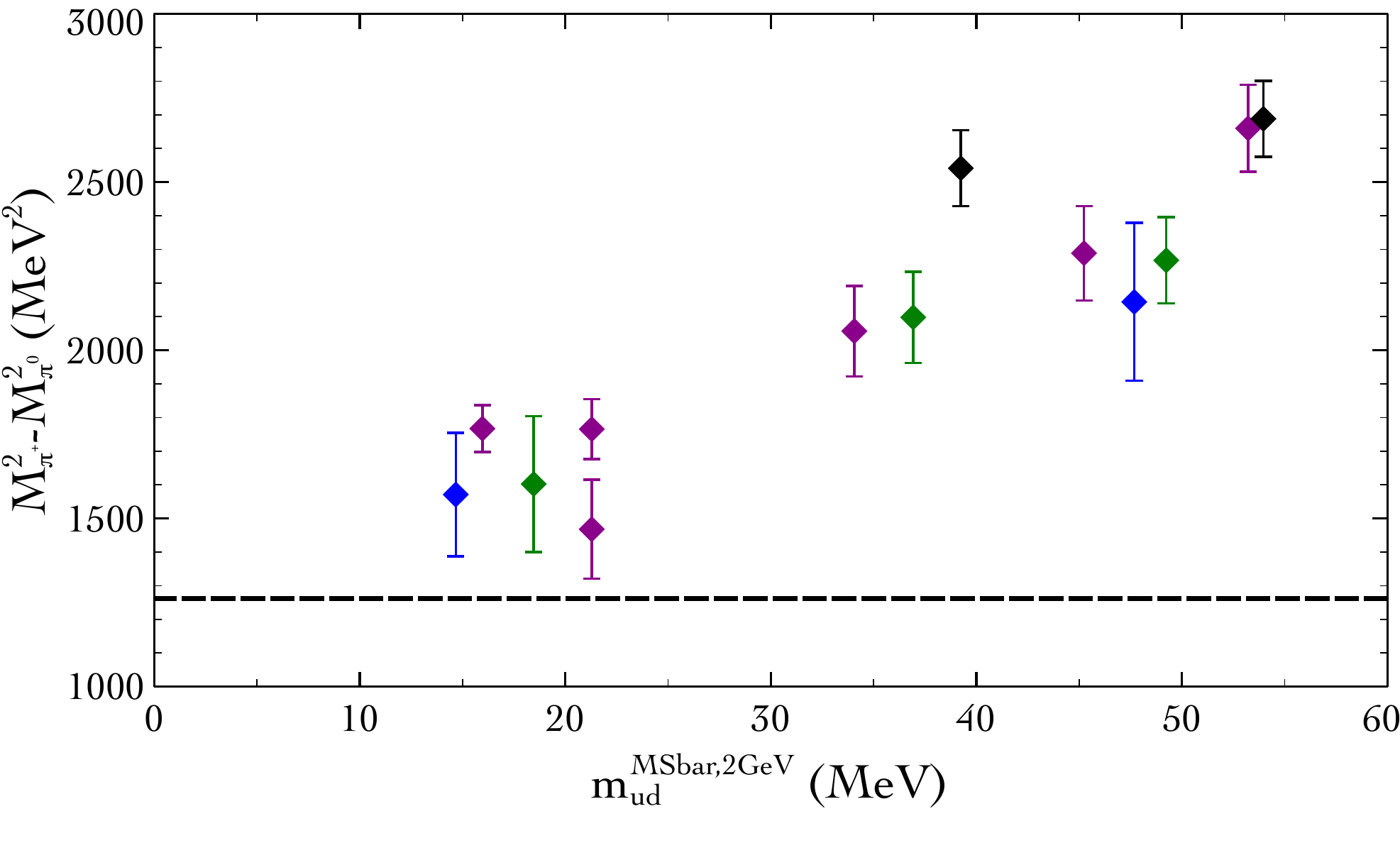}\hfill
\caption{\label{fig:PIchiral} \footnotesize
$M_{\pi^+}^2-M_{\pi^0}^2$ as a function of $\hat m_{ud}$ as extracted from lattice correlators and converted in physical units by using $a^0$.
Black points correspond to $\beta=3.80$, dark magenta points correspond to $\beta=3.90$, green points correspond to $\beta=4.05$ and blue points correspond to $\beta=4.20$ (see Appendix~\ref{sec:tm}). The dashed horizontal line correspond to the experimental value of $M_{\pi^+}^2-M_{\pi^0}^2$. The physical value of the symmetric combination of the light quark masses is $\hat m_{ud}(\overline{MS},2\ GeV)=3.6(2)\mbox{ MeV}$.
}
\end{center}
\end{figure}
Using eq.~(\ref{eq:pionmasseseq}), the pion mass difference extracted from the fits of Figure~\ref{fig:PIslopes} can be converted in the physical result for $M_{\pi^+}^2-M_{\pi^0}^2$ multiplying the results by the factor
\begin{eqnarray}
 e^2 = \hat e^2=4\pi\hat \alpha_{em}=\frac{4\pi}{137}
\end{eqnarray}
and by two powers of the inverse lattice spacing $a^0$ determined within the isosymmetric theory (see Table~\ref{tab:gaugeconfigs}). Indeed, as shown in the diagrammatic analysis of section~\ref{sec:pionstwopt}, the pion mass difference is a genuine isospin breaking effect and the change of the lattice spacing as well as the change of the average up--down quark mass enter at higher orders in the perturbative expansion.
In Figure~\ref{fig:PIchiral} we show the data for $M_{\pi^+}^2-M_{\pi^0}^2$ converted in physical units at the four different values of lattice spacings used in this paper and for the different values of $\hat m_{ud}$ used in the simulations. The horizontal black line in the figure corresponds to the experimental determination of the pion mass difference squared and the lattice data are very close to the experimental value. 

Our pions are heavier than the physical ones and our lattice data need to be extrapolated toward the chiral limit. Furthermore, QED is a long range interaction and we have to cope with the associated power--law finite volume effects. The chiral extrapolation and the removal of lattice artifacts from simulated numerical data is the subject of section~\ref{sec:extrapolations}.

\section{Tuning critical masses}
\label{sec:tuningk}
In order to extract physical informations from the expression for $M_{K^+}-M_{K^0}$, see eq.~(\ref{eq:kaonmasses}) above, we first need to obtain a numerical determination of the electromagnetic shift of the critical masses of the light quarks. Our results for the kaon mass splitting have been obtained within the electro--quenched approximation that, consistently, we employ in this section to calculate $\Delta m^{cr}_{u,d}$.

As discussed in section~\ref{sec:kmasses}, we can use two different conditions to obtain a numerical estimate of the divergent parameters $\Delta m^{cr}_{u,d}$. The first strategy, based on Dashen's theorem, consists in imposing the validity of the continuum $SU(3)$ chiral limit relations
\begin{eqnarray}
\lim_{ \hat m_f \mapsto 0}{M_{\pi^0}} \ = \ \lim_{ \hat m_f \mapsto 0}{M_{K^0}} \ =\ 0 \; ,
\end{eqnarray}
where $\hat m_f=\{\hat m_u,\hat m_d,\hat m_s\}$.
Relying on the determination of the QCD critical mass $m^{cr}_0$ performed in ref.~\cite{Baron:2009wt} and using eqs.~(\ref{eq:mpi0}) and~(\ref{eq:mk0}), we have that in the electro--quenched approximation the neutral pion and neutral kaon masses vanish for $\Delta m^{cr}_{f}$ given by 
\begin{eqnarray}
\Delta m^{cr}_f
&=&-\frac{e_f^2}{2} e^2
\lim_{ \hat m_f\mapsto 0}{
\frac{\partial_t\frac{\gdllexch}{\gdll}+
2\partial_t\frac{\gdllself+\gdllphtad}{\gdll}}
{\partial_t\frac{\gdlip}{\gdll}}} \; ,
\label{eq:kc1}
\end{eqnarray}
where $f=\{u,d,s\}$. From the numerical point of view, the parameters $\Delta m^{cr}_{f}$ have to be determined as accurately as possible because they are needed in order to cancel a linear ultraviolet divergence present in eq.~(\ref{eq:kaonmasses}). The numerical problem with eq.~(\ref{eq:kc1}) is that the associated determination of $\Delta m^{cr}_{f}$ requires a chiral extrapolation and this in turn introduces larger uncertainties compared to the alternative method discussed in section~\ref{sec:kmasses}, namely the numerical determination of the electromagnetic critical masses based on the use of the WTI of eq.~(\ref{eq:tmvwi}). 

\begin{figure}[!t]
\begin{center}
\includegraphics[width=0.49\textwidth]{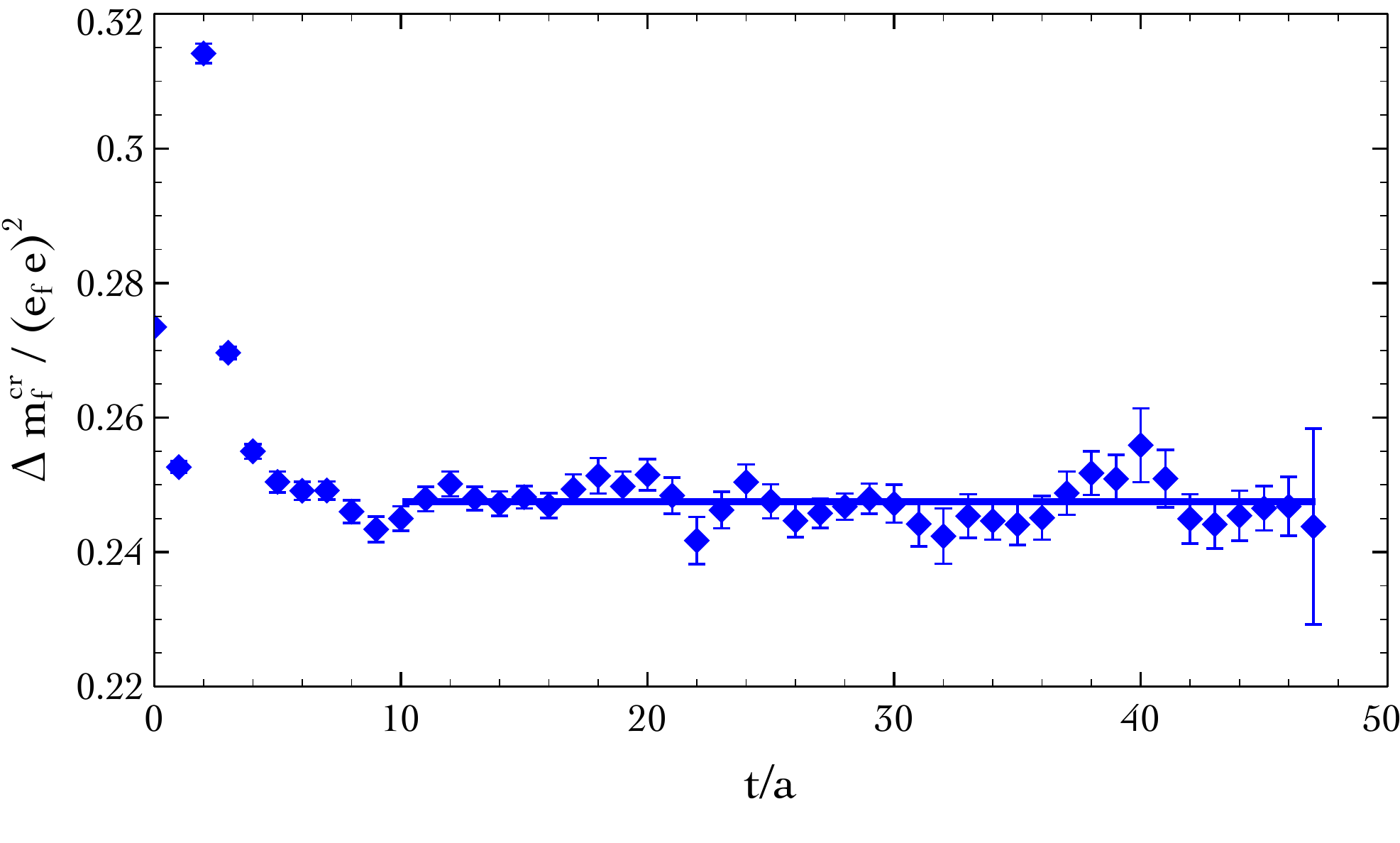}\hfill
\includegraphics[width=0.49\textwidth]{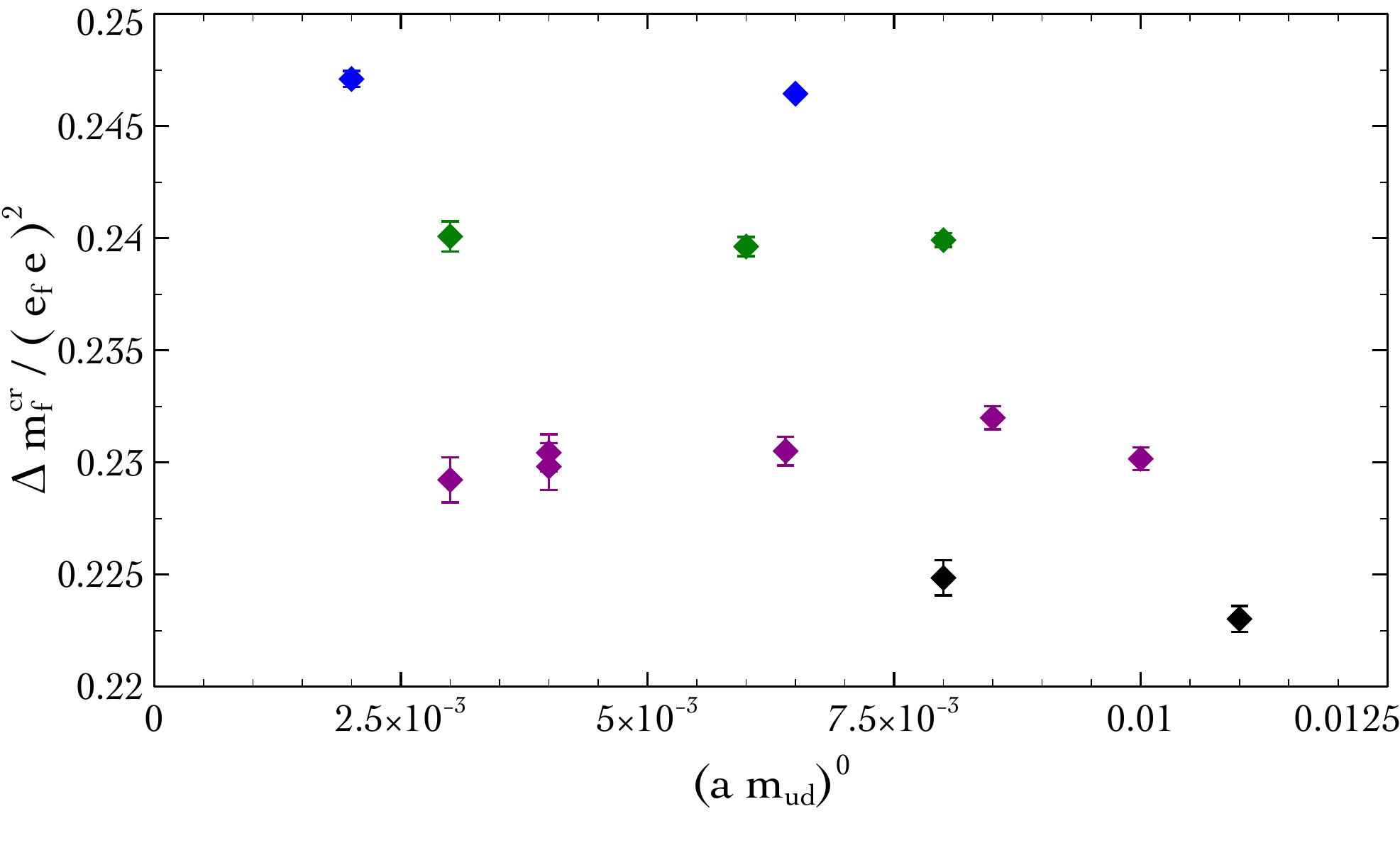}
\caption{\label{fig:KC} \footnotesize
{\it Left panel}: determination of $\Delta m^{cr}_f$ according to eq.~(\ref{eq:kc2}) for the simulation corresponding to $\beta=4.20$ and $(am_{ud})^0=0.0020$ (see Appendix~\ref{sec:tm}). As expected the combination of correlators appearing into eq.~(\ref{eq:kc2}) give a constant plateau in time from which we extract $\Delta m^{cr}_f$.
{\it Right panel}: numerical results for $\Delta m^{cr}_f$ for the different simulations. Black points correspond to $\beta=3.80$, dark magenta points correspond to $\beta=3.90$, green points correspond to $\beta=4.05$ and blue points correspond to $\beta=4.20$. As expected the critical mass counter--terms depend very mildly from the simulated symmetric light quark mass $(am_{ud})^0$: the small dependence is due to statistical fluctuations and (small) cutoff effects. 
}
\end{center}
\end{figure}
By applying the methods of section~\ref{sec:pathintegral} to the Ward--Takahashi identity $W_f(\vec g)=0$, i.e. by applying the differential operator $\mydelta $ to the full theory parity--odd correlator (l.h.s. of eq.~(\ref{eq:tmvwi}))
\begin{eqnarray}
W_f(\vec g)=
-\nabla_0 
\begin{array}{c}
+\\
\gvdll\\
-
\end{array}
= -\nabla_0\ \mbox{Tr}\left\{\; 
\gamma^0\; S_{f}^+[U,A; \vec g;t,\vec p=0] \;
\gamma^5\; S_{f}^-[U,A; \vec g;-t,\vec p=0] \;
\right\} =0 \; ,
\end{eqnarray}
one obtains the following alternative definition of $\Delta m^{cr}_{f}$
\begin{eqnarray}
\mydelta  W_f=0 \qquad \longrightarrow \qquad
\Delta m^{cr}_f
\; = \; -\frac{e_f^2}{2} e^2
\frac{ \nabla_0\left[ \gvdllexch +2\gvdllself+2\gvdllphtad \right]}
{\nabla_0\ \gvdlip}
\; .
\label{eq:kc2}
\end{eqnarray}
Note that the two definitions of eqs.~(\ref{eq:kc1}) and~(\ref{eq:kc2}) have the same ``structure" in terms of corrected correlators. Indeed, the Dashen's theorem is a consequence of the chiral WTI of the continuum theory and, concerning valence flavour doublets, eq.~(\ref{eq:tmvwi}) is the chirally twisted version of one of these relations. From the numerical point of view, however, the great advantage of eq.~(\ref{eq:kc2}) with respect to eq.~(\ref{eq:kc1}) is that the first does not require chiral extrapolations.

In the left panel of Figure~\ref{fig:KC} we show the combination of correlators appearing into eq.~(\ref{eq:kc2}) as a function of time for the simulation at $\beta=4.20$ and $(am_{ud})^0=0.0020$, see Appendix~\ref{sec:tm}. As expected, coming from a WTI, the numerical data exhibit a very long plateau from which we obtain a reliable determination of $\Delta m^{cr}_f$. We have similar results for the other values of quark masses and lattice spacings simulated in this paper. In the right panel of the same figure we show, for each lattice spacing, $\Delta m^{cr}_f$ as a function of $(a m_{ud})^0$. As expected the results are almost insensitive to $m_{ud}^0$. The tiny dependence that can be appreciated in the figure is due to statistical fluctuations and to (small) cutoff effects. In the following we use for each simulated value of $m_{ud}^0$ the corresponding determination of $\Delta m^{cr}_f$ and subsequently extrapolate the results for $M_{K^+}^2-M_{K^0}^2$ to continuum, infinite volume and chiral limits, see section~\ref{sec:extrapolations}. 

As a cross--check of our results, at $\beta=3.90$ where the number of simulated values of $m_{ud}^0$ do allow a reliable chiral extrapolation, we have compared the determination of $\Delta m^{cr}_f$ obtained by using eq.~(\ref{eq:kc1}) with the values in Figure~\ref{fig:KC}. Though the two determinations may differ because of cutoff effects, the two numerical values agree within the errors (that are two orders of magnitude larger for the value of $\Delta m^{cr}_f$ obtained from the chiral extrapolation).

\section{Kaon masses and separation of QED from QCD LIB effects}
\label{sec:separating}
With a reliable numerical determination of $\Delta m^{cr}_u$ and $\Delta m^{cr}_d$ we are now in the position of using the kaon mass difference formula of eq.~(\ref{eq:kaonmasses}) for physical applications. By defining
\begin{eqnarray}
&&R^m_{K}= \frac{\gdsi}{\gdsl} \;,
\qquad \qquad
R^{k}_{K}= \frac{\gdsip}{\gdsl} \;,
\nonumber \\
\nonumber \\
&&R^{exch}_{K}=\frac{\gdslexch}{\gdsl} \;,
\qquad \qquad
R^{self}_{K}= \frac{\gdslselfl+\gdslphtadl}{\gdsl} \; ,
\label{eq:Kratios}
\end{eqnarray}
in the electro--quenched approximation we have
\begin{eqnarray}
M_{K^+}-M_{K^0}&=&
-2\Delta  m_{ud}\ \partial_t R^m_{K}
-(\Delta m^{cr}_u-\Delta m^{cr}_d)\ \partial_t R^k_{K}
+(e_u^2-e_d^2)  e^2\ \partial_t\left[ R^{exch}_{K}
-R^{self}_{K}
\right]
\; .
\label{eq:kaonmasseseq}
\end{eqnarray}
The kaon mass splitting is a physical quantity and, since the electric charge does not renormalize at first order in $\hat \alpha_{em}$, the right hand side of eq.~(\ref{eq:kaonmasseseq}) can be made finite and equal to the physical value of $M_{K^+}-M_{K^0}$ by properly tuning the bare parameter $\Delta m_{ud}$. Afterward, the parameter $\Delta m_{ud}$ can be used in order to predict the mass splitting of other hadrons, as for example the neutron--proton mass difference.

Eq.~(\ref{eq:kaonmasseseq}) can also be used for introducing a renormalization prescription to separate QED from QCD isospin breaking corrections. This separation is not needed when, as in this paper, simulations are performed in the full theory but it may be useful in practice, as discussed in our previous work on the subject~\cite{deDivitiis:2011eh} or in ref.~\cite{Colangelo:2010et}. To this end, we need to express eq.~(\ref{eq:kaonmasseseq}) in terms of the renormalized light quark masses. Note that the bare parameters $m_{ud}$ and $\Delta m_{ud}$ of the full theory mix under renormalization because the two light quarks have different electric charges and, consequently, different renormalization constants $Z_{m_u}(\mu)$ and $Z_{m_d}(\mu)$. Specifically, we have
\begin{eqnarray}
\Delta m_{ud} = \frac{1}{2}\left(\frac{\hat m_d}{Z_{m_d}}-\frac{\hat m_u}{Z_{m_u}}\right)
&=&\frac{\Delta \hat m_{ud}}{Z_{ud}}+\frac{\hat m_{ud}}{\mathcal{Z}_{ud}} \; ,
\label{eq:deltamren}
\end{eqnarray}
where we have defined 
\begin{eqnarray}
\frac{1}{Z_{ud}}=\frac{1}{2}\left(\frac{1}{Z_{m_d}}+\frac{1}{Z_{m_u}}\right)\; ,
\qquad\qquad
\frac{1}{\mathcal{Z}_{ud}}=\frac{1}{2}\left(\frac{1}{Z_{m_d}}-\frac{1}{Z_{m_u}}\right)\; .
\end{eqnarray}
The mixing does not occur in the isosymmetric theory where the quarks are neutral with respect to electromagnetic interactions and we have
\begin{eqnarray}
\frac{1}{Z_{ud}^0}=Z_{\bar \psi \psi}^0 \; ,
\qquad
\qquad
\frac{1}{\mathcal{Z}_{ud}^0}=0 \;.
\end{eqnarray}
In the maximally twisted mass regularization used in this paper  $Z_{\bar \psi \psi}^0=Z_P^0$, see refs.~\cite{Frezzotti:2000nk,Dinter:2012tt}, and in our numerical analysis we have used the non--perturbative results for $Z_P^0(\beta^0)$ obtained in ref.~\cite{Constantinou:2010gr} and listed in Table~\ref{tab:gaugeconfigs}.
By neglecting all the contributions of $O(e^2 \Delta m_{ud})$, eq.~(\ref{eq:deltamren}) can be rewritten as
\begin{eqnarray}
\Delta m_{ud} = Z_{\bar \psi \psi}^0\ \Delta \hat m_{ud}+\frac{\hat m_{ud}}{\mathcal{Z}_{ud}} \; .
\label{eq:deltamren2}
\end{eqnarray}

The electromagnetic and strong contributions to the kaon mass difference are conveniently separated by plugging the previous expression into eq.~(\ref{eq:kaonmasseseq}) and by defining 
\begin{eqnarray}
&&\left[M_{K^+}-M_{K^0}\right]^{QED}(\mu)=
-2\hat m_{ud}\ \frac{\partial_t R^m_{K}}{\mathcal{Z}_{ud}}
-(\Delta m^{cr}_u-\Delta m^{cr}_d) \partial_t R^k_{K}
+
(e_u^2-e_d^2)  e^2\partial_t \left[ R^{exch}_{K}
-R^{self}_{K} \right]
,
\nonumber \\
\nonumber \\
&&\left[M_{K^+}-M_{K^0}\right]^{QCD}(\mu)=
-2\Delta  \hat m_{ud}\left( Z_{\bar \psi \psi}^0\ \partial_t R^{m}_{K} \right)\; ,
\nonumber \\
\nonumber \\
&&M_{K^+}-M_{K^0}= 
\left[M_{K^+}-M_{K^0}\right]^{QED}(\mu) + \left[M_{K^+}-M_{K^0}\right]^{QCD}(\mu) \; .
\label{eq:kaonmassessep}
\end{eqnarray}
This prescription was used in ref.~\cite{deDivitiis:2011eh} where we determined $\Delta \hat m_{ud}(\mu)$. 

Given a conventional separation between electromagnetic and strong isospin breaking corrections, violations to Dashen's theorem are parametrized in terms of ``small" parameters and, concerning the kaon mass difference, one has (see refs.~\cite{deDivitiis:2011eh,Colangelo:2010et})
\begin{eqnarray}
\varepsilon_\gamma(\mu) &=&
\frac{\left[M_{K^+}^2-M_{K^0}^2\right]^{QED}(\mu)-\left[M_{\pi^+}^2-M_{\pi^0}^2\right]^{QED}(\mu)}{M_{\pi^+}^2-M_{\pi^0}^2}
\nonumber \\
\nonumber \\
\nonumber \\
&=&
\frac{\left[M_{K^+}^2-M_{K^0}^2\right]^{QED}(\mu)}{M_{\pi^+}^2-M_{\pi^0}^2}-1 
\quad +\quad O(\hat \alpha_{em}\Delta \hat m_{ud})\; ,
\label{eq:prescription}
\end{eqnarray}
As observed in ref.~\cite{deDivitiis:2011eh} one can in principle specify a renormalization prescription by fixing a value for $\varepsilon_{\gamma}$ and by using eq.~(\ref{eq:prescription}) to compute the corresponding value of $\mathcal{Z}_{ud}(\mu)$.
This is not the strategy followed in this paper. Here, by relying on the perturbative result in the $\overline{MS}$ scheme that can be extracted from ref.~\cite{Aoki:1998ar},
\begin{eqnarray}
\frac{1}{\mathcal{Z}_{ud}(\overline{MS},\mu)}= \frac{(e_d^2-e_u^2)e^2}{32 \pi^2}\ 
\Big[\ 6 \log(a\mu)-22.596\dots\ \Big]\ Z_{\bar \psi \psi}^0 \; ,
\end{eqnarray}
we employ eq.~(\ref{eq:prescription}) to calculate $\varepsilon_\gamma(\overline{MS},\mu)$. 
The estimate provided for this quantity in ref.~\cite{Colangelo:2010et} and used in ref.~\cite{deDivitiis:2011eh} to calculate the leading QCD isospin breaking corrections to the $K_{\ell2}$ decay rate is $\varepsilon_\gamma \sim 0.7$. 

\begin{figure}[!t]
\begin{center}
\includegraphics[width=0.49\textwidth]{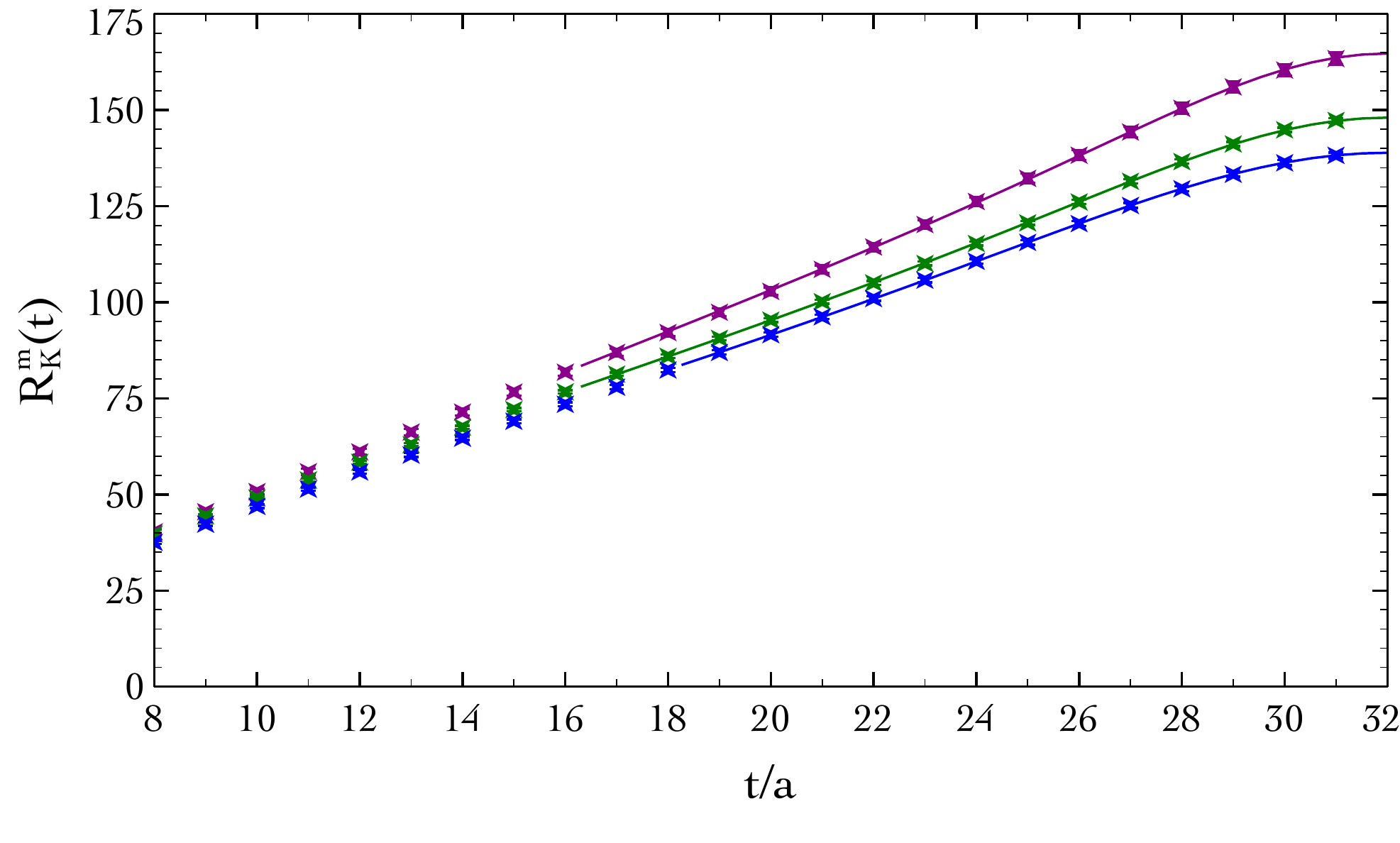}\hfill
\includegraphics[width=0.49\textwidth]{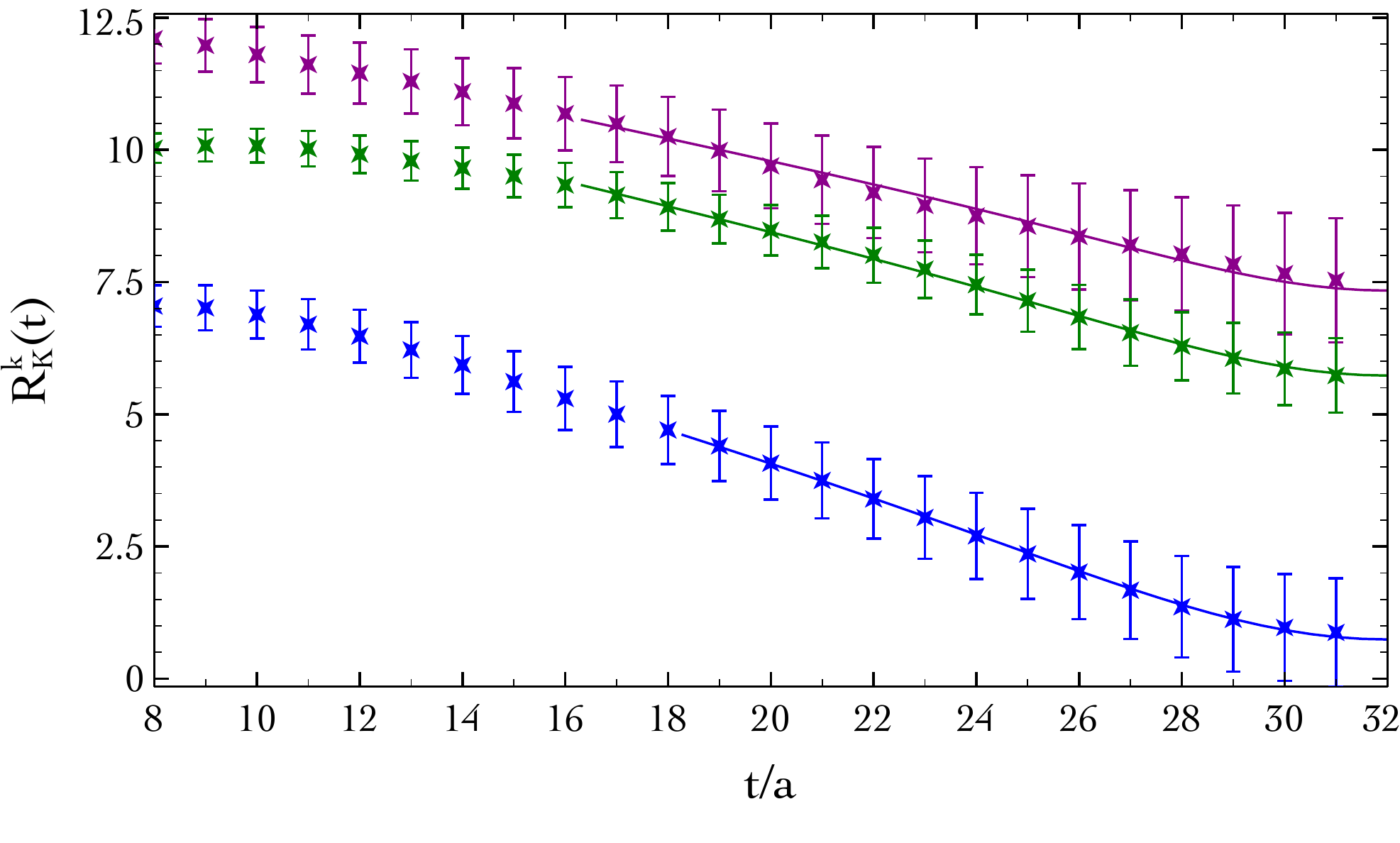}\hfill
\includegraphics[width=0.49\textwidth]{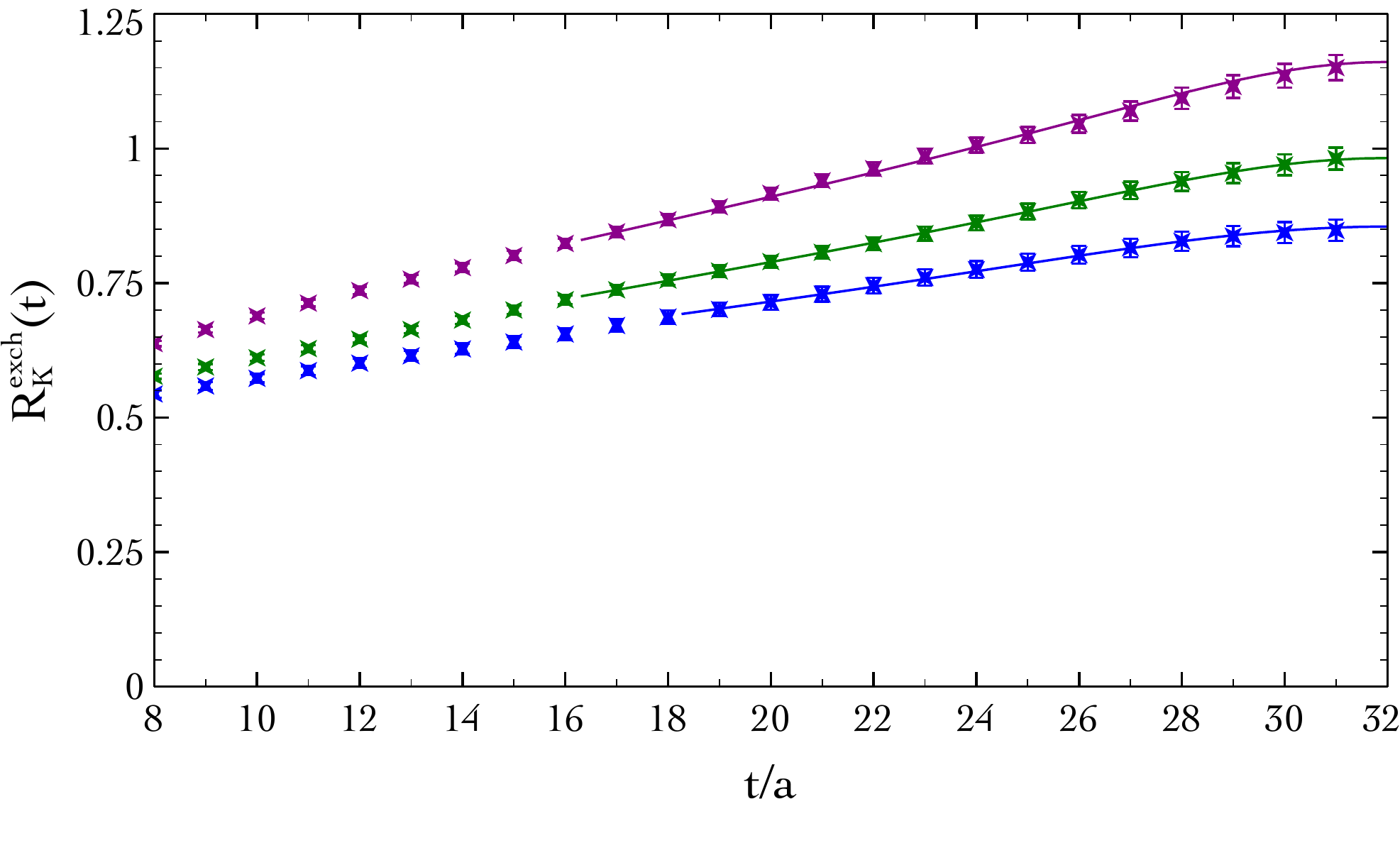}\hfill
\includegraphics[width=0.49\textwidth]{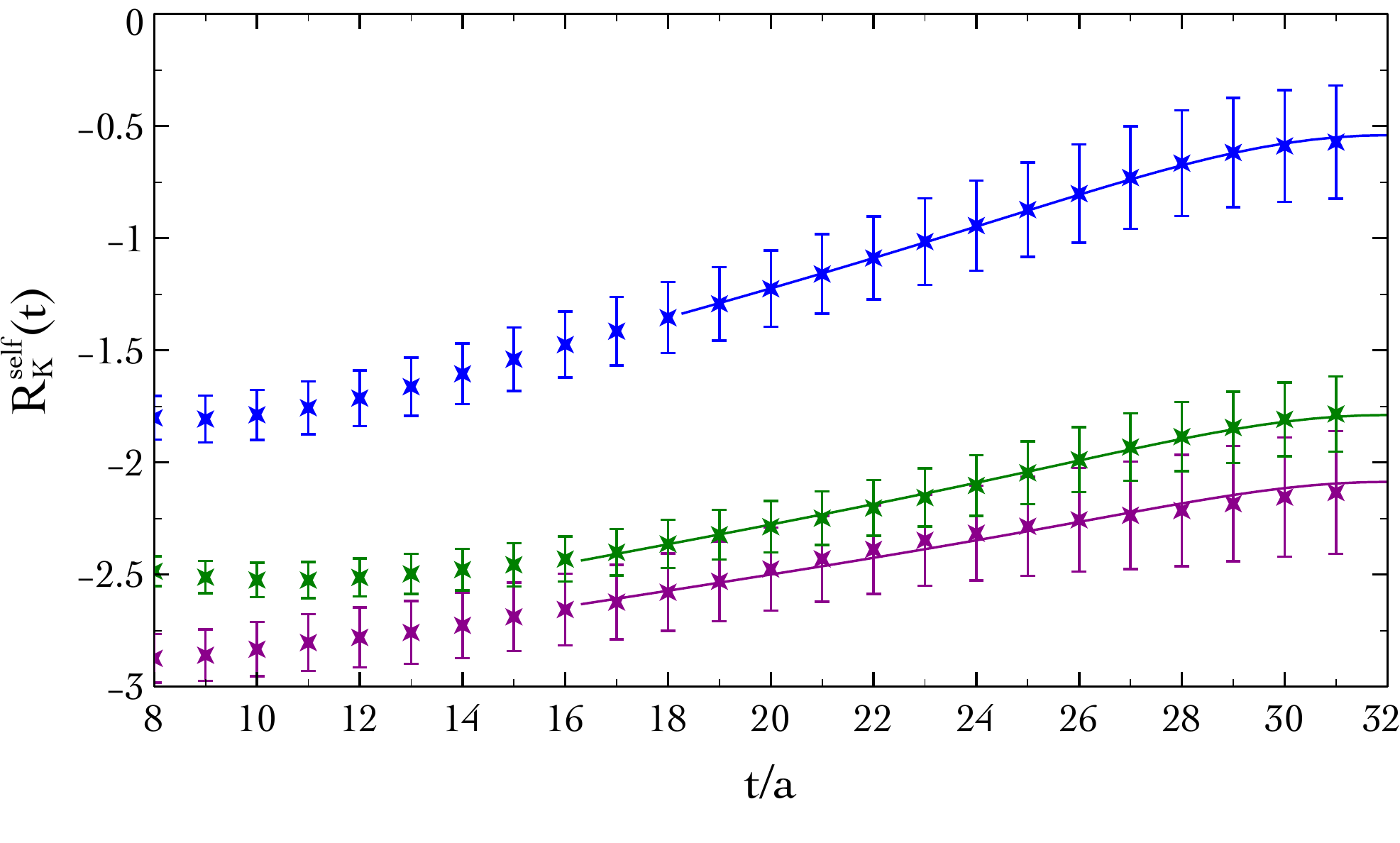}\hfill
\caption{\label{fig:Kslopes} \footnotesize
Fits of $R^{m}_{K}(t)$ (top--left), $R^{k}_{K}(t)$ (top--right), $R^{exch}_{K}(t)$ (bottom--left) and $R^{self}_{K}(t)$ (bottom--right) according to eq.~(\ref{eq:latticepartialt}). The dark magenta points correspond to $\beta=3.90$ and $(am_{ud})^0=0.0030$, the green points to $\beta=4.05$ and $(am_{ud})^0=0.0060$ while the blue points correspond to $\beta=4.20$ and $(am_{ud})^0=0.0065$ (see Appendix~\ref{sec:tm}). Data are in lattice units.
}
\end{center}
\end{figure}
\begin{figure}[!t]
\begin{center}
\includegraphics[width=0.6\textwidth]{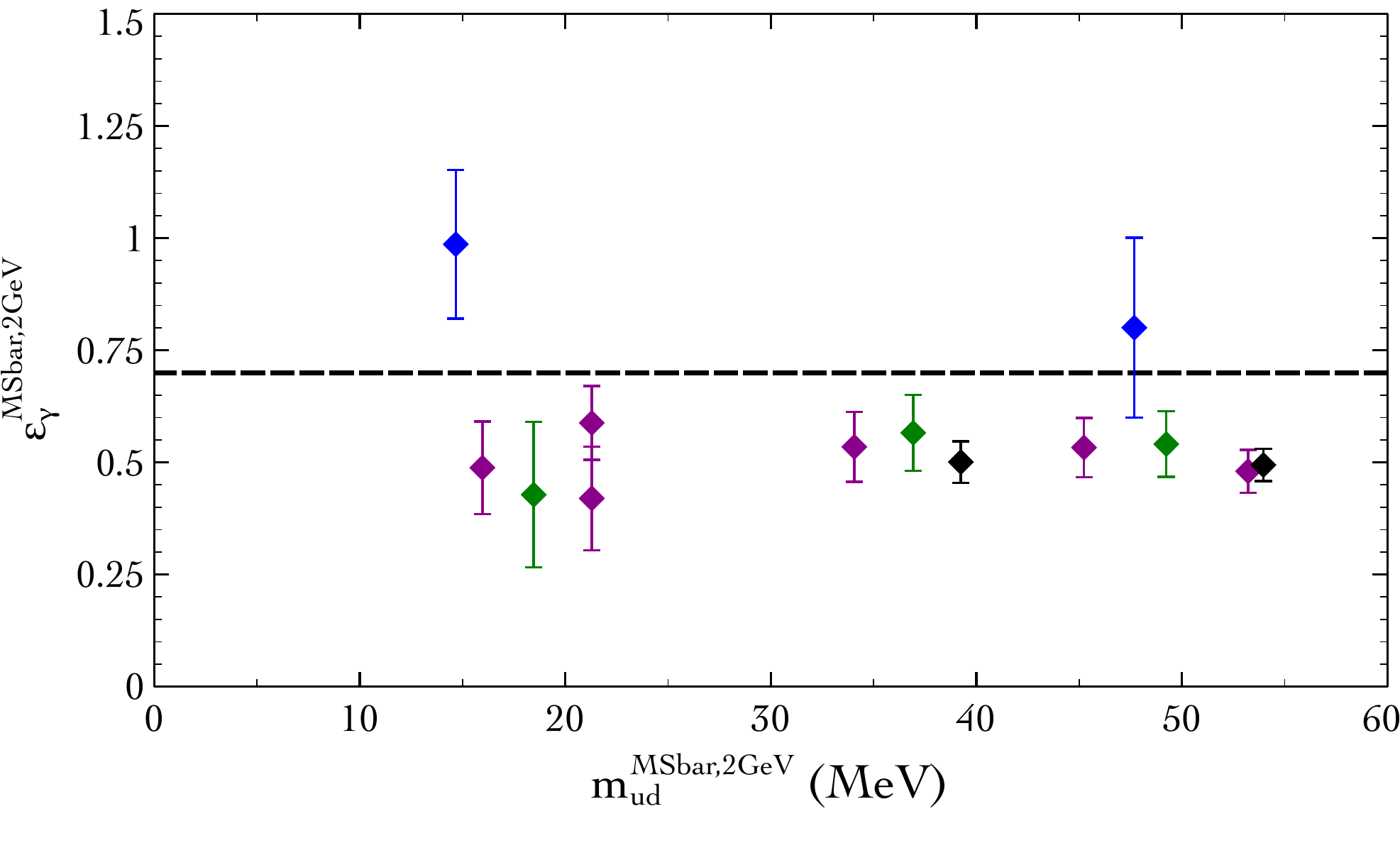}\hfill
\caption{\label{fig:MKscale} \footnotesize
$\varepsilon_{\gamma}(\overline{MS},2\mbox{ GeV})$ as a function of $\hat m_{ud}$ as extracted from lattice correlators. The dashed horizontal line correspond to the value $\varepsilon_\gamma=0.7$ used into ref.~\cite{deDivitiis:2011eh}.
Black points correspond to $\beta=3.80$, dark magenta points correspond to $\beta=3.90$, green points correspond to $\beta=4.05$ and blue points correspond to $\beta=4.20$ (see Appendix~\ref{sec:tm}).
}
\end{center}
\end{figure}
In Figure~\ref{fig:Kslopes} we plot the fits of the different ratios of corrected correlators defined into eqs.~(\ref{eq:Kratios}), performed according to eq.~(\ref{eq:latticepartialt}). In Figure~\ref{fig:MKscale} we plot $\varepsilon_{\gamma}(\overline{MS},2\mbox{ GeV})$ as obtained from our numerical simulations, i.e. by using our lattice results for both the numerator and the denominator of the ratio appearing in eq.~(\ref{eq:prescription}). The data obtained at unphysical values of $\hat m_{ud}$, at finite lattice spacing and at finite volume seem to confirm that $\epsilon_\gamma=0.7$ is a reasonable estimate for the Dashen's theorem breaking parameter. In the next section we discuss the chiral extrapolation and the removal of cutoff and finite volume effects from the data of Figure~\ref{fig:MKscale}.

\section{Chiral extrapolations and lattice artifacts}
\label{sec:extrapolations}
In this section we discuss the chiral extrapolation and the removal of cutoff and finite volume effects from $M_{\pi^+}^2-M_{\pi^0}^2$ (see section~\ref{sec:pionmasses} and Figure~\ref{fig:PIchiral}) and from $\varepsilon_{\gamma}$ (see previous section and Figure~\ref{fig:MKscale}).

We rely on the chiral formulae of ref.~\cite{Hayakawa:2008an} (see also ref.~\cite{Blum:2010ym}) where the authors studied the dependence of $M_{\pi^+}^2-M_{\pi^0}^2$ and $M_{K^+}^2-M_{K^0}^2$ on the quark masses together with finite volume corrections by using an effective chiral lagrangian at NLO with the inclusion of electromagnetic interactions. They defined QED on a finite volume by considering the same infrared regularization used in this paper. In the case of the pion mass difference the formulae corresponding to the $n_f=2$ theory with a quenched strange quark are
\begin{eqnarray}
&&f^{\chi\pi}[C,K]= \left[ M_{\pi^+}^2-M_{\pi^0}^2 \right]
=
2  \hat e^2 F_0^2 \left\{
C-(3+4C)\frac{M_\pi^2}{32\pi^2F_0^2}\left[\log\left(\frac{M_\pi^2}{\mu^2}\right)+K(\mu) \right] 
\right\} \; ,
\nonumber \\
\nonumber \\
\nonumber \\
&&f^{\chi\pi}_{L}[C]=
\left[M_{\pi^+}^2-M_{\pi^0}^2\right](L)-\left[M_{\pi^+}^2-M_{\pi^0}^2\right](\infty)
=
\frac{ \hat e^2}{4\pi L^2} \left[
H_2(M_\pi L)-4C H_1(M_\pi L)
\right] \; .
\label{eq:pionschiralfve}
\end{eqnarray}
The corresponding formulae for $\varepsilon_{\gamma}$ are
\begin{eqnarray}
\varepsilon_\gamma \ =\
\left[\frac{4}{3}+2e_u^{sea}+2e_d^{sea}+\frac{3}{C}\right]\bigg\{
&-&
\frac{M_K^2}{32\pi^2F_0^2}
\left[\log\left(\frac{M_K^2}{\mu^2}\right)+\tilde K^1(\mu) \right]
\nonumber \\
\nonumber \\
&+&
\frac{M_\pi^2}{32\pi^2F_0^2}
\left[\log\left(\frac{M_\pi^2}{\mu^2}\right)+\tilde K^2(\mu) \right] \ \
\bigg\}
\label{eq:kaonschiralfve1}
\end{eqnarray}
and 
\begin{eqnarray}
\varepsilon_{\gamma}(L) 
-
\varepsilon_{\gamma}(\infty)
&=&
\frac{1}{8\pi L^2 F_0^2 C} \Big[H_2(M_K L)-H_2(M_\pi L)\Big]
\nonumber \\
\nonumber \\
&-&\frac{1}{8\pi L^2 F_0^2}
\left(\frac{4}{3}+2e_u^{sea}+2e_d^{sea}\right)\Big[ H_1(M_K L)- H_1(M_\pi L)\Big]
 \; .
\label{eq:kaonschiralfve2}
\end{eqnarray}
In the previous expressions $M_\pi=2B_0  \hat m_{ud}$ and $M_K=B_0( \hat m_s+ \hat m_{ud})$ with $B_0$ and $F_0$ the QCD low energy constants entering the leading order chiral lagrangian. $F_0$ is normalized so that the decay constant of the physical pion is about $90$~MeV. Furthermore, $C$ 
is the single electromagnetic low energy constant entering the leading order lagrangian while $K(\mu)$, $\tilde K^1(\mu)$ and $\tilde K^2(\mu)$ are combinations of electromagnetic low energy constants at next--to--leading order. Note that the formulae for $\varepsilon_\gamma$ depend upon the charges $e_u^{sea}$ and $e_d^{sea}$ of the sea quarks that have to be set to zero in the electro--quenched approximation. 

The functions $H_{1,2}(x)$ entering the finite volume correction formulae are
\begin{eqnarray}
H_1(x)&=& \int_0^{+\infty}{du \frac{F(u)}{u^2}\ e^{-\frac{u x^2}{4\pi}}} \; ,
\nonumber \\
\nonumber \\
H_2(x)&=& -2\mathtt{k}-\int_0^{+\infty}{du 
\frac{S(u)}{u^2}\left[
e^{-\frac{u x^2}{4\pi}} +
x\sqrt{u} \mbox{ erf}\left(x\sqrt{\frac{u}{4\pi}}\right)
\right]
} \; ,
\label{eq:h1h2}
\end{eqnarray}
where
\begin{eqnarray}
F(x)=\left[\vartheta_3(0,i/x) \right]^3-1 \; ,
\qquad
S(x)=x^{3/2}-F(x) \; ,
\qquad
\mathtt{k}=\int_0^{+\infty}{du \frac{S(u)}{u^2}}=2.8373\dots \; ,
\end{eqnarray}
are expressed in terms of the Jacobi elliptic $\vartheta$--function,
\begin{eqnarray}
\vartheta_3(0,i/x)= \sqrt{x}\ \vartheta_3(0,ix)
=\sqrt{x}\left(
1+2\sum_{n=1}^{+\infty}e^{-\pi x n^2} 
\right)\; .
\end{eqnarray}
\begin{figure}[!t]
\begin{center}
\includegraphics[width=0.7\textwidth]{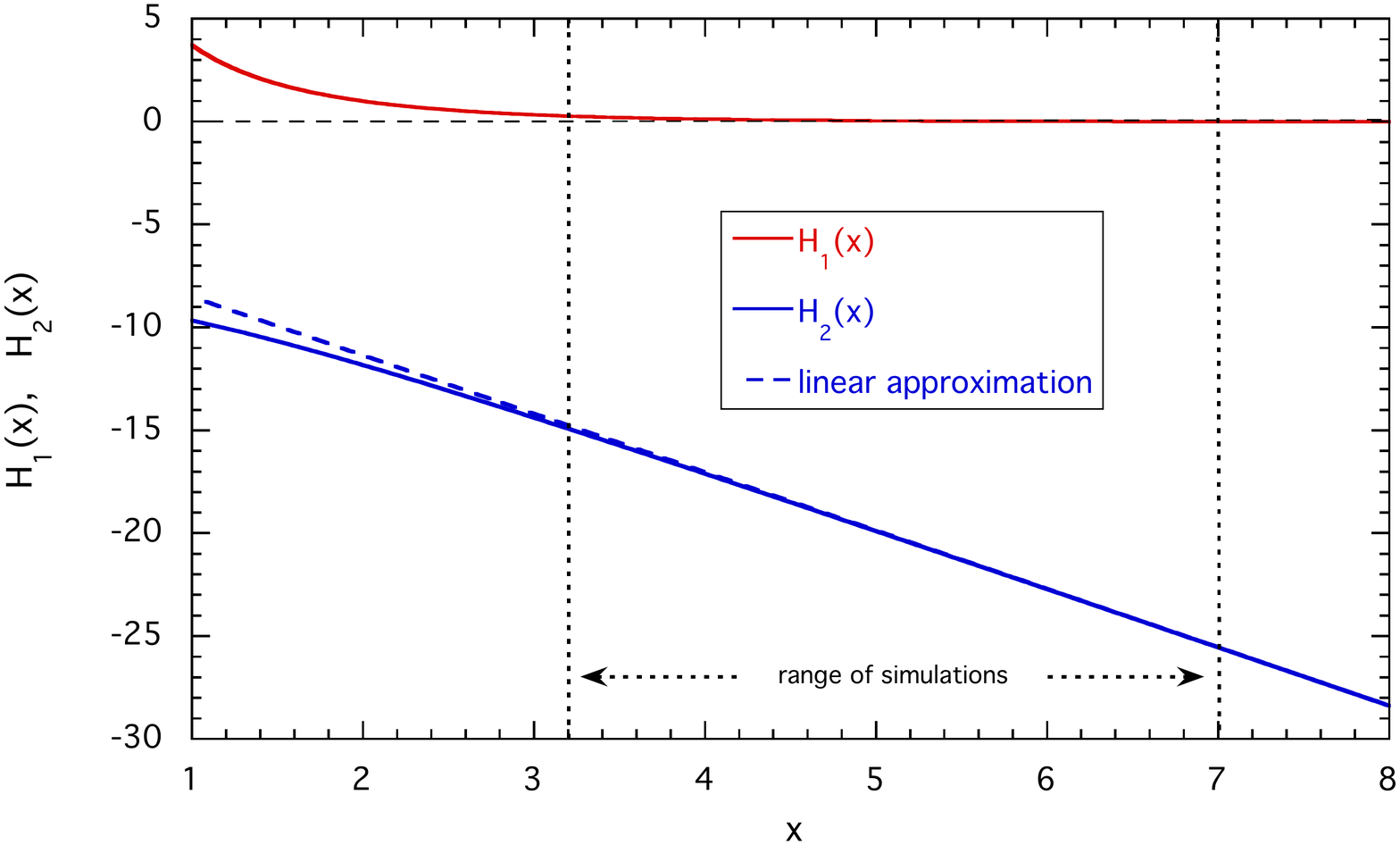}
\caption{\label{fig:H12} \footnotesize
Dependence of the functions $H_1(x)$ and $H_2(x)$ with respect to $x=M_{\pi,K} L$. In our numerical simulations we have $3.2< M_\pi L< 5.8$ and $5.1< M_K L< 7.0$. The linear approximation of the function $H_2(x)$ is explicitly given by $H_2(x) \sim -\mathtt{k} (2 + x)$.
}
\end{center}
\end{figure}
The formulae defining the functions $H_{1,2}(x)$ contain rather involved integral expressions but, in the range of values of $x=M_{\pi,K} L$ corresponding to the meson masses and lattice volumes simulated in this paper, i.e. $3.2 < x < 7.0$, the function $H_2(x)$ is almost a linear function of $x$ and $H_1(x)$ can be safely neglected (see Figure~\ref{fig:H12}). From the asymptotic expressions
\begin{eqnarray}
\left[M_{\pi^+}^2-M_{\pi^0}^2\right](L)-\left[M_{\pi^+}^2-M_{\pi^0}^2\right](\infty) 
&\sim& 
-\frac{ \hat e^2 \mathtt{k}}{4\pi}\left(\frac{M_\pi}{L}+\frac{2}{L^2}\right)
\; ,
\nonumber \\
\nonumber \\
\varepsilon_{\gamma}(L)
-
\varepsilon_{\gamma}(\infty) 
&\sim& 
-\frac{\mathtt{k}}{8\pi F_0^2 C}\frac{M_K-M_\pi}{L}
\; ,
\label{eq:finiteapprox}
\end{eqnarray}
one recognizes that QED, a long range interaction, introduces sizable power--law finite volume effects on hadron masses that have to be eliminated by extrapolating numerical data obtained on different physical volumes. These are predicted to be as large as $30$\% in our simulations and to be larger at heavier meson masses. 

\begin{figure}[!t]
\begin{center}
\includegraphics[width=0.49\textwidth]{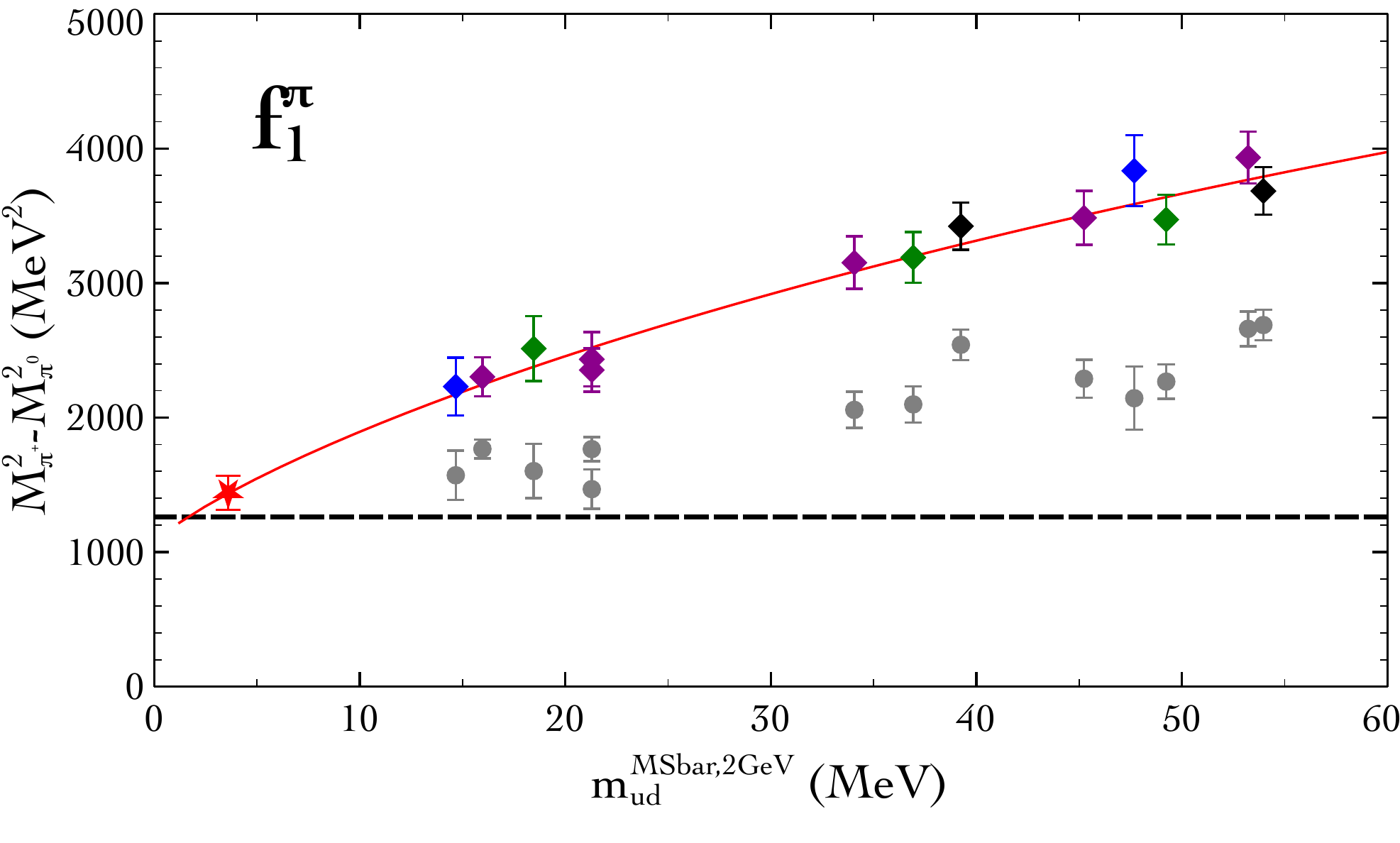}\hfill
\includegraphics[width=0.49\textwidth]{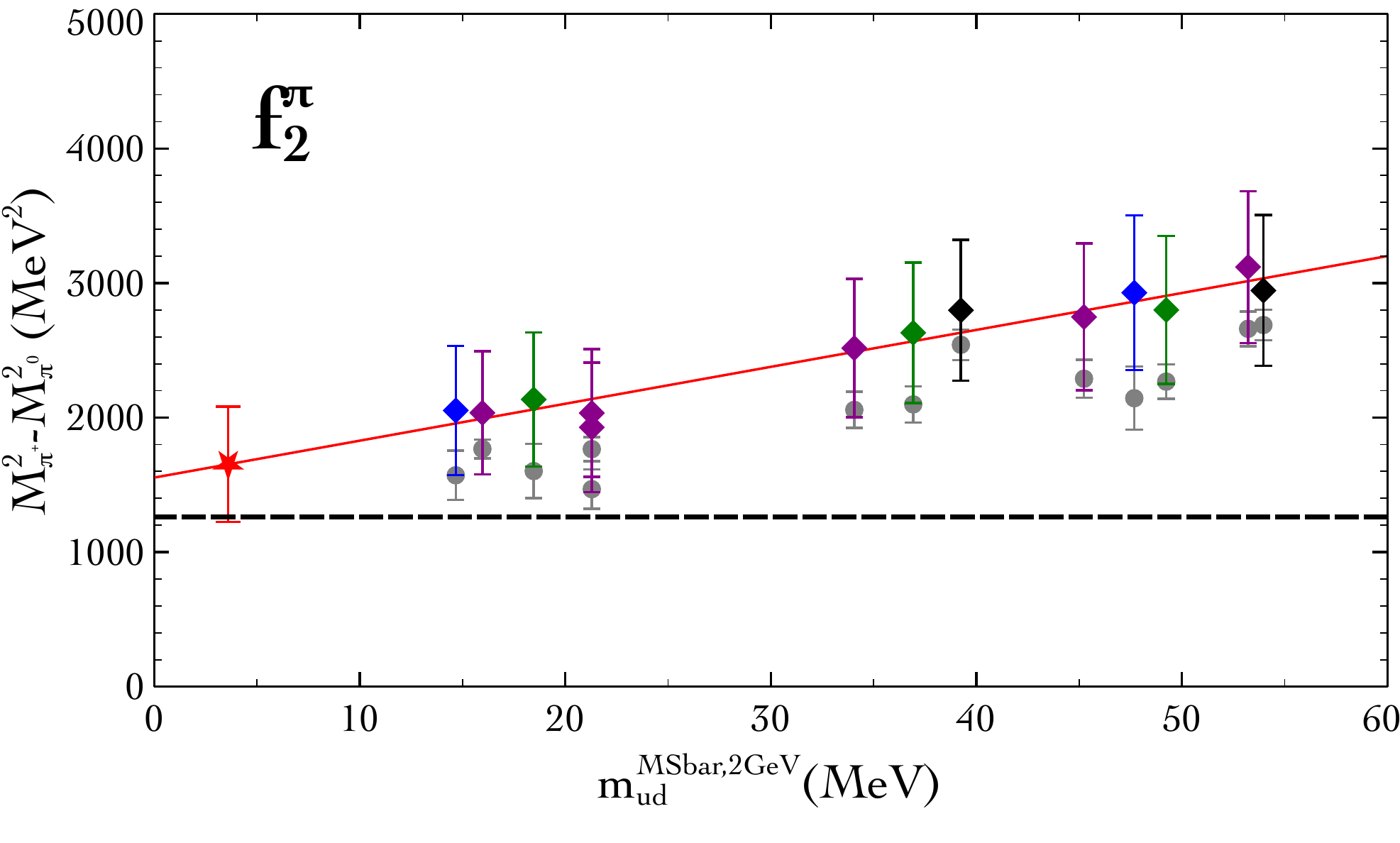}\hfill
\includegraphics[width=0.49\textwidth]{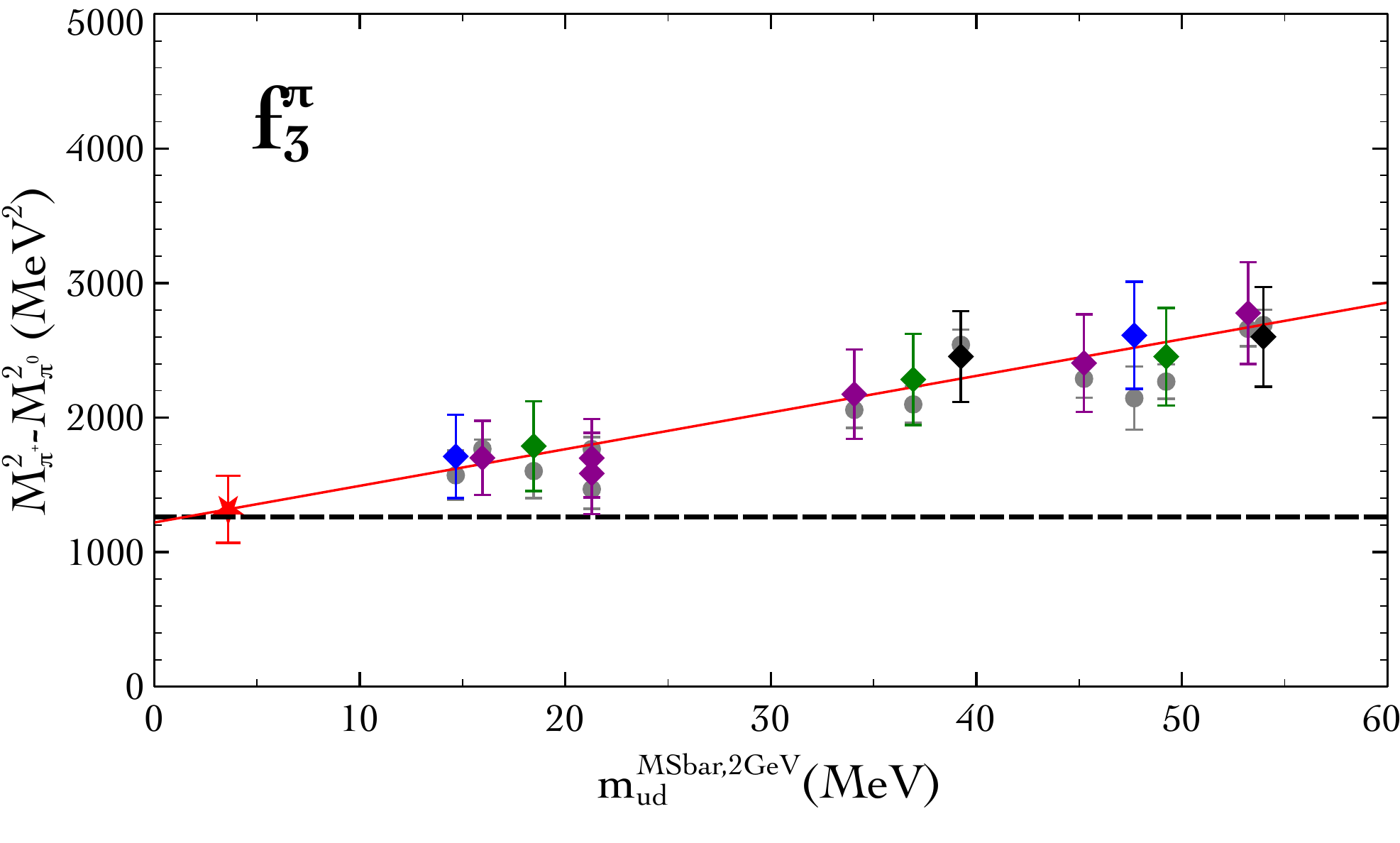}\hfill
\caption{\label{fig:PIextr} \footnotesize
Combined chiral, continuum and infinite volume extrapolations of $M_{\pi^+}^2-M_{\pi^0}$.  The grey points are the data as extracted from lattice simulations and converted in physical units. The red point and the solid red curve is the result of the extrapolation. Black points correspond to $\beta=3.80$, dark magenta points correspond to $\beta=3.90$, green points correspond to $\beta=4.05$ and blue points correspond to $\beta=4.20$ (see Appendix~\ref{sec:tm}). The dashed horizontal line corresponds to the experimental value of $M_{\pi^+}^2-M_{\pi^0}^2$. For each panel the colored points have been obtained by subtracting the fitted lattice artifacts. Note the two points at $\hat m_{ud}\sim 22$~MeV that have been obtained at $\beta=3.9$ and at the same value of sea quark mass but on different volumes: within quoted errors the two points differ at finite volume (grey) and coincide after the removal of discretization effects (dark magenta).}  
\end{center}
\end{figure}
In order to extrapolate lattice data for the pion mass difference $M_{\pi^+}^2-M_{\pi^0}^2$ we have performed three fits differing in both the dependence on the average light quark mass $\hat m_{ud}$ and the parametrization of the lattice artifacts. We have considered the following functions 
\begin{eqnarray}
f^{\pi}_1[C,K,A_\pi] &=&
f^{\chi\pi}[C,K]+
f^{\chi\pi}_{L}[C] +
A_\pi\ [a^0]^2\; ,
\nonumber \\
\nonumber \\
f^{\pi}_2[C,K,A_\pi,B_\pi] &=&
C + K\ \hat m_{ud}+
\frac{B_\pi}{L}+
A_\pi\ [a^0]^2 \; ,
\nonumber \\
\nonumber \\
f^{\pi}_3[C,K,A_\pi,B_\pi] &=&
C + K\ \hat m_{ud} +
\frac{B_\pi^2}{L^2}+
A_\pi\ [a^0]^2 \; .
\end{eqnarray}
The function $f^{\pi}_1$ correspond to the $SU(2)$ chiral fit and has been obtained by using the chiral perturbation theory results of eqs.~(\ref{eq:pionschiralfve}) for the dependence on $\hat m_{ud}$ and the finite volume corrections. The other two functions correspond to phenomenological fits performed assuming a linear $\hat m_{ud}$ dependence and parametrizing finite volume effects with a term either proportional to $1/L$ or to $1/L^2$. In all cases we have added a term proportional to $(a^0)^2$ in order to estimate cutoff effects.

In Figure~\ref{fig:PIextr} we show the combined chiral, continuum and infinite volume extrapolations corresponding to the three fitting functions defined above. In each plot the solid red curve represents the fitted function $f_i^\pi$ evaluated at $a^0=0$ and at $1/L=0$, i.e. to the result of the continuum and infinite volume extrapolations. The grey points are our lattice results for $M_{\pi^+}^2-M_{\pi^0}^2$, already shown in Figure~\ref{fig:PIchiral}. The red point is the result of the extrapolation at the physical value $\hat m_{ud}(\overline{MS},2\ GeV)=3.6(2)\mbox{ MeV}$ determined within the isosymmetric theory in ref.~\cite{Blossier:2010cr}. The remaining colored points have been obtained from the corresponding grey points by subtracting out the lattice artifacts as determined by the fit. As the colored points are the results of the continuum and infinite volume extrapolations they have larger errors compared to the corresponding grey points. The three fits give consistent results for $M_{\pi^+}^2-M_{\pi^0}^2$, though with different errors. In particular, the fit that include a $1/L$ finite volume term gives larger errors on the extrapolated points than the other fitting functions. By comparing the $f_1^\pi$ panel with the other two, we see that the finite volume effects obtained from the fits $f_2^\pi$ and $f_3^\pi$ are considerably smaller than the one--loop chiral perturbation theory prediction. Similar results for finite volume effects, parametrized by a $1/L^2$ term and fitted to lattice data, have been obtained by the authors of ref.~\cite{Basak:2013iw}. 
The value of $\chi^2/dof$ corresponding to the fit $f_1^\pi$ is $1.2$, while for the fits $f_2^\pi$ and $f_3^\pi$ we have $\chi^2/dof=1.0$. 

We obtain our final estimate of the pion mass difference by taking as central value and as statistical error the results of the $SU(2)$ chiral fit, i.e. $f_1^\pi$, while we estimate our systematic error by taking half of the difference between the $f_2^\pi$ and $f_3^\pi$ results. We get
\begin{eqnarray}
M_{\pi^+}^2-M_{\pi^0}^2=\deltampi \;.
\label{eq:physmpi}
\end{eqnarray}
This result compares nicely with the experimental determination
\begin{eqnarray}
\left[M_{\pi^+}^2-M_{\pi^0}^2\right]^{exp}=1.2612(1) \times 10^{3} \ \mbox{MeV}^2 \;,
\end{eqnarray}
suggesting, a posteriori, that the effect of having neglected the disconnected contribution of $O(\hat \alpha_{em} \hat m_{ud})$ appearing in eq.~(\ref{eq:pionmasses}) is smaller or of the same order of magnitude as the other uncertainties affecting our result.

\begin{figure}[!t]
\begin{center}
\includegraphics[width=0.49\textwidth]{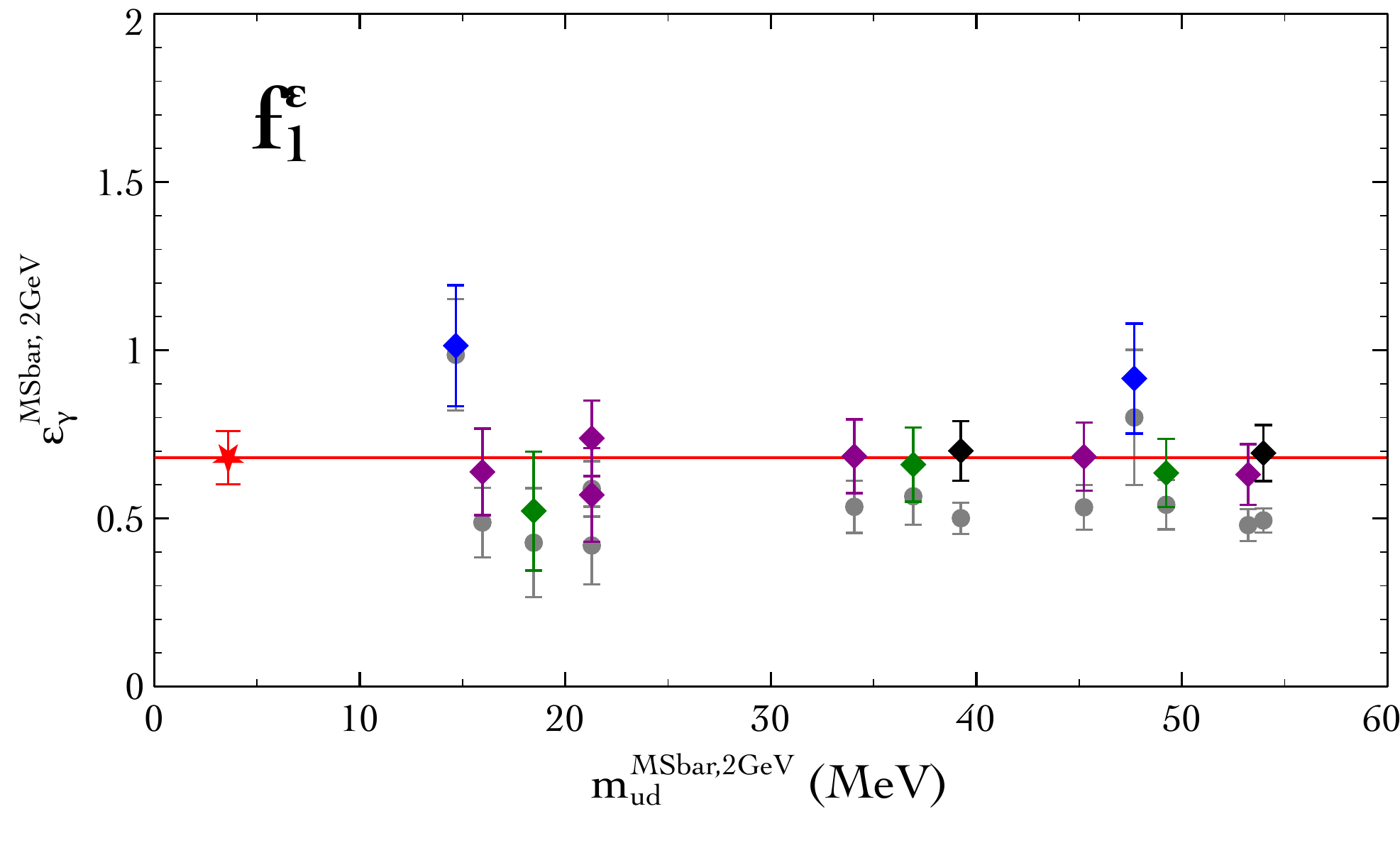}\hfill
\includegraphics[width=0.49\textwidth]{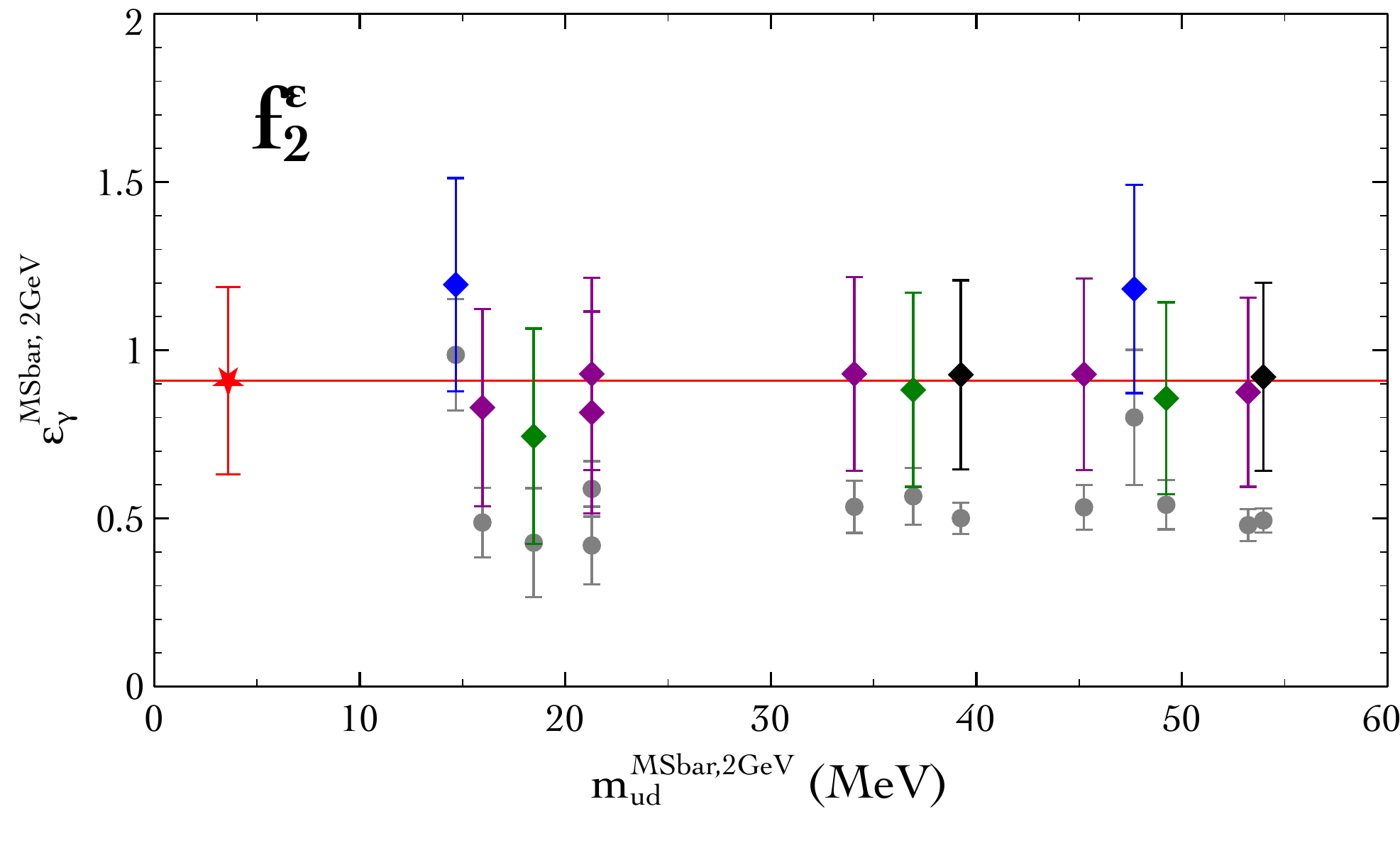}\hfill
\includegraphics[width=0.49\textwidth]{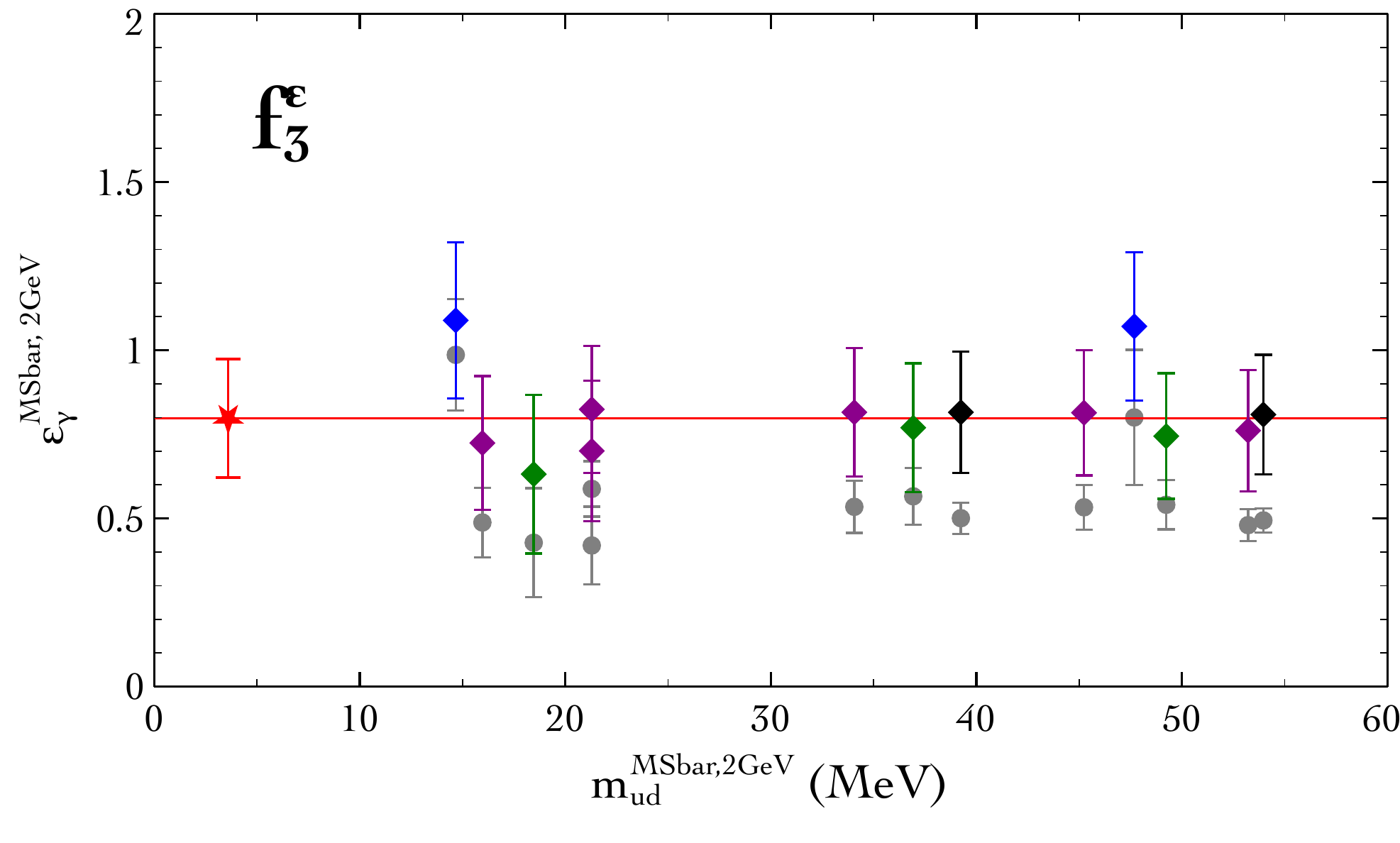}\hfill
\caption{\label{fig:Kextr} \footnotesize
Combined chiral, continuum and infinite volume extrapolations of $\varepsilon_\gamma(\overline{MS},2\mbox{ GeV})$.  The grey points are the data as extracted from lattice simulations. The red point and the solid red curve is the result of the extrapolation. Black points correspond to $\beta=3.80$, dark magenta points correspond to $\beta=3.90$, green points correspond to $\beta=4.05$ and blue points correspond to $\beta=4.20$ (see Appendix~\ref{sec:tm}). For each panel the colored points have been obtained by subtracting the fitted lattice artifacts.}  
\end{center}
\end{figure}

We now discuss the extrapolations of the results for $\varepsilon_\gamma(\overline{MS},2\mbox{ GeV})$. In general we expect reduced lattice artifacts for $\varepsilon_{\gamma}$ with respect to $M_{\pi^+}^2-M_{\pi^0}^2$ because of possible cancellations of systematics effects between the numerator and the denominator in the ratio of eq.~(\ref{eq:prescription}). Within the quoted errors, the lattice data shown in Figure~\ref{fig:MKscale} are fairly flat in $\hat m_{ud}$ so that we have not attempted a $SU(3)$ chiral extrapolation using eqs.~(\ref{eq:kaonschiralfve1}) and~(\ref{eq:kaonschiralfve2}). Linear chiral extrapolations lead to vanishing slopes and errors on the fitted results of the same order of magnitude of the ones shown in Figure~\ref{fig:Kextr}, corresponding to constant chiral extrapolations. More precisely, the fit functions shown in Figure~\ref{fig:Kextr} are
\begin{eqnarray}
f_1^{\varepsilon}[E,A_\varepsilon]&=& E + A_\varepsilon\ [a^0]^2 \; ,
\nonumber \\
\nonumber \\
f_2^{\varepsilon}[E,A_\varepsilon,B_\varepsilon]&=& E+\frac{B_\varepsilon}{L}+ A_\varepsilon\ [a^0]^2\; ,
\nonumber \\
\nonumber \\
f_3^{\varepsilon}[E,A_\varepsilon,B_\varepsilon]&=& E +\frac{B_\varepsilon^2}{L^2} + A_\varepsilon\ [a^0]^2\; ,
\end{eqnarray}
In order to obtain an estimate of the systematic errors associated with our result, we have parametrized cutoff effects with a term proportional to $(a^0)^2$ and finite volumes effects with terms vanishing as $1/L$ and as $1/L^2$. 

For each plot in Figure~\ref{fig:Kextr}, the solid red line corresponds to the fitted function $f_i^\varepsilon$ evaluated at $a^0=0$ and at $1/L=0$, i.e. to the result of the continuum and infinite volume extrapolations. The grey points are our lattice results for $\varepsilon_\gamma(\overline{MS},2\mbox{ GeV})$ already shown in Figure~\ref{fig:MKscale}. The red point is the result of the extrapolation at the physical value $\hat m_{ud}(\overline{MS},2\ GeV)=3.6(2)\mbox{ MeV}$ determined in the isosymmetric theory in ref.~\cite{Blossier:2010cr}. The remaining colored points have been obtained from the corresponding grey points by subtracting out the lattice artifacts as determined by the fit. As the colored points are the results of the continuum and infinite volume extrapolations they have larger errors compared to the corresponding grey points. The three fits give consistent results for $\varepsilon_\gamma(\overline{MS},2\mbox{ GeV})$. The $\chi^2/dof$ of all the three fits is $1.0$. 

We obtain our final result of $\varepsilon_\gamma(\overline{MS},2\mbox{ GeV})$ by taking as central value the average of the maximum ($f_2^\varepsilon$) and minimum ($f_1^\varepsilon$) central values and as statistical error the one obtained from the $f_3^\varepsilon$ fit. 
In order to estimate the systematic error associated with the fits we take half of the difference between the $f_2^\varepsilon$ and $f_1^\varepsilon$ results ($13$\%). By taking the deviation of our result for $M_{\pi^+}^2-M_{\pi^0}^2$ from the experimental value we obtain a rough estimate of the error associated with the neglected contributions, i.e. the $M_{\pi^0}$ disconnected diagram and all the terms of $O(\hat \alpha_{em}\Delta \hat m_{ud})$ or higher, and add in quadrature a $15$\% uncertainty to the systematic error. The uncertainty associated with the electro--quenched approximation is estimated by using the chiral formula for $\varepsilon_\gamma$ given in eq.~(\ref{eq:kaonschiralfve1}). More precisely, by taking $C$ from the fit $f_1^\pi$ and by neglecting the variation of the terms $\tilde K^1$ and $\tilde K^2$, we evaluate the ratio $\varepsilon_\gamma(e_u^{sea}+e_d^{sea}=1/3)/\varepsilon_\gamma(e_u^{sea}+e_d^{sea}=0)$ and add in quadrature the resulting $12$\% uncertainty to the systematic error. We get
\begin{eqnarray}
\varepsilon_\gamma(\overline{MS},2\mbox{ GeV})=\epsilongamma\;.
\end{eqnarray}
This result corresponds to the following separation of QED from QCD isospin breaking corrections to the kaons mass difference (see eqs.~(\ref{eq:kaonmassessep}) above)
\begin{eqnarray}
\left[M_{K^+}^2-M_{K^0}^2\right]^{QED}(\overline{MS},2\mbox{ GeV})&=&\ \
\deltamkQED \;,
\nonumber \\
\nonumber \\
\left[M_{K^+}^2-M_{K^0}^2\right]^{QCD}(\overline{MS},2\mbox{ GeV}) &=&
\deltamkQCD \;,
\end{eqnarray}
where the QCD contribution is obtained by imposing the experimental constraint
\begin{eqnarray}
\left[M_{K^+}^2-M_{K^0}^2\right]^{exp}=-3.903(3)  \times 10^{3} \ \mbox{MeV}^2 \;.
\end{eqnarray}
The experimental input for the kaon mass splitting also allows a determination of the up--down mass difference and, from the second of eqs.~(\ref{eq:kaonmassessep}), we obtain
\begin{eqnarray}
[\hat m_d - \hat m_u](\overline{MS},2\mbox{ GeV}) \ =\ 2\Delta \hat m_{ud}(\overline{MS},2\mbox{ GeV})&=& \deltamq \;,
\nonumber \\
\nonumber \\
\frac{\hat m_u}{\hat m_d}(\overline{MS},2\mbox{ GeV}) &=& \rmq \;,
\label{eq:qmasses1}
\end{eqnarray}
The results for the light quark mass ratio has been obtained by combining the determination of $\Delta \hat m_{ud}$ performed in this paper with the result $\hat m_{ud}(\overline{MS},2\ GeV)=3.6(2)\mbox{ MeV}$  obtained in ref.~\cite{Blossier:2010cr}. The authors of ref.~\cite{Blossier:2010cr} have extracted the symmetric light quark mass $\hat m_{ud}^0$, the strange quark mass $\hat m_{s}^0$ and the lattice spacing $a^0$ by performing simulations of the isosymmetric theory and using the necessary number of hadronic inputs to calibrate the lattice. By following this procedure their result for $\hat m_{ud}^0$ differs from $\hat m_{ud}$, as defined in eqs.~(\ref{eq:matching}) at $\mu=2\mbox{ GeV}$, by isosymmetric  $O((e_u^2+e_d^2)\hat \alpha_{em})$ contributions. These terms do not affect our results for the quark mass ratios since $\hat m_u/\hat m_d$ is a function of $\Delta \hat m_{ud}/\hat m_{ud}$ only and, in turn,
\begin{eqnarray}
\frac{\Delta \hat m_{ud}}{\hat m_{ud}}&=&\frac{\Delta \hat m_{ud}}{\hat m_{ud}^0}\
+\ O(\hat \alpha_{em} \Delta \hat m_{ud}) \; ,
\end{eqnarray}
but represent a systematic error for the determinations of $\hat m_u$ and $\hat m_d$ which follow from eqs.~(\ref{eq:qmasses1}), namely
\begin{eqnarray}
\hat m_u(\overline{MS},2\mbox{ GeV}) &=& \mqu \;,
\nonumber \\
\nonumber \\
\hat m_d(\overline{MS},2\mbox{ GeV}) &=& \mqd \;.
\label{eq:qmasses2}
\end{eqnarray}
The argument used in the case of $\hat m_{u}/\hat m_{d}$ also applies to the flavour symmetry breaking parameters 
\begin{eqnarray}
R(\overline{MS},2\mbox{ GeV})=\left[\frac{\hat m_s -\hat m_{ud}}{\hat m_d - \hat m_u}\right](\overline{MS},2\mbox{ GeV})
 &=& \Rsu \;,
\nonumber \\
\nonumber \\
\nonumber \\
Q(\overline{MS},2\mbox{ GeV})=\left[\sqrt{\frac{\hat m_s^2 -\hat m_{ud}^2}{\hat m_d^2 - \hat m_u^2}}\right](\overline{MS},2\mbox{ GeV})
 &=& \Qsu \;,
\end{eqnarray}
that have been calculated starting from the relations
\begin{eqnarray}
\frac{1}{R}&=&
\frac{2\Delta \hat m_{ud}}{\hat m_{s}^0-\hat m_{ud}^0}
\ +\ O(\hat \alpha_{em} \Delta \hat m_{ud}) \; ,
\nonumber \\
\nonumber \\
\nonumber \\
\frac{1}{Q^2}&=&
\frac{4 \Delta \hat m_{ud}\ \hat m_{ud}^{0}}{(\hat m_s^0)^2-(\hat m_{ud}^0)^2}
\ +\ O(\hat \alpha_{em} \Delta \hat m_{ud}) \; ,
\end{eqnarray}
with the result $\hat m_{s}^0(\overline{MS},2\ GeV)=95(6)\mbox{ MeV}$  obtained in ref.~\cite{Blossier:2010cr}.

Using our own determination of $\varepsilon_\gamma$ we are here in position of updating the results obtained in ref.~\cite{deDivitiis:2011eh} for the QCD isospin breaking effects on the $K_{\ell2}$ decay rate and the neutron--proton mass difference. We get
\begin{eqnarray}
\left[ \frac{F_{K^+}/F_{\pi^+}}{F_K/F_\pi}-1 \right]^{QCD}(\overline{MS},2\mbox{ GeV}) &=& 
\deltaf \; ,
\nonumber \\
\nonumber \\
\left[ M_n - M_p \right]^{QCD}(\overline{MS},2\mbox{ GeV}) &=& \deltamp \;,
\label{eq:update}
\end{eqnarray}
where $F_{K^+}$ and $F_{\pi^+}$ are the charged kaon and charged pion decay constants in QCD with $\hat m_d\neq \hat m_u$ and $F_K$ and $F_\pi$ are the corresponding quantities in the isosymmetric theory. Analogously, $M_n$ and $M_p$ are respectively the masses of the neutron and of the proton in QCD with $\hat m_d\neq \hat m_u$. The results in eqs.~(\ref{eq:update}) have now been obtained from a self consistent non--perturbative lattice calculation.

\section{Conclusions}
\label{sec:theend}

In this paper we have shown that leading isospin breaking effects on hadron masses can be conveniently calculated on the lattice by starting from simulations of the underlying isosymmetric QCD theory and by expanding the path--integral in powers of $\hat \alpha_{em}$ and $\hat m_d- \hat m_u$. We have discussed all the details necessary for applying our method to the calculation of isospin breaking corrections to any observable and discussed the renormalization of the corrected correlation functions.

In particular, we have shown how the ultraviolet divergences generated by the contact--terms of the two electromagnetic currents can be absorbed in a redefinition of the bare parameters of the full theory with respect to the corresponding values of isosymmetric QCD. We have also shown that the linear divergences associated with the shift of the critical masses of the quarks, a problem to be addressed when the fermion action is discretized by using a Wilson term, can be subtracted out by determining the associated counter--terms with high numerical precision. 

By using the proposed method we have derived theoretical predictions for the pion mass splitting and for the up and down quark masses. We have also implemented a well defined renormalization prescription to separate QED from QCD isospin breaking corrections to hadron masses allowing to determine the electromagnetic contribution to the kaon mass splitting and the associated value of the Dashen's theorem breaking parameter $\varepsilon_\gamma$. These results have been used in order to update our previous determinations~\cite{deDivitiis:2011eh} of the QCD contributions to the neutron--proton mass splitting and the $K_{\ell2}$ decay rate.

The results obtained in this paper are affected by systematic errors.  Particularly important are those associated with the chiral extrapolation required because our pions are heavier than the physical ones. Another important source of systematics errors comes from finite volume effects. These are not peculiar to our method. We estimate finite volume effects by using the results of effective field theory calculations and by fitting them to lattice data obtained at different physical volumes. The finite volume effects arising from the fits turn out to be considerably smaller than chiral perturbation theory predictions, though we cannot make this statement more quantitative until we have results at the physical pion masses and on larger physical volumes. Finally, our results have been obtained in the $n_f=2$ theory and neglecting certain quark disconnected diagrams. Concerning the pion mass splitting, we have neglected the disconnected diagram appearing in eq.~(\ref{eq:pionmasses}), an $O(\hat \alpha_{em} \hat m_{ud})$ correction to $M_{\pi^0}$ that phenomenologically is expected to be of the same order of magnitude of the other $O(\hat \alpha_{em} \Delta \hat m_{ud})$ contributions neglected in this paper. The results for the kaon mass difference have been obtained by relying on the electro--quenched approximation that consists in treating dynamical quarks as electrically neutral particles. We have provided all the formulae necessary to remove these approximations. This will be the subject of future work.

\section*{Acknowledgements}
We warmly thank P.~Dimopoulos and C.~Tarantino for their collaboration at the early stages of this work and for valuable suggestions. Illuminating discussions with M.~Testa are also gratefully acknowledged.
We thank the members of the ETMC collaboration for having generated and made publicly available the gauge configurations
used for this study. Part of this work has been completed thanks to the allocation of CPU time on the BlueGene/Q -Fermi at Cineca for the specific initiative INFN-RM123 under the agreement between INFN and CINECA and we thank the Cineca HPC staff for their support. V.L., G.M., S.S. thank MIUR (Italy) for partial support under the contract PRIN 2010-2011 and G.M.deD., R.F., R.P. and G.C.R. thank MIUR (Italy) for partial support under the contract PRIN 2009-2010. G.M. also acknowledges partial support from ERC Ideas Advanced Grant n. 267985 ``DaMeSyFla''.

\begin{appendices}

\section{Gauge ensembles}
\label{sec:tm}
%
\begin{table}[!h]
\begin{center}
\begin{tabular}{cccccccc}
$\quad$ $\beta^0$    $\quad$ & 
$\quad$ $k_0$    $\quad$ & 
$\quad$ $(am_{ud})^0$  $\quad$ & 
$\quad (am_s)^0 \quad$ & 
$L/a$  & 
$N_{conf}$ & 
$a^0$~(fm) & 
 $Z_P^0(\overline{MS},2\mbox{ GeV})$ \tabularnewline
\hline 
\tabularnewline
3.80 & 0.164111 & 0.0080 & 0.0194 & 24 & 240 & 0.0977(31) & 0.411(12)\tabularnewline
     &          & 0.0110 &        & 24 & 240 &            &          \tabularnewline[4ex]

3.90 & 0.160856 & 0.0030 & 0.0177 & 32 & 150 & 0.0847(23) & 0.437(07)\tabularnewline
     &          & 0.0040 &        & 32 & 150 &            &          \tabularnewline
     &          & 0.0040 &        & 24 & 240 &            &          \tabularnewline
     &          & 0.0064 &        & 24 & 240 &            &          \tabularnewline
     &          & 0.0085 &        & 24 & 240 &            &          \tabularnewline
     &          & 0.0100 &        & 24 & 240 &            &          \tabularnewline[4ex]

4.05 & 0.157010 & 0.0030 & 0.0154 & 32 & 150 & 0.0671(16) & 0.477(06)\tabularnewline
     &          & 0.0060 &        & 32 & 150 &            &          \tabularnewline
     &          & 0.0080 &        & 32 & 150 &            &          \tabularnewline[4ex]

4.20 & 0.154073 & 0.0020 & 0.0129 & 48 & 100 & 0.0536(12) & 0.501(20)\tabularnewline
     &          & 0.0065 &        & 32 & 150 &            &          \tabularnewline[4ex]

\hline 
\end{tabular}
\end{center}
\label{tab:gaugeconfigs}
\caption{{\footnotesize Gauge ensembles used in this work. The gauge configurations have been generated within the isosymmetric theory with $N_f=2$ dynamical flavours of maximally twisted quarks of mass $(am_{ud})^0$. The strange quark mass $(am_s)^0$ has been used for valence propagators. The hopping parameter $k_0$ is related to the critical mass parameter $m^{cr}_0$ appearing in the main body of the paper by the relation $k_0=1/(2m^{cr}_0+8)$.}}
\end{table}
In this work we have used the $n_f=2$ dynamical gauge ensembles generated and made publicly available by the European Twisted Mass Collaboration (see Table~\ref{tab:gaugeconfigs}). These gauge configurations have been generated within the isosymmetric theory, by using the so--called Twisted Mass lattice discretization of the QCD action~\cite{Frezzotti:2000nk,Frezzotti:2003ni}, see eqs.~(\ref{eq:diracoperator1}) and~(\ref{eq:diracoperator2}). 
For the different gauge ensembles used in this work, the values of 
the critical hopping parameter $k_0=1/(2m^{cr}_0+8)$ (ref.~\cite{Baron:2009wt}), lattice spacing $a^0$ (ref.~\cite{Blossier:2010cr}), strange valence quark mass $(a m_s)^0$ (ref.~\cite{Blossier:2010cr}), renormalization constant $Z_P^0$ (ref.~\cite{Constantinou:2010gr}) are given in Table~\ref{tab:gaugeconfigs}.

In the following we give a dictionary to translate in the operator language the diagrammatic notation used in the main body of the paper. To this end, it is convenient to define the following local operators
\begin{eqnarray}
S_{fg}^{\pm\pm}(x)&=&\bar \psi^\pm_f(x)\psi^\pm_g(x) \; ,
\nonumber \\
\nonumber \\ 
\nonumber \\
P_{fg}^{\pm\pm}(x)&=&\bar \psi^\pm_f(x) \gamma^5 \psi^\pm_g(x) \; ,
\nonumber \\
\nonumber \\ 
\nonumber \\
\left[V_{fg}^{++}\right]^\mu(x)&=&i\left[
\bar \psi^+_f(x)\frac{+i\gamma_5-\gamma_\mu}{2}U_\mu(x)\psi^+_g(x+\mu)
-
\bar \psi^+_f(x+\mu)\frac{+i\gamma_5+\gamma_\mu}{2}U_\mu^\dagger(x)\psi^+_g(x) 
\right]\; ,
\nonumber \\
\nonumber \\ 
\nonumber \\
\left[V_{fg}^{--}\right]^\mu(x)&=&
i\left[\bar \psi^-_f(x)\frac{-i\gamma_5-\gamma_\mu}{2}U_\mu(x)\psi^-_g(x+\mu)
-
\bar \psi^-_f(x+\mu)\frac{-i\gamma_5+\gamma_\mu}{2}U_\mu^\dagger(x)\psi^-_g(x) \right]\; ,
\nonumber \\
\nonumber \\ 
\nonumber \\
\left[T_{fg}^{++}\right]^\mu(x)&=&
\bar \psi^+_f(x)\frac{+i\gamma_5-\gamma_\mu}{2}U_\mu(x)\psi^+_g(x+\mu)
+
\bar \psi^+_f(x+\mu)\frac{+i\gamma_5+\gamma_\mu}{2}U_\mu^\dagger(x)\psi^+_g(x) \; ,
\nonumber \\
\nonumber \\ 
\nonumber \\
\left[T_{fg}^{--}\right]^\mu(x)&=&
\bar \psi^-_f(x)\frac{-i\gamma_5-\gamma_\mu}{2}U_\mu(x)\psi^-_g(x+\mu)
+
\bar \psi^-_f(x+\mu)\frac{-i\gamma_5+\gamma_\mu}{2}U_\mu^\dagger(x)\psi^-_g(x) \; .
\nonumber \\
\end{eqnarray}
In terms of them we can now translate in local operator language the diagrammatic relations appearing in the main text. For example
\begin{eqnarray}
-
\begin{array}{c}
+ \\
\gdll \\
-  
\end{array}
&=&
T\langle\; P_{12}^{+-}(x)\; P_{21}^{-+}(0)\; \rangle \; ,
\end{eqnarray}
where, as a general rule, if two different quark lines have the same color they have the same mass (i.e. $m_1=m_2$). As further examples we have
\begin{eqnarray}
-\gdsi &=& \sum_y
T\langle\; P_{12}^{+-}(x)\; S_{23}^{--}(y)\; P_{31}^{-+}(0)\; \rangle \; ,
\nonumber \\
\nonumber \\
\nonumber \\
-\gdsip &=& i\sum_y
T\langle\; P_{12}^{+-}(x)\; P_{23}^{--}(y)\; P_{31}^{-+}(0)\; \rangle \; .
\end{eqnarray}
Examples involving the insertion of the electromagnetic currents are given here below
\begin{eqnarray}
-\gdslselfl &=& \sum_{yz}
T\langle\; P_{12}^{+-}(x)\; 
\left[V_{23}^{--}\right]^\mu(y)\; \left[V_{34}^{--}\right]^\nu(z)\; 
P_{41}^{-+}(0)\; \rangle \; D_{\mu\nu}(y,z)\;,
\nonumber \\
\nonumber \\
\nonumber \\
-\gdslexch &=& \sum_{yz}
T\langle\; P_{12}^{+-}(x)\; 
\left[V_{23}^{--}\right]^\mu(y)\; P_{34}^{-+}(0)\;
\left[V_{41}^{++}\right]^\nu(z)\; 
\rangle \;  D_{\mu\nu}(y,z)\;,
\nonumber \\
\nonumber \\
\nonumber \\
-\gdllphtad &=& \frac{1}{2}\sum_{y}
T\langle\; P_{12}^{+-}(x)\; 
P_{23}^{-+}(0)\; \left[T_{31}^{++}\right]^\mu(y)\; 
\rangle \;  D_{\mu\mu}(y,y)\;.
\end{eqnarray}
We finish with an example concerning a quark disconnected diagram,
\begin{eqnarray}
\gdlltadf &=& \sum_{yz}
T\langle\; P_{12}^{+-}(x)\; 
P_{23}^{-+}(0)\;
\left[V_{31}^{++}\right]^\mu(y)\;   
\left[V_{44}^{\pm\pm}\right]^\nu(z)\;
\rangle \; D_{\mu\nu}(y,z) \;.
\end{eqnarray}
All the other diagrams appearing in the text can be translated into the operator language following the correspondence illustrated above.

\end{appendices}

\end{document}